\documentclass[useAMS,usenatbib]{mn2e}
\usepackage{epstopdf,amsmath,amsfonts,amssymb,graphicx,subfigure,enumitem,epstopdf}

\usepackage{natbib}
\bibpunct{(}{)}{;}{a}{,}{,}

\usepackage{graphicx}
\usepackage{helvet}
\usepackage{natbib}
\usepackage{setspace}
\usepackage{amssymb}
\usepackage{longtable}

\title[Variable stars in NGC\,6823]
{Photometric variable stars in the young open cluster NGC\,6823}

\author[Sneh Lata et al.]
       {Sneh Lata$^1$\thanks{E-mail: sneh@aries.res.in}, W.~P. Chen$^{2,3}$, J. C. Pandey$^1$, Athul Dileep$^1$, Zhong-Han Ai$^3$,
\newauthor Alisher S. Hojaev$^4$, Neelam Panwar$^1$, Santosh Joshi$^1$, Soumen Mondal$^5$, 
\newauthor Siddhartha Biswas$^5$, B. C. Bhatt$^6$   \\
       $^1$Aryabhatta Research Institute of Observational Sciences, Manora Peak, Nainital 263002, Uttarakhand, India \\
       $^2$Graduate Institute of Astronomy, National Central University, 300 Zhongda Road, Zhongli 32001 Taoyuan, Taiwan \\
       $^3$Department of Physics, National Central University, 300 Zhongda Road, Zhongli 32001 Taoyuan, Taiwan \\
       $^4$Ulugh Beg Astronomical Institute, Uzbekistan Academy of Sciences, Tashkent, Republic of Uzbekistan \\
       $^5$S. N. Bose National Centre for Basic Sciences, Kolkata 700106, India\\
       $^6$Indian Institute of Astrophysics, Koramangala, Bangalore-560034, India \\
       }

\date{Accepted ---------.
      Received ---------;
      }

\pagerange{\pageref{firstpage}--\pageref{lastpage}}

\def\LaTeX{L\kern-.36em\raise.3ex\hbox{a}\kern-.15em
    T\kern-.1667em\lower.7ex\hbox{E}\kern-.125emX}

\begin{document}

\label{firstpage}

\maketitle

\label{firstpage}
\begin{abstract}
We present stellar variability towards the young open cluster NGC\,6823. Time
series $V$- and $I$-band CCD photometry led to identification and characterization 
	of 88 variable stars, of which only 14 have been previously recognized.  We ascertain 
the membership of each variable with optical $UBVI$ and infrared photometry, and 
with Gaia EDR3 parallax and proper motion data.  Seventy two variable stars are 
found to be cluster members, of which 25 are main sequence stars and 48 are 
pre-main-sequence stars. The probable cluster members collectively suggest
an isochrone age of the cluster to be about 2~Myrs based on the GAIA photometry. 
With the color and magnitude, as well as the shape of the light curve, we have classified
the main sequence variables into $\beta$~Cep, $\delta$~Scuti, slowly pulsating B type,
and new class variables. Among the pre-main-sequence variables, eight are classical
T Tauri variables, and four are Herbig Ae/Be objects, whereas the remaining belong to the 
weak-lined T Tauri population.  The variable nature of 32 stars is validated with TESS light
curves.  Our work provides refined classification of variability of pre-main-sequence and
main-sequence cluster members of the active star-forming complex, Sharpless\,86.
Despite no strong evidence of the disk-locking mechanism in the present sample of TTSs, one TTS with larger $\Delta(I-K)$ 
is found to be slow rotator. 
\end{abstract}

\begin {keywords} 
open clusters and associations: individual NGC 6823, Hertzsprung-Russell and color-magnitude diagram, stars: pre-main-sequence, 
stars: variables: T Tauri, Herbig Ae/Be
\end {keywords}

\section{Introduction}

Young open clusters serve as useful tools for the studies of the star formation mechanism and early stellar evolution.  
For example, young star clusters are used to trace the Galactic spiral structure.  In particular, variability of 
young stellar members provides diagnostics on the sporadic (accretion or occultation) or periodical (rotation) properties of the stars, and of their relation to the circumstellar environments (Morales-Calderon et al. 2011).

Pre-main-sequence (PMS) objects are categorized on the basis of the spectral energy distribution in the infrared wavelengths: Class\,0 , Class\,I, Class\,II, and Class\,III (Lada 1987; Andre et al. 1993) with the classification sequence roughly corresponding to the evolutionary status.  Namely, a Class\,0 object signifies a clump of dust and gas heavily enshrouded in the molecular envelope, and is detected only in far-infrared wavelengths or longer.  A Class\,I object is more evolved, now emerging from the cloud to become visible in near- and mid-infrared.   A Class\,I object is in the protostellar stage and derives the luminosity from mass accretion. 

A Class\,II object, corresponding to a classical T~Tauri star (TTS), has dispersed much of the envelope of gas and dust but retains a circumstellar disk within which planets may condense or are being formed.  Inside the optically thick but geometrically thin disk, the dust grains absorb the starlight and re-emit in the infrared, manifest as infrared excess seen typically in a classical TTS.  Accretion from the disk onto the star, while matter is partly lost as bipolar jets/outflows, leads to strong emission lines in the spectrum.  As the inner disk is dissipated (or going into planet formation), the PMS object then evolves to Class\,III, now with negligible infrared excess and with weak emission lines, if any, due to surface chromospheric activity.  A Class\,III object hence is called a weak-lined TTS (Joy 1945, Appenzeller \& Mundt 1989).   Variability of PMS objects hence serves as an important diagnosis to understand the earliest PMS stellar evolution, e.g., the accretion (Johnstone et al. 2018), rotation (Herbst et al. 1994), or dust properties (Huang et al. 2019).  

Here we report the variability study of the Galactic young open cluster NGC\,6823.  At a distance of about 2~kpc, the cluster is associated with the prominent H\,II region, Sharpless\,86.  This cluster has been investigated by several authors (Turner 1979, Stone 1988, Bica et al. 2008, Sagar and Joshi 1981; Guetter 1992; Massey et al. 1995; 
Pigulski et al. 2000; Hojaev et al. 2003, Zahajkiewicz 2012). Using optical and $JHK$ photometric observations Riaz et al. (2012) found a large population of young stellar sources in the region, including two $\delta$\,Scuti variables of PMS nature, and 13 other variables such as eclipsing binaries, slowly pulsating B candidates and UX Ori type variables.  In the line of sight to the cluster the reddening has been found to be from 0.7 to 1.1~mag following a normal reddening law (Rangwal et al. 2017).  The aim of the present work is to identify variables in a relatively large field of $\sim 14\arcmin \times 14\arcmin$ of the member versus nonmember variable stellar populations in the region.  Particularly, photometric rotation periods of PMS members are derived to add to the data inventory for the study of the angular momentum evolution of low-mass stars.      

We describe in Section~2 observations, data reduction procedure, detection of variables, and period determinations. In Section~3, membership of the identified variable candidates is discussed using Gaia proper motion data, photometric two-color diagrams (TCDs) and color-magnitude diagrams (CMDs). Section~4 then presents the nature of known and newly identified variable stars, while Section~5 deals with TESS light curves. We discuss correlation between amplitude and rotation periods of TTSs along with their color excess in Section 6. The results are summarized in Section~7.   
\begin{figure}
\includegraphics[width=9cm]{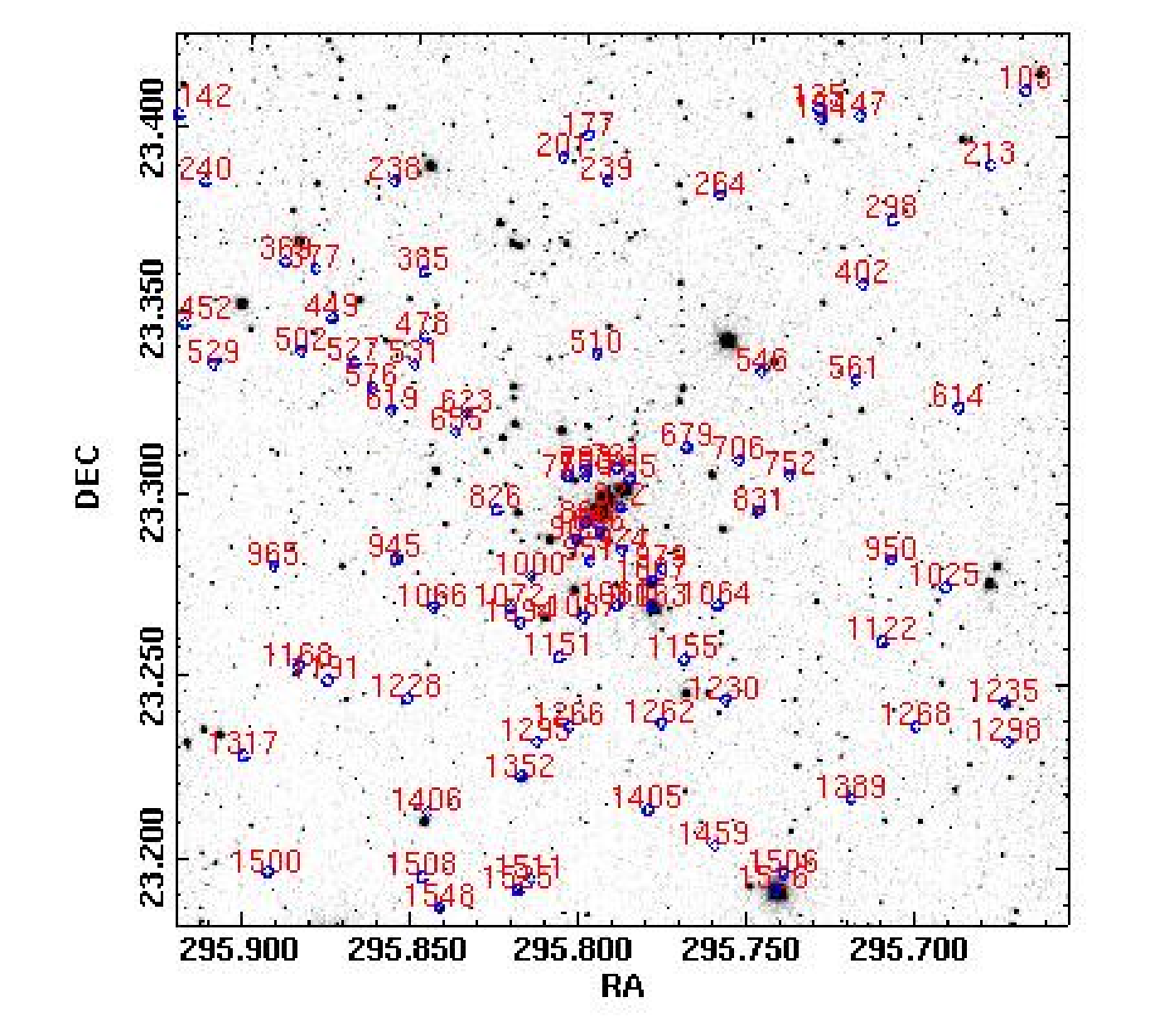}
\caption{ The observed region of open cluster NGC\,6823. Each variable star identified in this work is encircled.
 }
 \label{fig:findchart}
\end{figure}
\begin{figure}
\includegraphics[width=8cm]{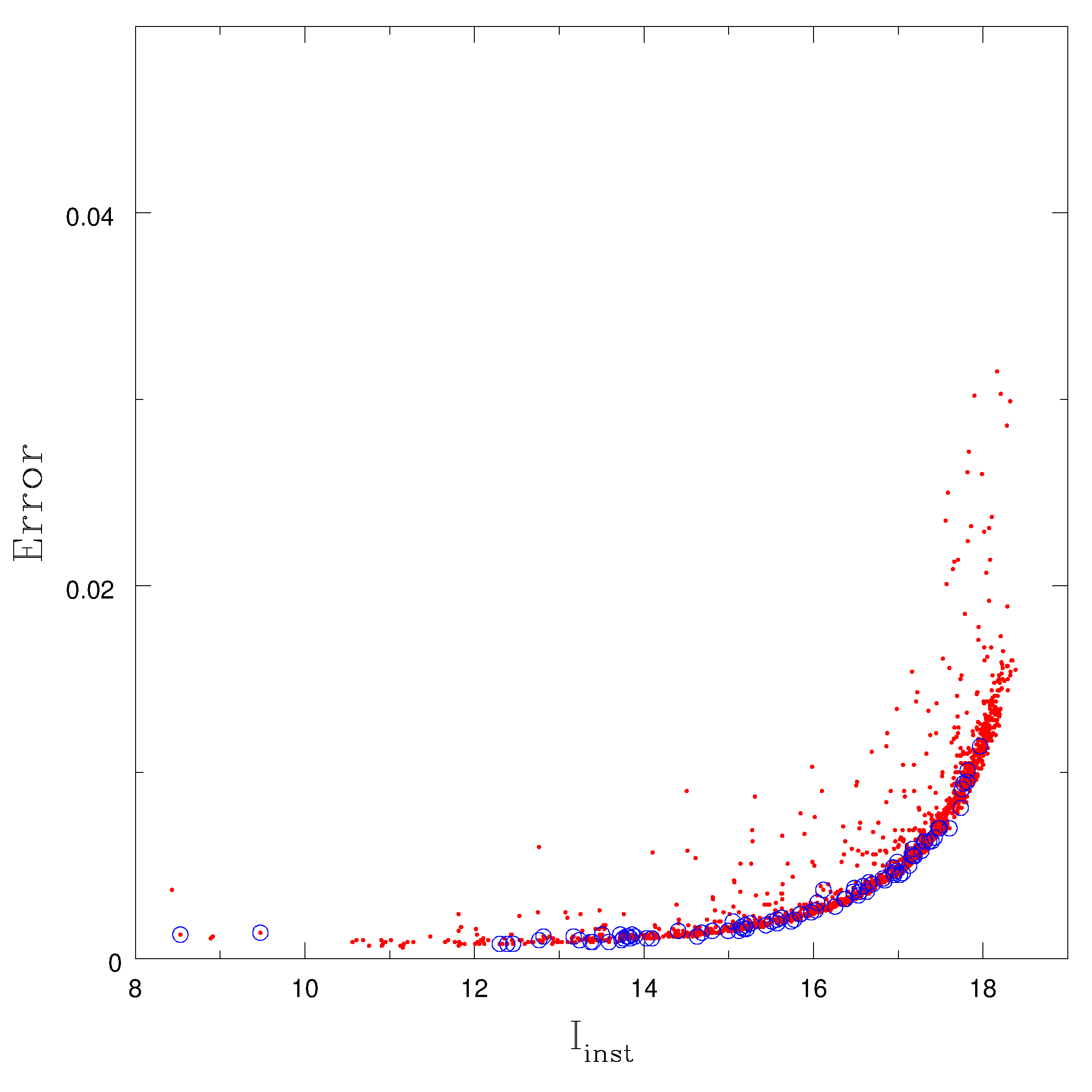}
\caption{Photometric errors as a function of instrumental magnitude in $I$ band.  Open circles represent 
             the variables stars identified in the present work.}
\end{figure}

\begin{figure}
\includegraphics[width=9.cm, height=9cm]{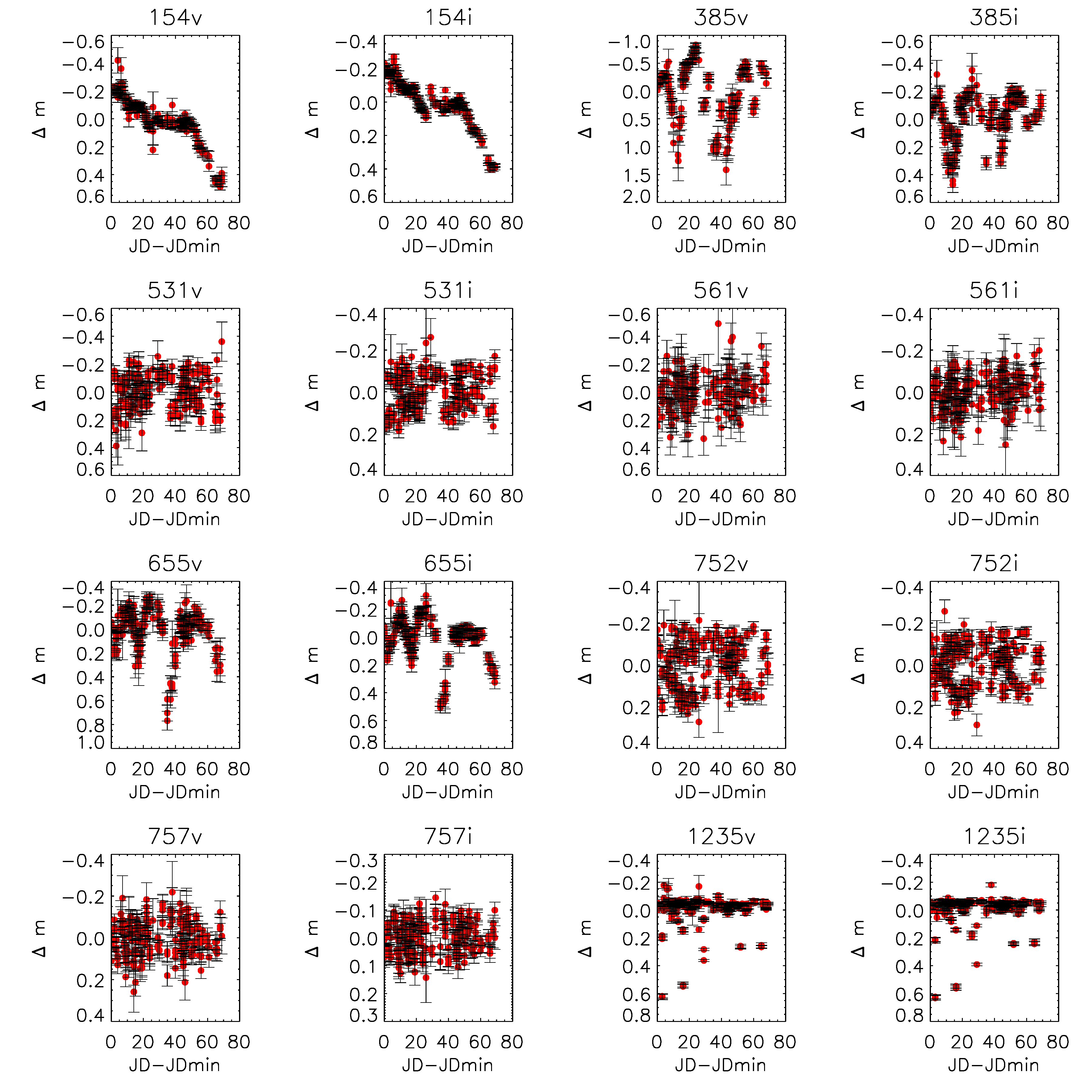}
\includegraphics[width=9.cm, height=9cm]{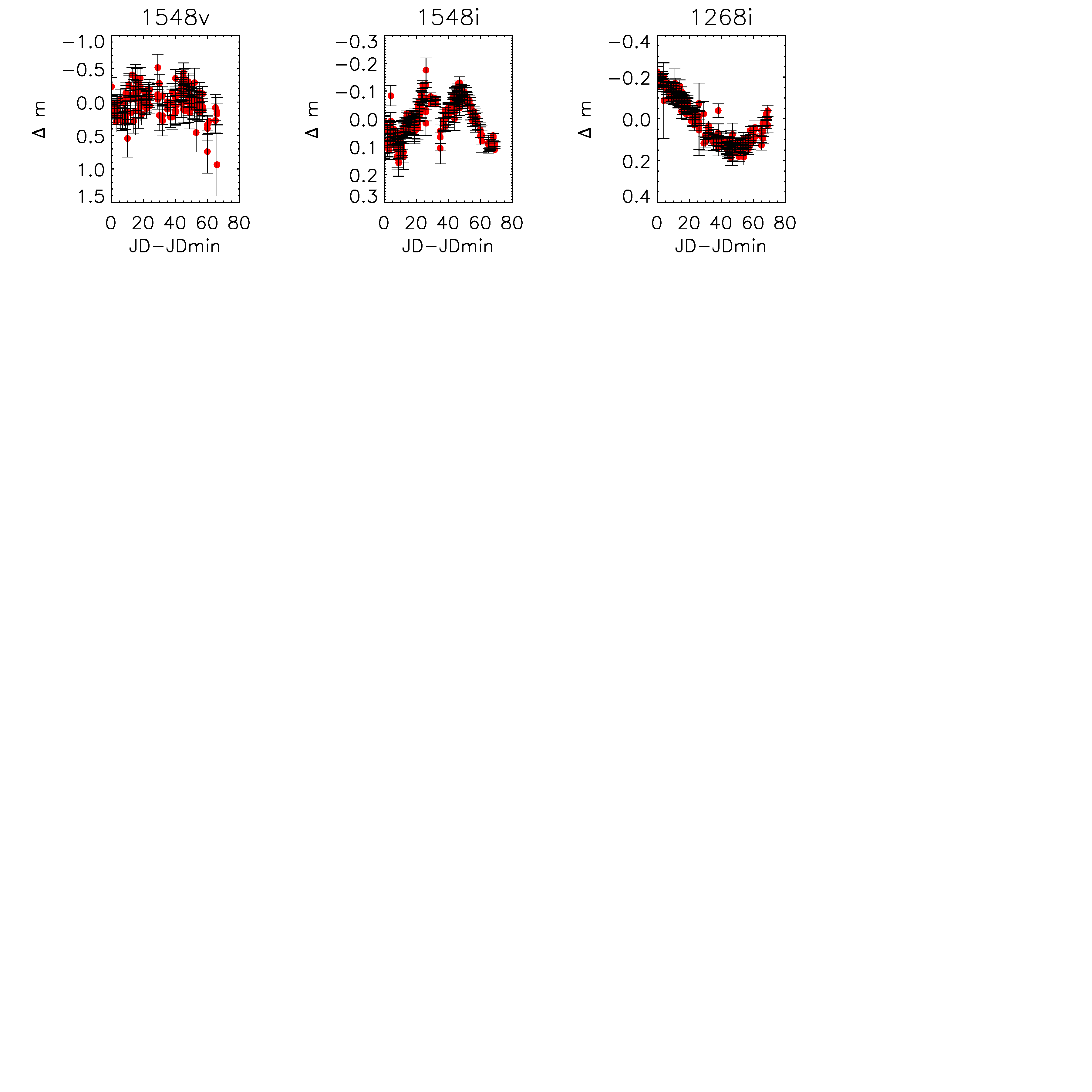}
\vspace{-7.0cm}
\caption{The $V$ and $I$ band sample light curves of a few variables identified in the
present work where $\Delta$m represents the differential magnitude.}
\end{figure}

\begin{table}
\tiny
\caption{Log of the observations of NGC 6823. N and Exp. represent number of frames obtained and exposure time in seconds, respectively.} 
\begin{tabular}{llll}
\hline
S. No.&Date of        &{\it I}                &{\it V}                               \\
      &Observations   &(N$\times$Exp.)        &(N$\times$Exp.) \\
\hline
1 &13 oct 2012 &6$\times$ 30 & 6 $\times$  50 \\
 2 &15 oct 2012 &3$\times$ 30 & 3 $\times$  50 \\
 3 &16 oct 2012 &6$\times$ 30 & 5 $\times$  50 \\
 4 &17 oct 2012 &6$\times$ 30 & 6 $\times$  50 \\
 5 &19 oct 2012 &3$\times$ 30 & 5 $\times$  50 \\
 6 &20 oct 2012 &6$\times$ 30 & 6 $\times$  50 \\
 7 &21 oct 2012 &6$\times$ 30 & 4 $\times$  50 \\
 8 &22 oct 2012 &6$\times$ 30 & 6 $\times$  50 \\
 9 &23 oct 2012 &3$\times$ 30 & 2 $\times$  50 \\
 10& 24 oct 2012 &6$\times$ 30 & 6 $\times$  50 \\
 11& 25 oct 2012 &6$\times$ 30 & 5 $\times$  50 \\
 12& 26 oct 2012 &5$\times$ 30 & 5 $\times$  50 \\
 13& 27 oct 2012 &6$\times$ 30 & 6 $\times$  50 \\
 14& 28 oct 2012 &6$\times$ 30 & 6 $\times$  50 \\
 15& 29 oct 2012 &6$\times$ 30 & 6 $\times$  50 \\
 16& 30 oct 2012 &6$\times$ 30 & 6 $\times$  50 \\
 17& 31 oct 2012 &6$\times$ 30 & 6 $\times$  50 \\
 18& 01  nov 2012 &6$\times$ 30 & 4 $\times$  50 \\
 19& 02  nov 2012 &-            & 2 $\times$  50 \\
 20& 03  nov 2012 &6$\times$ 30 & 5 $\times$  50 \\
 21& 04  nov 2012 &6$\times$ 30 & 5 $\times$  50 \\
 22& 05  nov 2012 &5$\times$ 30 & 5 $\times$  50 \\
 23& 06  nov 2012 &6$\times$ 30 & 6 $\times$  50 \\
 24& 08  nov 2012 &6$\times$ 30 & 6 $\times$  50 \\
 25& 11 nov 2012 &5$\times$ 30 & 5 $\times$  50 \\
 26& 12 nov 2012 &2$\times$ 30 & 3 $\times$  50 \\
 27& 14 nov 2012 &6$\times$ 30 & 6 $\times$  50 \\
 28& 17 nov 2012 &3$\times$ 30 & 3 $\times$  50 \\
 29& 19 nov 2012 &3$\times$ 30 & 3 $\times$  50 \\
 30& 20 nov 2012 &6$\times$ 30 & 6 $\times$  50 \\
 31& 22 nov 2012 &6$\times$ 30 & 6 $\times$  50 \\
 32& 25 nov 2012 &6$\times$ 30 & 6 $\times$  50 \\
 33& 27 nov 2012 &6$\times$ 30 & 6 $\times$  50 \\
 34& 28 nov 2012 &6$\times$ 30 & 6 $\times$  50 \\
 35& 29 nov 2012 &5$\times$ 30 & 3 $\times$  50 \\
 36& 30 nov 2012 &6$\times$ 30 & 6 $\times$  50 \\
 37& 01 dec 2012  &3$\times$ 30 & 3 $\times$  50 \\
 38& 02 dec 2012  &3$\times$ 30 & 3 $\times$  50 \\
 39& 03 dec 2012 &3$\times$ 30 & 3 $\times$  50 \\
 40& 04 dec 2012 &3$\times$ 30 & 3 $\times$  50 \\
 41& 05 dec 2012  &3$\times$ 30 & 3 $\times$  50 \\
 42& 06 dec 2012 &3$\times$ 30 & 3 $\times$  50 \\
 43& 07 dec 2012 &3$\times$ 30 & 3 $\times$  50 \\
 44& 08 dec 2012 &3$\times$ 30 & 3 $\times$  50 \\
 45& 09 dec 2012 &3$\times$ 30 & 3 $\times$  50 \\
 46& 10 dec 2012 &3$\times$ 30 & 1 $\times$  50 \\
 47& 11 dec 2012 &2$\times$ 30 & 2 $\times$  50 \\
 48& 12 dec 2012 &3$\times$ 30 & 3 $\times$  50 \\
 49& 13 dec 2012 &3$\times$ 30 & 3 $\times$  50 \\
 50& 17 dec 2012 &3$\times$ 30 & 3 $\times$  50 \\
 51& 18 dec 2012 &3$\times$ 30 & 3 $\times$  50 \\
 52& 20 dec 2012 &3$\times$ 30 & 3 $\times$  50 \\
 53& 21 dec 2012 &3$\times$ 30 & 3 $\times$  50 \\
 54& 26 dec 2012 &6$\times$ 30 & -              \\
\hline                                          
\end{tabular}                                   
\end{table}

\begin{table*}
\caption{The $VI$ and $JHK$ 2mass data, amplitude and period of variables identified towards NGC\,6823. 
	The 2MASS data were obtained from the 2mass catalog (Cutri et al. 2003). The first column with 
	an asterisk symbol represents a known variable. 
}
\tiny
\begin{tabular}{llllllll}
\hline
 ID &       RA     &          Dec&                    V&               $V-I$&      $J$    &             $H$  &   $K$           \\ 
    &              &             &                (mag)&               (mag)&     (mag)   &            (mag) &  (mag)                    \\
\hline
103&    295.668444&   23.412417 &  18.115$\pm$0.101 &   2.190$\pm$0.048 &14.231$\pm$0.041 & 13.140$\pm$0.036 & 12.508$\pm$0.034      \\
135&    295.730111&   23.406833 &  18.454$\pm$0.124 &  2.428$\pm$0.054 &14.324$\pm$0.056 & 12.758$\pm$0.081 & 11.586$\pm$0.044 \\
142&    295.921528&   23.403194 &                 - &      -           & 9.997$\pm$0.022 &  8.072$\pm$0.034 &  7.137$\pm$0.020      \\ 
147&    295.717833&   23.404972 &  18.251$\pm$0.102 &  2.442$\pm$0.043 &13.925$\pm$0.029 & 12.915$\pm$0.037 & 12.452$\pm$0.031     \\
154&    295.729444&   23.404194 &  15.305$\pm$0.021 &  2.514$\pm$0.015 &10.509$\pm$0.021 &  9.561$\pm$0.030 &  8.621$\pm$0.024     \\
177&    295.798972&   23.398806 &                 - &      -           &14.623$\pm$0.040 & 13.622$\pm$0.040 & 13.154$\pm$0.036   \\
201&    295.806361&   23.392806 &  18.651$\pm$0.129 &  1.686$\pm$    - &               - &                - &                -     \\
213&    295.678611&   23.391889 &  18.583$\pm$0.133 &  1.531$\pm$0.099 &15.949$\pm$0.093 & 15.302$\pm$0.128 & 14.913$\pm$0.129     \\
238&    295.856583&   23.385806 &                 - &      -           & 9.096$\pm$0.022 &  7.248$\pm$0.033 &  6.386$\pm$0.027     \\
239&    295.793083&   23.386500 &  17.564$\pm$0.063 &  2.161$\pm$0.036 &13.328$\pm$    - & 12.411$\pm$0.044 & 11.853$\pm$0.034      \\
240&    295.913278&   23.384861 &  16.320$\pm$0.033 &  1.113$\pm$0.034 &14.480$\pm$0.035 & 14.178$\pm$0.051 & 14.074$\pm$0.052  \\
264&    295.759250&   23.382944 &                 - &      -           &14.953$\pm$0.040 & 14.097$\pm$0.043 & 13.771$\pm$0.045   \\
298&    295.707806&   23.376667 &  17.367$\pm$0.056 &  1.772$\pm$0.040 &14.216$\pm$0.034 & 13.319$\pm$0.040 & 12.605$\pm$0.030     \\
369&    295.889056&   23.363222 &                 - &      -           &13.520$\pm$0.026 & 12.672$\pm$0.033 & 12.243$\pm$0.031      \\
377&    295.880000&   23.361417 &  18.610$\pm$0.127 &  2.027$\pm$    - &                -&                 -&                -     \\
385&    295.847556&   23.360861 &  16.904$\pm$0.040 &  1.763$\pm$0.029 &13.554$\pm$0.028 & 12.719$\pm$0.033 & 12.086$\pm$0.027      \\
402&    295.716611&   23.358944 &                 - &      -           &15.213$\pm$0.058 & 14.159$\pm$0.063 & 13.558$\pm$0.056   \\
449&    295.874806&   23.347806 &  14.985$\pm$0.017 &  1.571$\pm$0.017 &12.262$\pm$0.023 & 11.665$\pm$0.028 & 11.521$\pm$0.024      \\
452&    295.918972&   23.345917 &                 - &      -           &14.529$\pm$0.056 & 13.475$\pm$0.055 & 13.008$\pm$0.043   \\
478&    295.847167&   23.343222 &                 - &      -           &14.474$\pm$0.042 & 13.418$\pm$0.046 & 13.002$\pm$0.043      \\
502&    295.883889&   23.339000 &  18.593$\pm$0.122 &  2.086$\pm$0.067 &15.050$\pm$0.051 & 14.019$\pm$0.053 & 13.695$\pm$0.055   \\
510&    295.795694&   23.339139 &  14.154$\pm$0.014 &  1.061$\pm$0.020 &12.211$\pm$0.021 & 11.957$\pm$0.031 & 11.770$\pm$0.027    \\
527&    295.868056&   23.335778 &  17.461$\pm$0.055 &  2.126$\pm$0.032 &13.688$\pm$0.037 & 12.840$\pm$0.040 & 12.605$\pm$0.038     \\
529&    295.910194&   23.335083 &                 - &      -           &14.243$\pm$0.039 & 12.844$\pm$0.040 & 11.984$\pm$0.033   \\
531&    295.850028&   23.335583 &  17.400$\pm$0.055 &  2.102$\pm$0.031 &13.562$\pm$0.032 & 12.740$\pm$0.036 & 12.460$\pm$0.031     \\
546&    295.746583&   23.334750 &  17.928$\pm$0.082 &  1.702$\pm$0.060 &14.461$\pm$    - & 13.676$\pm$    - & 14.144$\pm$0.102     \\
561&    295.718194&   23.332694 &  18.193$\pm$0.100 &  2.100$\pm$0.051 &14.652$\pm$    - & 13.775$\pm$    - & 13.376$\pm$0.041   \\
576&    295.862722&   23.329111 &                 - &      -           &14.668$\pm$0.042 & 13.423$\pm$0.039 & 12.865$\pm$0.034    \\
614&    295.687444&   23.325333 &  15.851$\pm$0.023 &  1.277$\pm$0.023 &13.709$\pm$0.028 & 13.328$\pm$0.041 & 13.109$\pm$0.036    \\
619&    295.856806&   23.323000 &  15.920$\pm$0.022 &  1.721$\pm$0.018 &12.883$\pm$0.024 & 12.309$\pm$0.031 & 12.076$\pm$0.026     \\
623*&   295.834472&   23.322306 &  13.173$\pm$0.009 &  1.202$\pm$0.012 &11.141$\pm$0.019 & 10.662$\pm$0.030 & 10.519$\pm$0.024     \\
655*&   295.837528&   23.317306 &  17.282$\pm$0.052 &  2.287$\pm$0.028 &12.735$\pm$0.025 & 11.397$\pm$0.031 & 10.313$\pm$0.023  \\
679&    295.768611&   23.313583 &  13.725$\pm$0.014 &  0.967$\pm$0.018 &11.556$\pm$0.021 & 10.530$\pm$0.030 &  9.546$\pm$0.024  \\
706&    295.752944&   23.310389 &  17.376$\pm$0.056 &  1.963$\pm$0.036 &13.936$\pm$0.037 & 13.149$\pm$0.046 & 12.880$\pm$0.041 \\
731&    295.789306&   23.307667 &  18.493$\pm$0.128 &  2.239$\pm$0.060 &14.424$\pm$0.043 & 13.458$\pm$0.051 & 13.086$\pm$0.044 \\
733*&   295.798444&   23.307361 &  15.211$\pm$0.018 &  1.224$\pm$0.023 &12.935$\pm$0.029 & 12.515$\pm$0.045 & 12.256$\pm$0.044 \\
752&    295.737667&   23.306472 &  16.172$\pm$0.028 &  1.398$\pm$0.028 &13.743$\pm$0.028 & 13.207$\pm$0.040 & 12.978$\pm$0.037   \\
753*&   295.798472&   23.305694 &                 - &      -           &14.578$\pm$0.048 & 13.491$\pm$0.062 & 12.736$\pm$0.058  \\
757*&   295.803833&   23.305278 &  16.899$\pm$0.042 &  2.077$\pm$0.029 &13.186$\pm$0.028 & 12.420$\pm$0.035 & 12.138$\pm$0.031 \\
765&    295.785417&   23.304806 &  17.592$\pm$0.124 &  1.533$\pm$0.075 &14.421$\pm$    - & 14.039$\pm$0.103 & 13.848$\pm$0.076  \\
822*&   295.787917&   23.297000 &  14.529$\pm$0.015 &  1.225$\pm$0.020 &12.396$\pm$0.026 & 12.024$\pm$0.030 & 11.823$\pm$0.032  \\
826&    295.825139&   23.295944 &                 - &      -           &14.661$\pm$0.044 & 13.740$\pm$0.051 & 13.406$\pm$0.045 \\
831*&   295.746556&   23.296667 &                 - &      -           &11.266$\pm$0.033 &  8.761$\pm$0.030 &  7.326$\pm$0.020    \\
860&    295.798389&   23.292583 &  17.563$\pm$0.121 &  1.880$\pm$0.072 &14.754$\pm$0.055 & 13.748$\pm$0.045 & 13.055$\pm$0.041  \\
886*&   295.794056&   23.290250 &  14.448$\pm$0.015 &  1.065$\pm$0.018 &12.637$\pm$0.022 & 12.379$\pm$0.031 & 12.207$\pm$0.029 \\
903*&   295.801167&   23.288028 &  14.382$\pm$0.014 &  1.036$\pm$0.017 &12.568$\pm$    - & 12.141$\pm$    - & 11.928$\pm$0.033 \\
924*&   295.787583&   23.285444 &  18.146$\pm$0.097 &  2.210$\pm$0.048 &14.154$\pm$    - & 13.363$\pm$    - & 13.035$\pm$0.046 \\
945&    295.855139&   23.282167 &  14.524$\pm$0.014 &  1.206$\pm$0.015 &12.395$\pm$0.033 & 12.051$\pm$0.040 & 11.913$\pm$0.033  \\
950&    295.707250&   23.283611 &                 - &      -           &12.424$\pm$0.022 & 11.060$\pm$0.026 & 10.310$\pm$0.021    \\
951&    295.797167&   23.282278 &   17.150$\pm$0.051 &  2.019$\pm$0.032 &13.664$\pm$0.030 & 12.913$\pm$0.043 & 12.626$\pm$0.038 \\
965&    295.891500&   23.279944 &   14.600$\pm$0.017 &  1.125$\pm$0.018 &12.661$\pm$0.023 & 12.150$\pm$0.031 & 12.050$\pm$0.030   \\
979*&   295.775389&   23.280333 &                 - &      -           &15.201$\pm$0.083 & 14.179$\pm$0.079 & 13.432$\pm$0.057    \\
1000&   295.814639&   23.277861 &  13.574$\pm$0.011 & 0.579 $\pm$0.014 &12.544$\pm$0.021 & 12.472$\pm$0.033  &12.342$\pm$0.032  \\
1007*&  295.778417&   23.277111 &  14.624$\pm$0.016 & 1.181 $\pm$0.020 &12.508$\pm$0.026 & 12.220$\pm$0.039  &12.003$\pm$0.033    \\
1025&   295.690889&   23.276194 &                 - &     -            &15.839$\pm$0.088 & 14.503$\pm$0.058  &13.751$\pm$0.052 \\
1061*&  295.788972&   23.270083 &                  -&     -            &14.267$\pm$0.033 & 12.456$\pm$    -  &12.081$\pm$    -   \\
1063&   295.778528&   23.270139 &   9.702$\pm$0.018  & 0.631$\pm$0.018  &8.785$\pm$0.019 &  8.712$\pm$0.029  & 8.652$\pm$0.024  \\
1064&   295.758861&   23.270389 &  14.451$\pm$0.014  & 1.047$\pm$0.018 &12.514$\pm$0.022 & 12.088$\pm$0.035  &11.874$\pm$0.029  \\
1066&   295.843528&   23.269278 &  16.715$\pm$0.037  & 3.602$\pm$0.014 &10.303$\pm$0.022 &  9.170$\pm$0.028  & 8.693$\pm$0.025  \\
1072&   295.820667&   23.269139 &  14.714$\pm$0.014  & 1.098$\pm$0.017 &12.751$\pm$0.022 & 12.523$\pm$0.033  &12.317$\pm$0.030   \\
1087*&  295.798806&   23.266806 &                 -  &     -            &9.945$\pm$0.019 &  7.952$\pm$0.042  & 7.059$\pm$0.040  \\
1094&   295.817722&   23.265111 &  16.634$\pm$0.034  & 1.949$\pm$0.023 &13.251$\pm$0.022 & 12.572$\pm$0.031  &12.283$\pm$0.029 \\
1122&   295.709611&  23.261028  &  13.788$\pm$0.011  & 0.824$\pm$0.014 &12.308$\pm$0.022 & 12.085$\pm$0.032  &11.956$\pm$0.028   \\
1151&   295.806028&  23.255889  &                 -  &     -           &15.007$\pm$0.059 & 13.956$\pm$0.041  &13.596$\pm$0.046  \\
1155&   295.768556&  23.255472  &  18.407$\pm$0.117  & 2.004$\pm$0.062 &14.928$\pm$0.039 & 14.169$\pm$0.049  &13.918$\pm$0.055 \\
1168&   295.883528&  23.252556  &                 -  &     -           &15.273$\pm$0.057 & 14.244$\pm$0.061  &13.974$\pm$0.059 \\
1191&   295.875000&  23.248667  &                 -  &     -           &15.444$\pm$0.065 & 13.630$\pm$    -  &12.955$\pm$    - \\
1228&   295.851056&  23.243861  &                 -  &     -           &15.146$\pm$0.055 & 13.810$\pm$0.063  &13.109$\pm$0.038 \\
1230&   295.756194&  23.244750  &  14.580$\pm$0.014  & 0.897$\pm$0.017 &12.948$\pm$0.026 & 12.700$\pm$0.036  &12.604$\pm$0.033   \\
1235&   295.672972&  23.244722  &  12.983$\pm$0.012  & 0.937$\pm$0.012 &11.555$\pm$    - & 11.322$\pm$    -  &11.205$\pm$0.037  \\
1262&   295.775250&  23.238000  &  17.173$\pm$0.050  & 2.100$\pm$0.030 &13.351$\pm$0.034 & 12.474$\pm$0.041  &11.998$\pm$0.032 \\
1266&   295.802833&  23.236806  &  18.656$\pm$0.134  & 2.123$\pm$0.069 &14.808$\pm$0.038 & 14.078$\pm$0.049  &13.694$\pm$0.045 \\
1268&   295.699250&  23.237972  &                 -  &     -            &9.148$\pm$0.019 &  7.211$\pm$0.034  & 6.303$\pm$0.026  \\
1295&   295.812361&  23.232472  &                 -  &     -           &14.982$\pm$0.043 & 13.783$\pm$0.040  &12.963$\pm$0.032  \\
1298&   295.671972&  23.234000  &  18.572$\pm$0.133  & 2.430$\pm$0.059 &14.280$\pm$0.049 & 13.287$\pm$0.055  &12.611$\pm$0.042  \\
1317&   295.899889&  23.227722  &                 -  &     -           &15.778$\pm$0.081 & 14.299$\pm$0.063  &13.344$\pm$0.038  \\
1352&   295.816861&  23.222944  &  12.586$\pm$0.011  & 0.693$\pm$0.012 &11.411$\pm$0.022 & 11.168$\pm$0.030  &11.106$\pm$0.026 \\
1389&   295.718444&  23.218000  &  16.432$\pm$0.031  & 1.683$\pm$0.025 &13.25 $\pm$0.027 & 12.387$\pm$0.034  &11.659$\pm$0.027  \\
1405&   295.778889&  23.214194  &  18.752$\pm$0.142  & 1.817$\pm$0.090 &15.644$\pm$0.069 & 14.713$\pm$0.071  &14.516$\pm$0.090 \\
1406&   295.844722&  23.213083  &  17.981$\pm$0.088  & 1.589$\pm$0.063 &15.113$\pm$0.122 & 14.442$\pm$0.237  &13.856$\pm$0.283 \\
1459&   295.758944&  23.204833  &                 -  &     -           &15.208$\pm$0.056 & 14.355$\pm$0.055  &13.932$\pm$0.060 \\
1500&   295.892056&  23.195889  &  15.993$\pm$0.033  & 1.365$\pm$0.031 &13.631$\pm$0.038 & 13.176$\pm$0.051  &13.053$\pm$0.042 \\
1506&   295.738167&  23.197333  &  15.499$\pm$0.024  & 1.237$\pm$0.021 &13.540 $\pm$   -  & 13.123$\pm$0.036  &13.088$\pm$   -  \\
1508&   295.846250&  23.195111  &                 -  &     -           &14.502$\pm$0.043 & 13.425$\pm$0.050  &12.769$\pm$0.036 \\
1511&   295.813722&  23.194889  &  16.513$\pm$0.034  & 1.795$\pm$0.026 &13.734$\pm$0.028 & 13.124$\pm$0.037  &12.874$\pm$0.033 \\
1525&   295.817722&  23.191972  &  13.517$\pm$0.015  & 1.170$\pm$0.015 &11.520 $\pm$0.023 & 11.221$\pm$0.028  &11.071$\pm$0.026  \\
1526&   295.740750&  23.192917  &   8.846$\pm$0.024  & 0.721$\pm$0.019  &7.610 $\pm$0.029 &  7.327$\pm$0.036  & 7.256$\pm$0.024 \\
1548&   295.840667&  23.186917  &   18.460$\pm$0.141  & 6.181$\pm$0.018  &8.177$\pm$0.023 &  6.609$\pm$0.017  & 5.875$\pm$0.020 \\
\hline
\end{tabular}
\end{table*}

\section{Observations and Data Reduction}

We have observed NGC\,6823 with the 0.81-m f/7 Ritchey-Chretien Tenagra automated telescope in southern Arizona,  equipped with a $1024\times1024$ pixel SITe camera. Each pixel corresponds to $0.87\arcsec$, which yields a field of view of $\sim 14.8\arcmin \times 14.8\arcmin$.  The observations were carried out from 2012 early October to 2012 December.  
In total, data were acquired on 54 nights in two passbands, with 232 frames in $V$ band and 243 frames in $I$ band, with typical seeing of 2--3\arcsec.  Bias and twilight flats were taken every observing night. The observed region of the cluster in $I$ band is shown in Fig.~\ref{fig:findchart}.  The log of the observations is given in Table~1.

The observed images were processed using standard IRAF tasks: zerocombine, flatcombine and CCDPROC.
We have performed aperture as well as point spread function (PSF) photometry to derive the magnitude of stars. 
The PSF photometry was obtained using program ALLSTAR (Stetson 1987).  To match the stars between different photometric files we used the daomatch routine of DAOPHOT (Stetson 1992) whereas daomaster was used to match the point sources, and to obtain a file having corrected magnitude of stars from all the files. The daomaster program also removes the flux variation of stars in different frames due to exposure time and airmass. This program makes the magnitudes of stars in each photometry file equal to that of reference file by applying a constant value.

We have used the $V$ and $I$ observations of Massey et al. (1995) for conversion of the present instrumental magnitudes to the standard ones.  For this, the mean instrumental magnitudes in $V$ and $I$ bands given by DAOMASTER (Stetson 1992) have been converted into standard ones with the  
following transformation equations.

\begin{eqnarray}
V = v+(-0.042\pm0.001)\times (V-I) + 0.818\pm0.014   \nonumber\\
V-I =(0.982\pm0.004)\times (v-i)+1.185\pm0.002,    \nonumber
\end{eqnarray}
where $v$ and $i$ are the instrumental magnitudes, and $V$ and $I$ refer to the standard magnitudes of stars in $V$ and $I$ filters. 
The estimated photometric error as a function of the mean instrumental magnitude is shown in Fig.~2.


\subsection{Variables identification}

To identify variable stars, we first produced the light curves of all the stars cross-matched in different CCD frames. The light curves were obtained by plotting the differential magnitudes ($\Delta$m) of stars (variable minus the comparison star) against the given Julian date (JD). 
We used the Lomb-Scargle periodogram (Lomb 1976; Scargle 1982) to derive the periods and produced phased light curves accordingly to ascertain their most probable periods. A few variables seem to show periodic variability but their periodic nature was not obvious in their observed light curves. The phased light curves of all stars were inspected, and we adopted the period value which produces the most consistent phased light curve. The light curves of a few variables are shown in Fig.~3 as examples.  The phased light curves of variables identified in both $V$ and $I$ bands are presented in Fig.~4 and Fig.~5, whereas Fig.~6 shows variables identified in the $I$ band only. 

By eye inspection and periodogram analysis, we have detected 88 variables.
We have listed optical and near-infrared (NIR) data of the variable stars in Table~2, including an identification number, coordinates, and optical as well as NIR photometric data.  These are the star ID numbers labelled in Fig.~1.
  These 88 variable stars include 14 known variables, with periods varying from $\sim$0.03~days to more than 60~days.  
We have plotted in Fig.~7 the root mean square (RMS) scatter of each star to confirm their variability.  The observed RMS scatter includes both the intrinsic variability and the mean photometric error.  The larger circles in Fig.~7 show the variables identified in the present work, indicating large RMS values for variables. Some stars have large RMS values but do not show noticeable brightness variation. Some of these objects are found to be close to the edge of the detector whereas a few stars contains spurious data points. 
The derived periods of stars are given in Table 2.

\section{Cluster membership of variable stars}

For each variable star, its $UBVI$ plus 2MASS photometry along with Gaia EDR3 proper motion and parallax have been used to assess the likelihood of cluster membership.  The $UBV$, $JHK$, and mid infrared (MIR) data at wavelengths 3.6, 4.5, 5.8 and 8 micron, are taken from Massey et al. (1995), Cutri et al. (2003), and GLIMPSE survey, respectively.

\subsection{Gaia Characterization of the Variable Stars}

The 88 variable stars reported in this work have been characterized with the Gaia EDR3 parallax and proper motion measurements. Fig.~\ref{fig:rade} plots the sky positions of all the Gaia sources (in gray) within $30\arcmin$ toward NGC\,6823.  This covers the field of the Tenagra images (variable stars marked in black crosses) and is much wider than the cluster's angular size of $\sim3\farcm6$ (red circle) (Morales et al. 2013).  The stellar density is clearly enhanced toward the center.  

Each variable star was matched with Gaia counterparts within a radius of $2\farcs5$ as the compromise of the seeing of the Tenagra images, leading to 91 Gaia sources.  Fig.~\ref{fig:pm} presents the proper motion vector plot of all the stars (gray) and those within $4\arcmin$ nominal cluster region (black small circles, 1294 stars) for which the members should be concentrated, serving as the sample of cluster members.  This $4\arcmin$ (positional) sample has a mean of $\mu_{\alpha}\approx -1.7$~mas~yr$^{-1}$ and $\mu_{\delta} \approx -5.3$~mas~yr${-1}$, which agrees well with the literature values (Cantat-Gaudin \& Anders 2020).  Shown in the bottom panel are the proper motions for variable stars (in black with error bars).  One sees that the majority of our variable stars share the same proper motion ranges.  

Gaia measures repeatedly the astrometry of a source from which the parallax and proper motion are solved simultaneously.  Parallax, however, does not serve as a constraint for membership as stringently as the proper motion, because given the uncertainties, negative average values may result, rendering the reciprocal to estimate the distance possible only if a statistical inference is exercised (Bailer-Jones et al. 2021).  For our work the parallax value was used directly.  The parallax of the $4\arcmin$ sample exhibits a peak around 0.45~mas, indeed consistent with the literature value (Cantat-Gaudin \& Anders 2020), and so does the variable star sample, as demonstrated in Fig.~\ref{fig:plx}.  If the $4\arcmin$ sample is further divided by proper motion ranges, one finds no star within 1--2~mas~yr$^{-1}$ from the cluster's mean having parallax between 0.4~mas and 0.6~mas. This signifies the sufficiency as membership criteria of (1)~a radius of 1~mas~yr$^{-1}$ in the proper motion from the cluster average proper motion, and (2)~a parallax value 0.35--0.55~mas.  A variable satisfying both (1) and (2) is therefore considered a ``highly probable'' member, whereas one that fulfills only (1) or (2) is classified as a ``possible'' member.  Table~4 lists information about the  proper motions, parallax, and magnitudes for the 88 variables identified in the present work. 

Fig.~\ref{fig:cmd} shows the Gaia $G$ versus $BP-RP$ CMD for the highly probably members (in red) and possible members (in blue).  Overlapped in the diagram is the PARSEC isochrones of 1, 2, and 4~Myr, respectively, each shifted by a distance modulus of 11.753 (parallax of 0.446~mas) and reddening of $E(B-V)=0.8$ (Sagar \& Joshi 1981) adopting the reddening law of $A_V=3.1\,E(B-V)$, 
$A_G=0.83627\, A_V$, $A_{BP}=1.08337\, A_V$, and $A_{RP}=0.63439\, A_V$.  The highly probable members indicate an age of roughly 2~Myr.  

Three variables have ambiguous Gaia counterparts within the matching radius. Star No.~478 has two possible matches, equally faint thereby with relatively large uncertainties in Gaia data but either one is consistent with being a member. The star was not detected in our $V$ band image and appears  progressively brighter from $I=14.47$~mag, to 2MASS $J=13.42$~mag, $H=13.42$~mag, and $Ks=13.00$~mag. 

Star No.~1063 is the second brightest star in our variable list, with the brightest one (No.~1526) being clearly not a member.  The star also has two Gaia matches, with contrasting brightness ($G=9.72$~mag versus $12.87$~mag).  Given its $V=9.70$~mag, the fainter one, having an outlying parallax of 0.282~mas, is eliminated.  The other counterpart, however, has a negative parallax value with a large uncertainty.  This compromises its membership determination.  Its optical and NIR colors both suggest an early-type star and its position in the CMD suggests a main-sequence member.  

Star No.~1262 has $V=17.173$~mag, 2MASS $J=13.351$~mag, $H=12.474$~mag, and $Ks=11.998$~mag. The brighter Gaia match has $G=16.379$~mag but has a negative parallax value and inconsistent proper motion. The other Gaia star is faint, with $G=19.291$ and no measurements in the other two Gaia bands, has parallax and proper motion values well consistent with being a member. 

\subsection{Colors and Magnitudes} 

\begin{figure*}
\hbox{
\includegraphics[width=9cm, height=9cm]{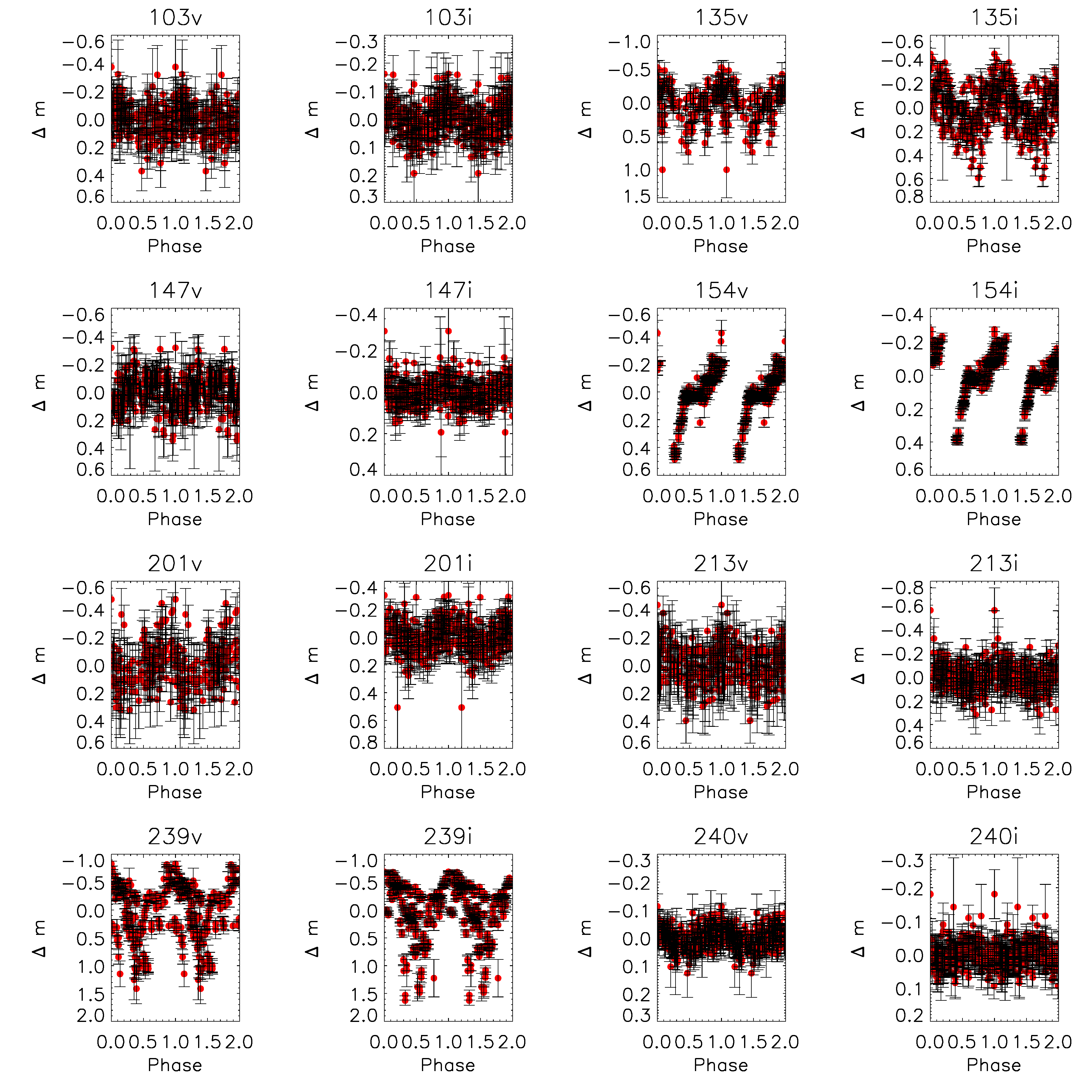}
\includegraphics[width=9cm, height=9cm]{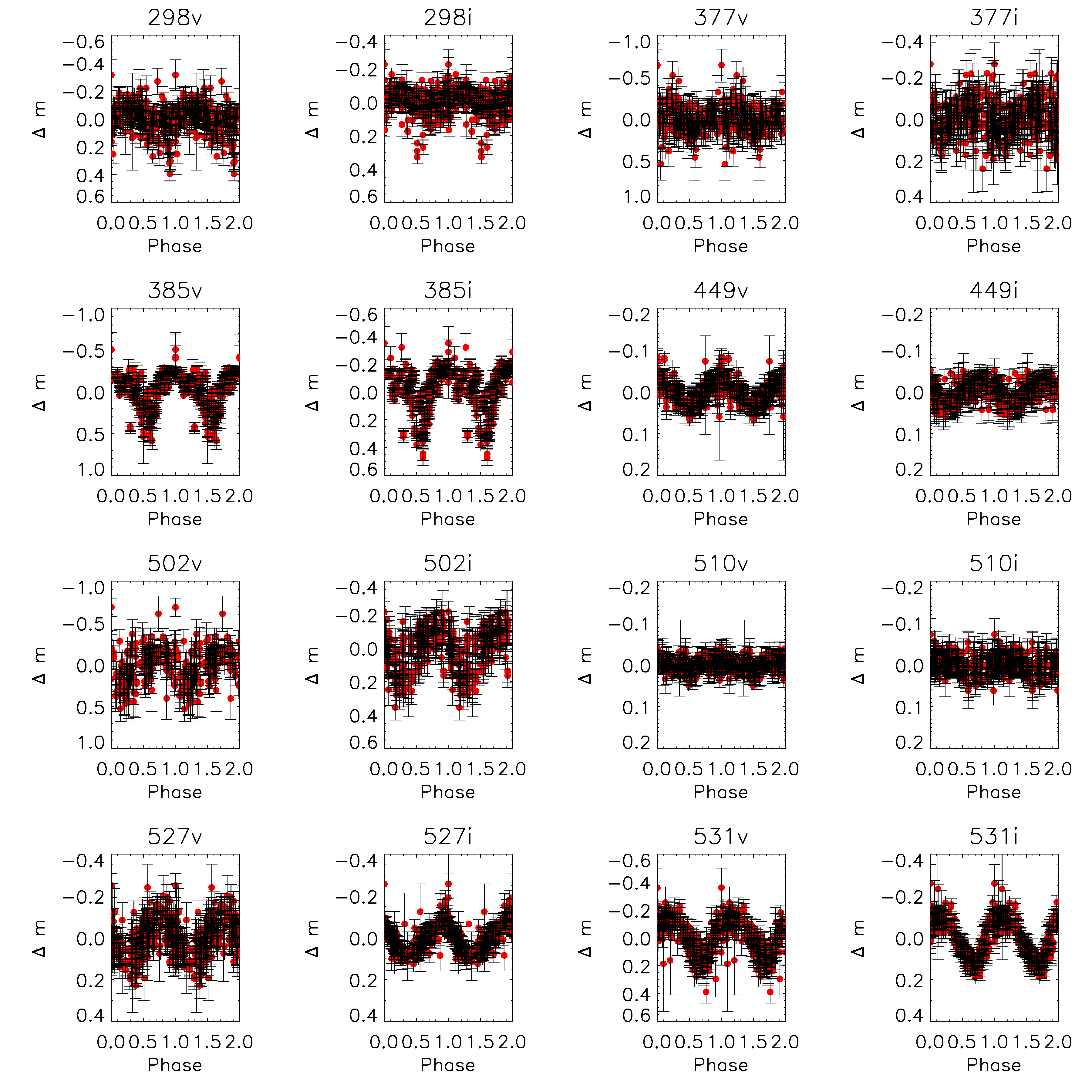}
}
\hbox{
\includegraphics[width=9cm, height=9cm]{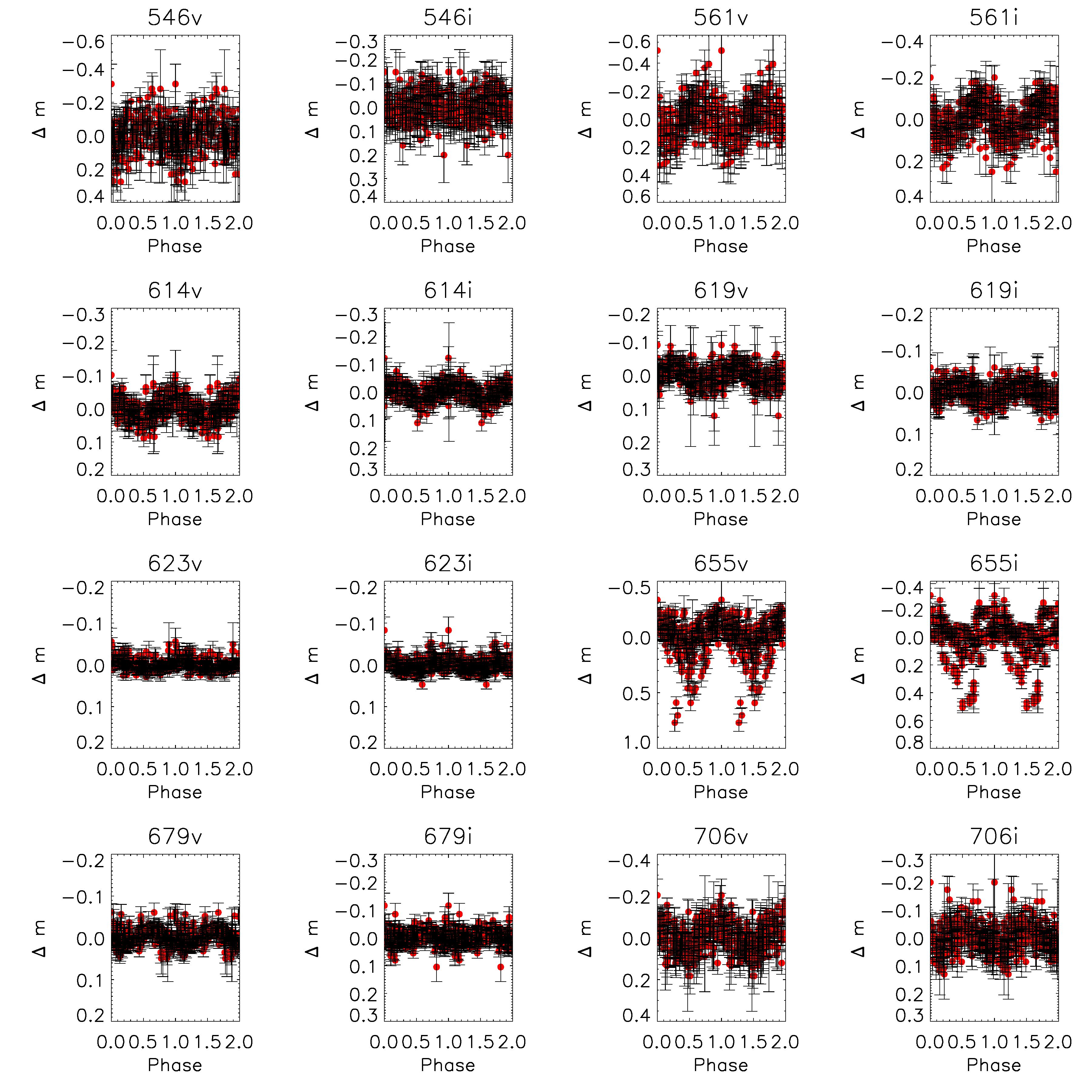}
\includegraphics[width=9cm, height=9cm]{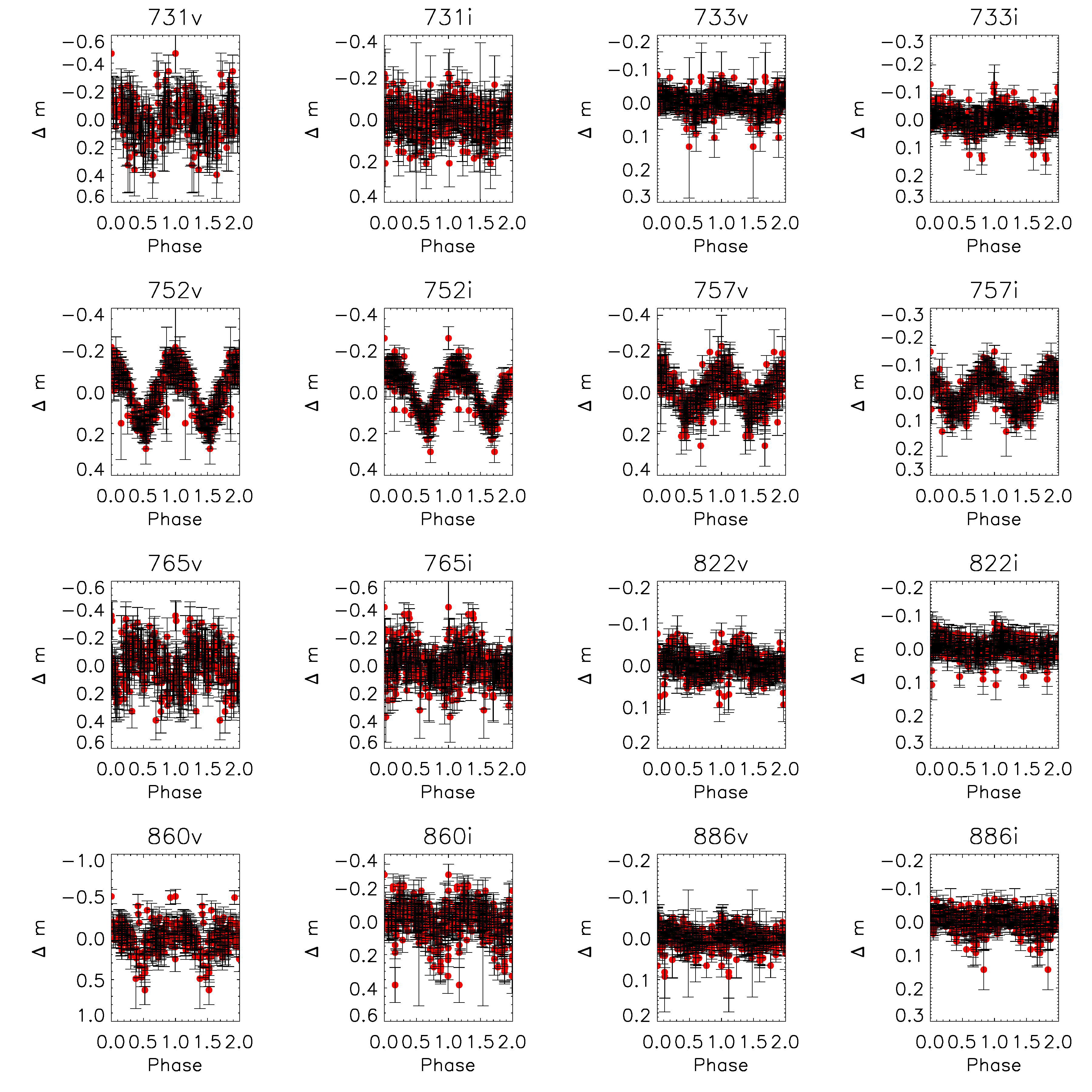}
}
\caption{The $I$ and $V$ band phased light curves of variable stars identified in the region of NGC 6823. }
\end{figure*}

\begin{figure*}
\hbox{
\includegraphics[width=9cm, height=9cm]{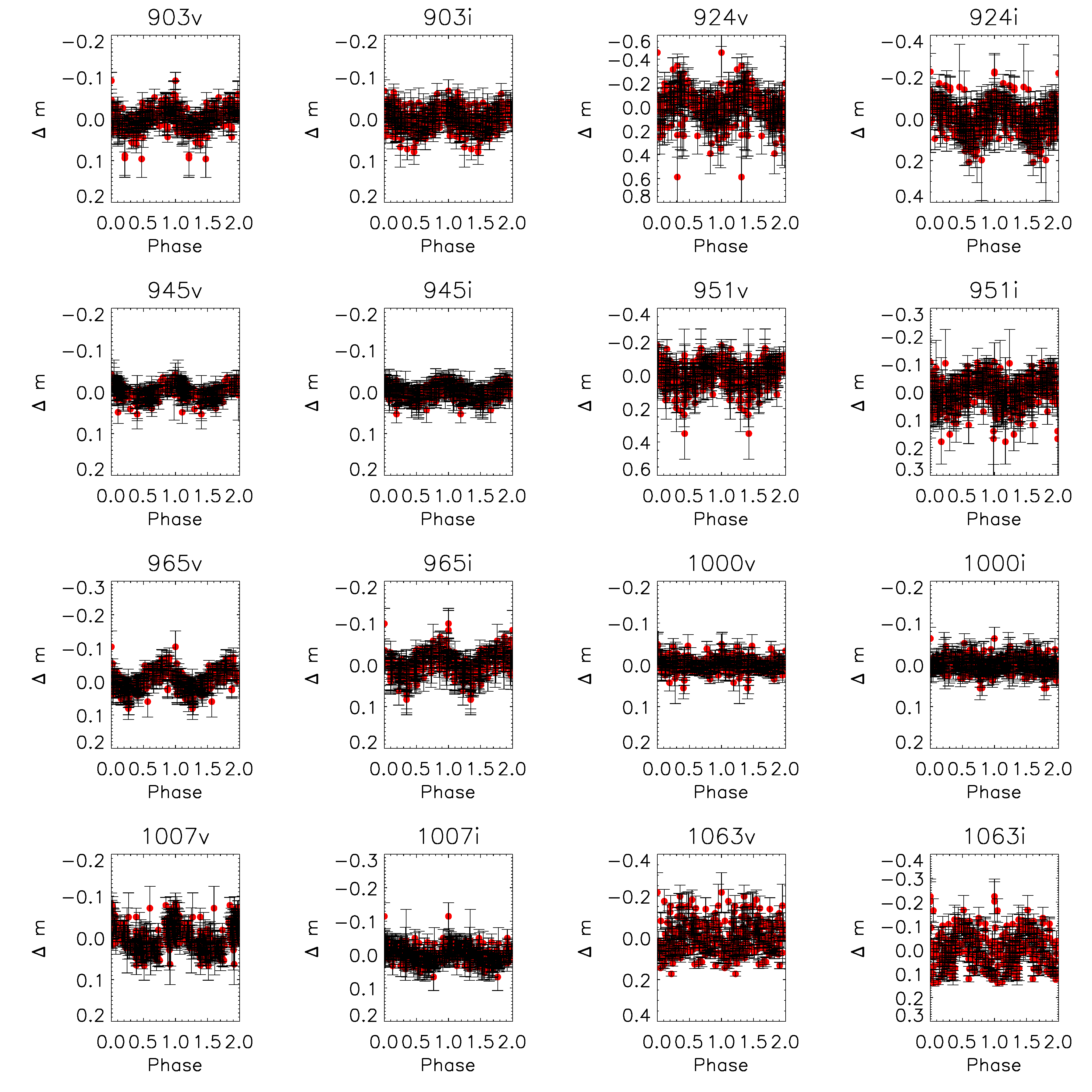}
\includegraphics[width=9cm, height=9cm]{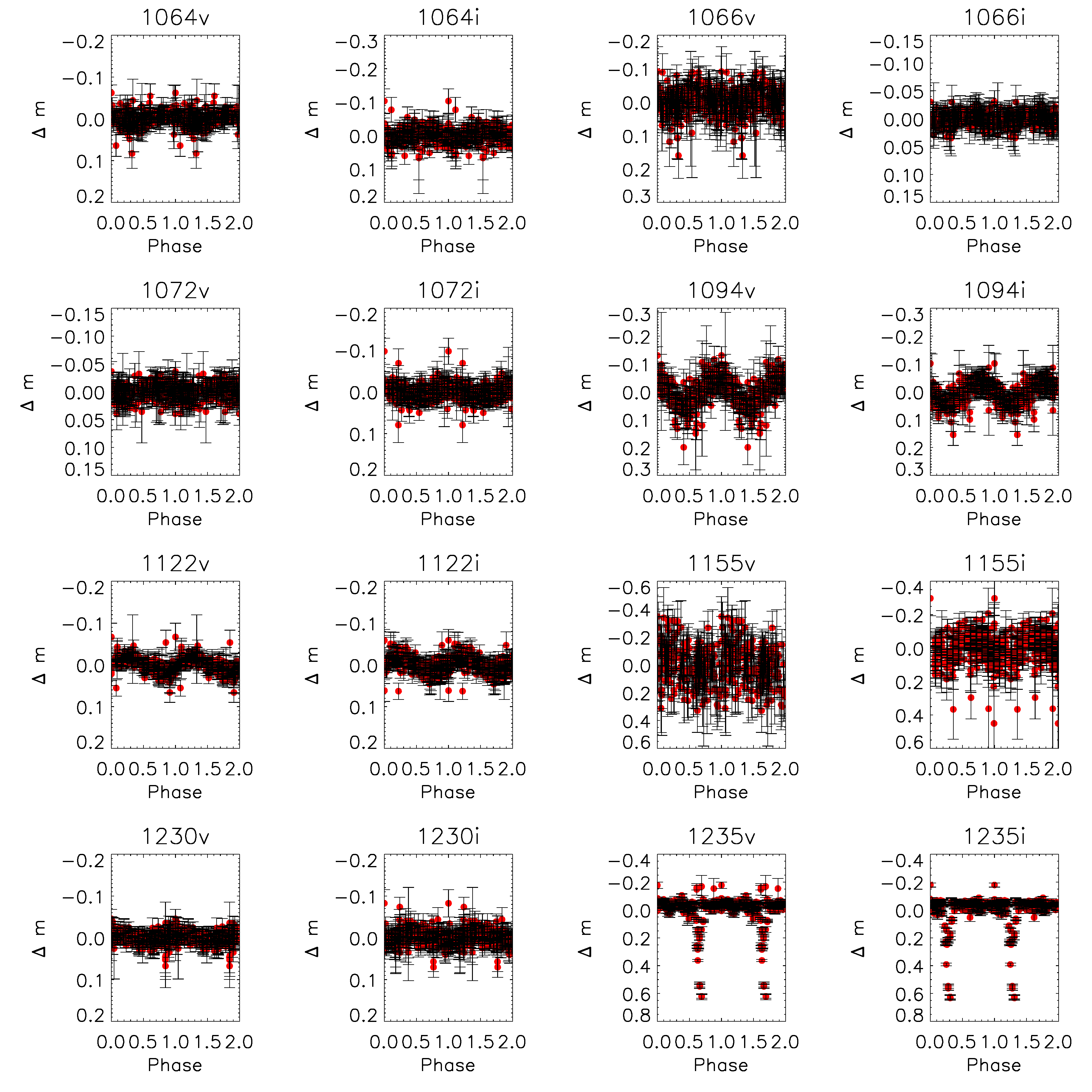}
}
\hbox{
\includegraphics[width=9cm, height=9cm]{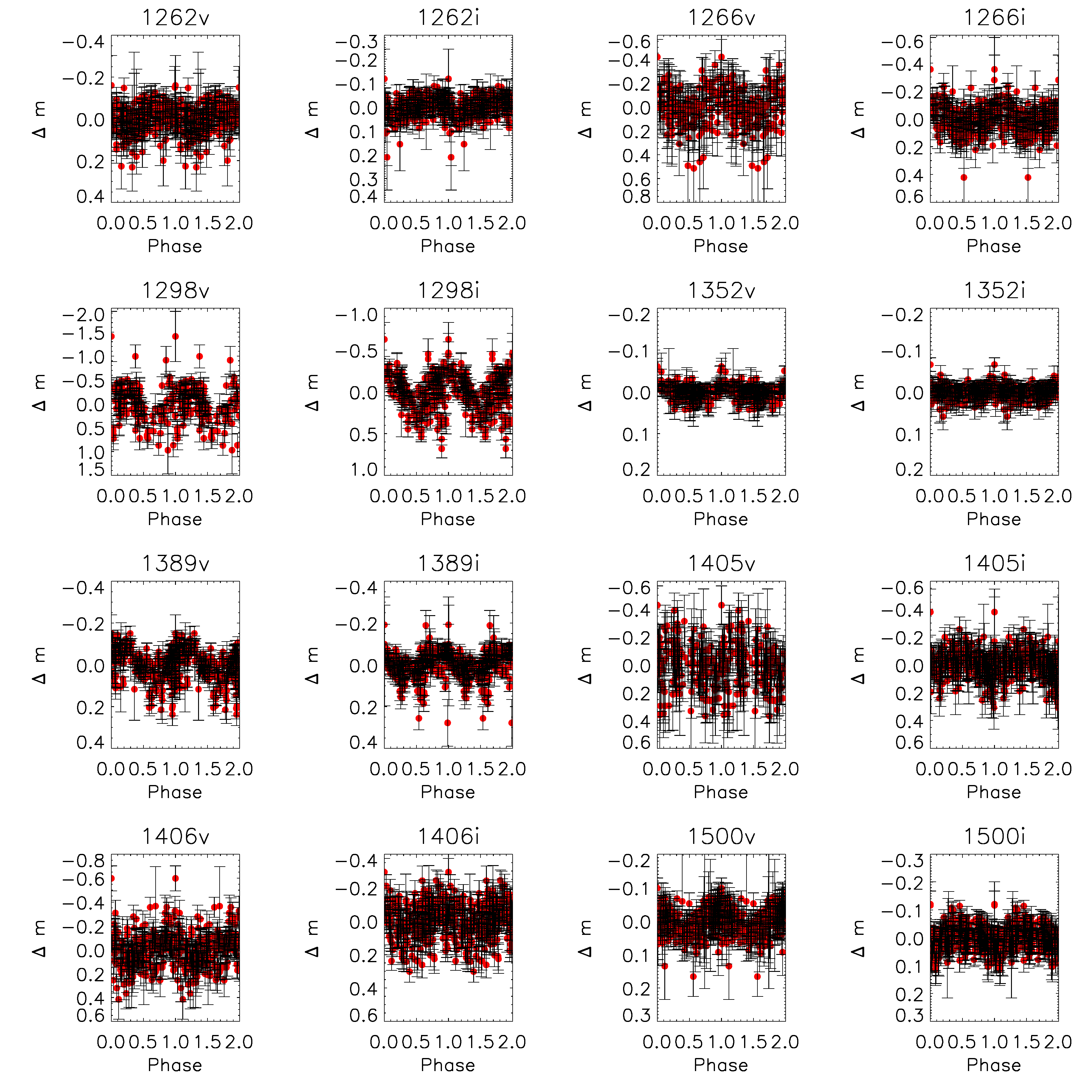}
\includegraphics[width=9cm, height=9cm]{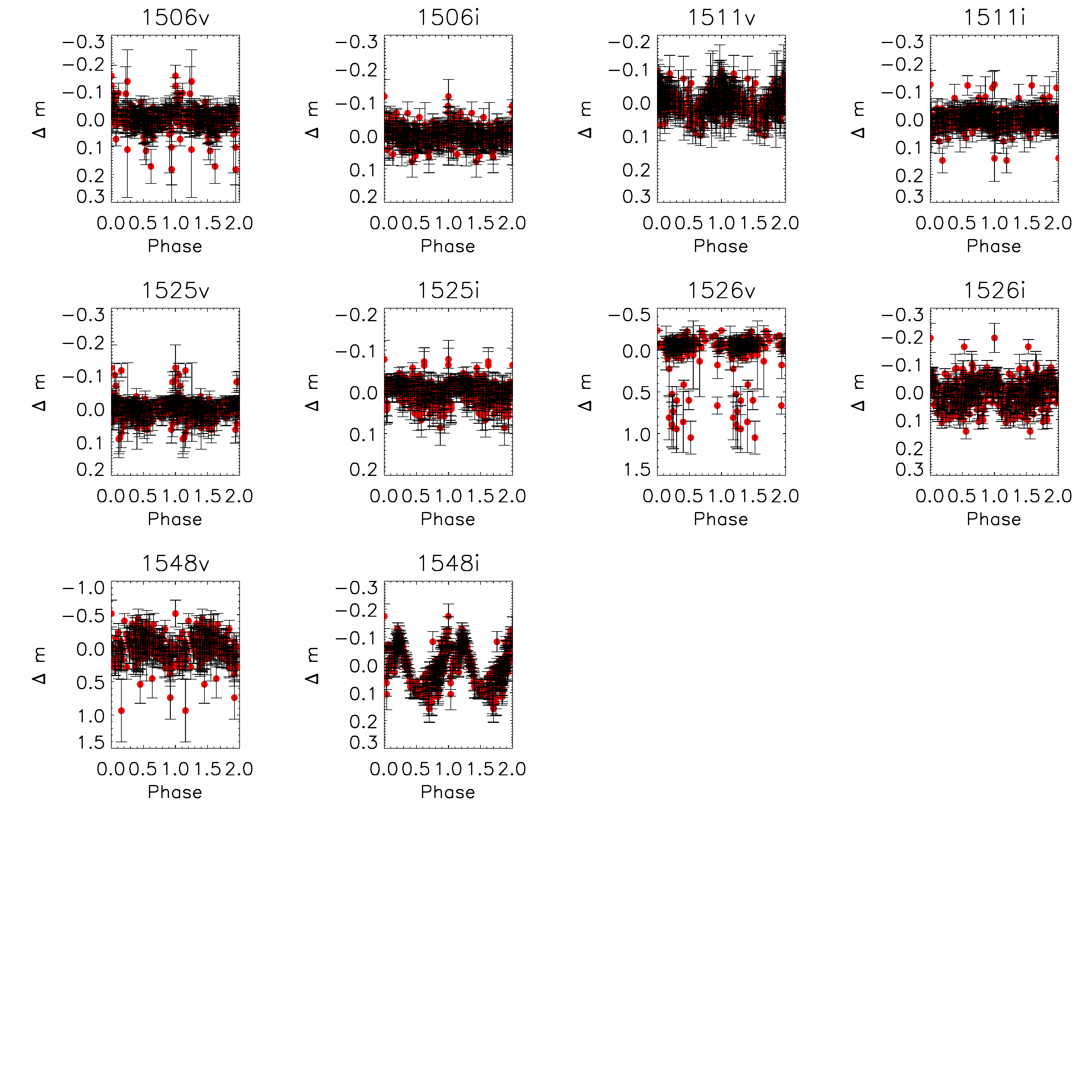}
}
\caption{Continued. }
\end{figure*}

\begin{figure*}
\hbox{
\includegraphics[width=9cm, height=9cm]{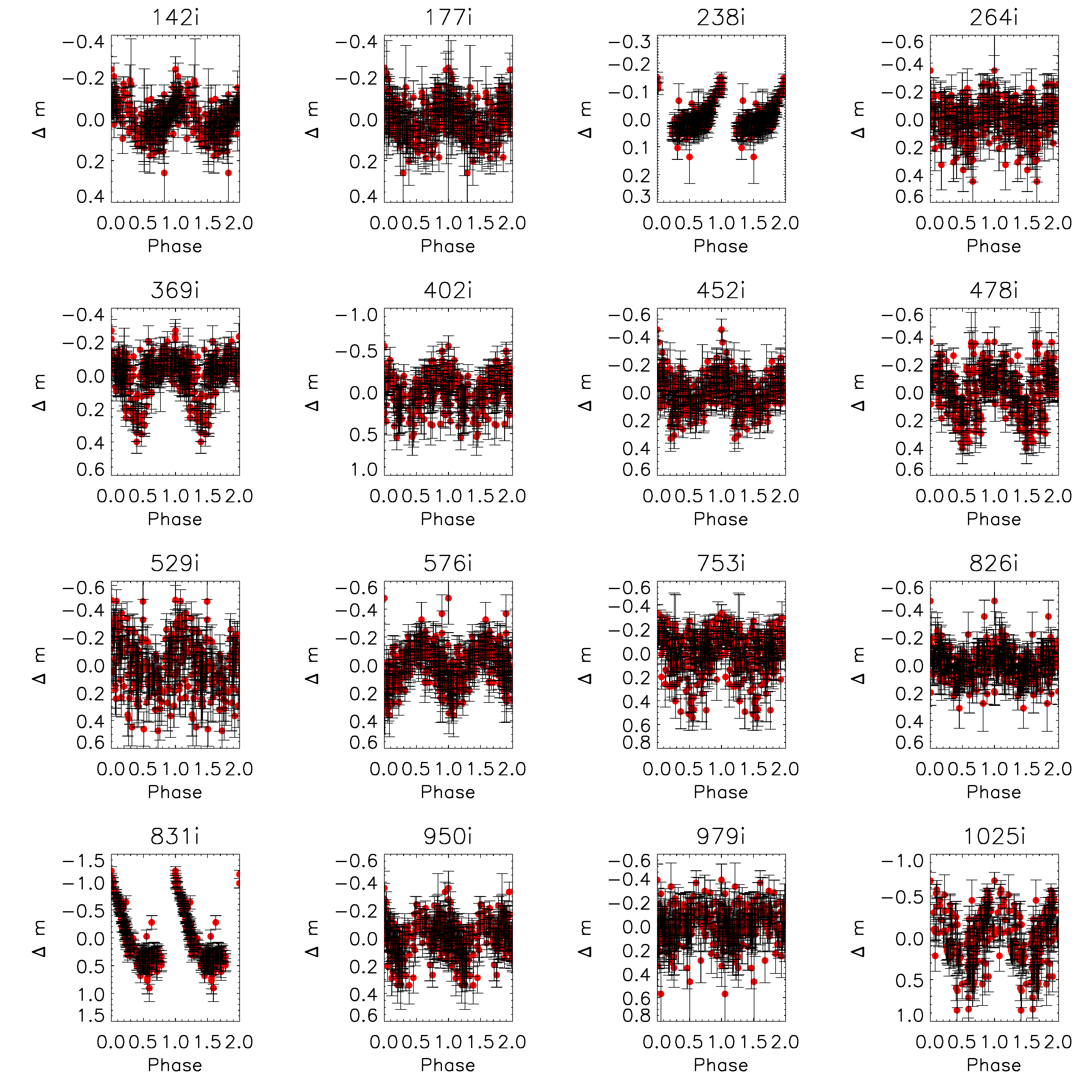}
\includegraphics[width=9cm, height=9cm]{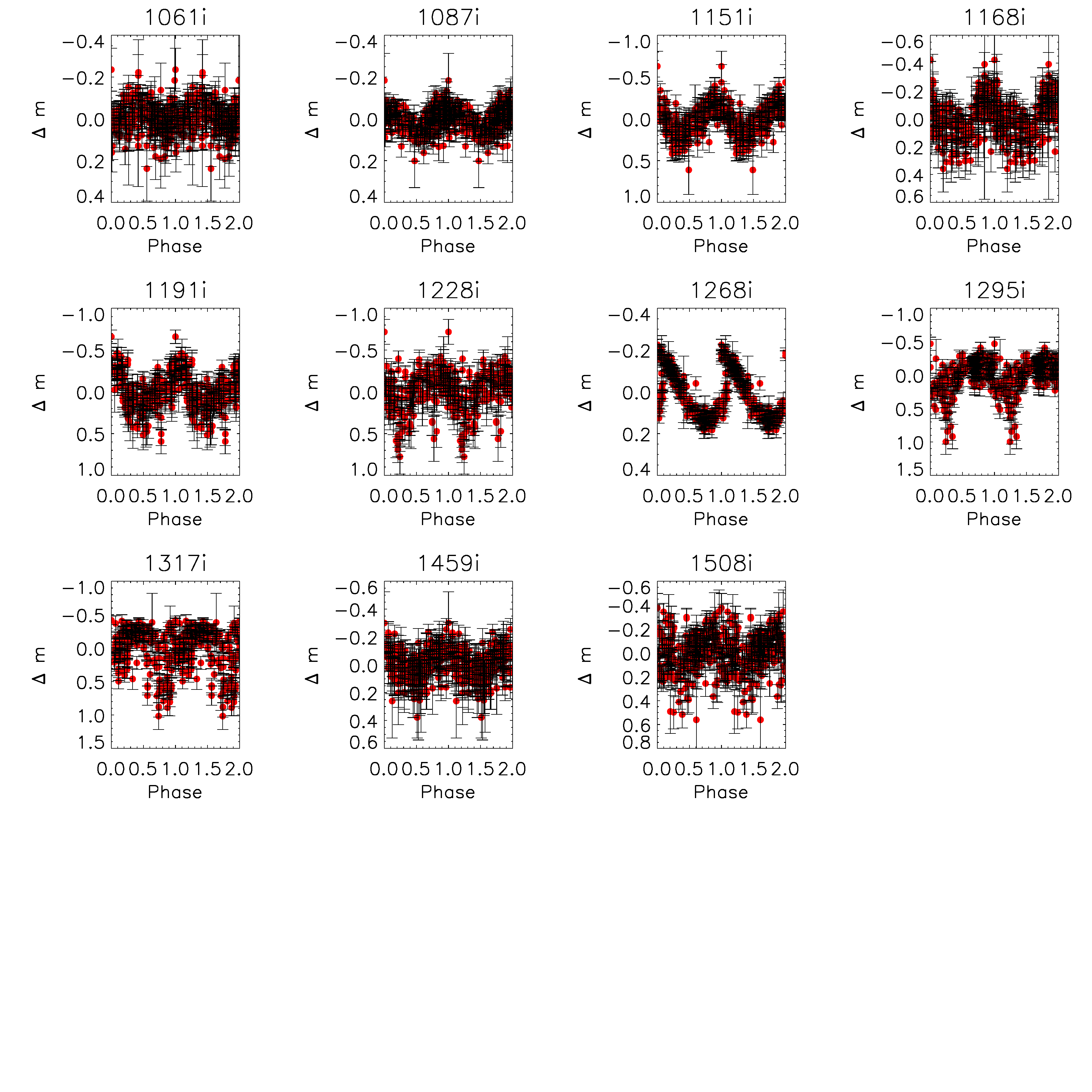}
}
\caption{The $I$ band phased light curves of probable variable candidates around the cluster. }
\end{figure*}
\subsubsection{$U-B$ vs $B-V$ TCD}
To identify probable MS variables, we have plotted in Fig.~12 the $U-B$ versus $B-V$ for variable stars identified in the cluster region with the photometric data of 22 stars found in Massey et al. (1995). Reddening in terms of color excess $E(B-V)$ ranges from 0.7 to 1.1~mag (Erickson 1971; Guetter 1992; Massey et al. 1995; Pigulski et al. 2000).  Pigulski et al. (2000) recognized the highest extinction in the eastern part of the cluster where a trapezium system, i.e., of O, B spectral types, is located. The eastern part of their observed field is the direction to the reflection nebula NGC\,6820, and their study suggested more than half of the total absorption to arise from nearby interstellar matter toward NGC\,6823. It is inferred that there is significant differential reddening within the cluster, manifest that the cluster is located behind at least $A_{V}$ = $\sim 3$~mag (Riaz et al. 2012).  Rangwal et al. (2017) studied the interstellar extinction of open clusters and found that NGC\,6823 follows a normal extinction law in optical as well as in the NIR wavelengths.  The $U-B$ versus $B-V$ TCD shows that the stars exhibiting within $E(B-V) = 0.7$--1.1~mag could be MS members of the cluster, indicating a nonuniform reddening across the cluster.  The reddened theoretical ZAMS of Girardi et al. (2002) is fitted to the TCD. The value of color excess $E(V-I)$ was taken as 0.88~mag which has been calculated using the minimum reddening value of $E(B-V)$=0.70 mag. 

\subsubsection{$J-H$ vs $H-K$ TCD}
Fig.~13 shows the $J-H$ versus $H-K$ TCD for NGC\,6823.  Only 86 stars were cross-matched between the present sample of variable stars and the 2MASS catalog, with the $JHK$ counterparts of two stars No.~201 and No.~377 not found during the match.  
In the 2MASS TCD, the ``F'' and ``T'' regions are locations of probable Class III/field stars and Class\,II sources, respectively.  The filled squares plotted in blue and green colors represent, respectively, probable PMS and MS members.  Circles in the diagram represents field stars.  Riaz et al. (2012) from their NIR CCD found early-type MS dwarfs concentrating close to $(H-Ks)\sim 0.7$~mag, and $(J-H) \sim 1.5$~mag, having extinction $A_{V} \ge 10$. These authors have noticed another population near the classical TTS locus close to $(H-Ks) \sim 0.6$~mag and $(J-H) \sim 1.0$~mag, presumably being young disk-bearing members.  In the 2MASS TCD, about half the detected variables are in the ``T'' or ``F'' regions hence could be T~Tauri variables.  
A few PMS stars located below the TTS locus are probably Herbig Ae/Be stars. We note that star No.~679 occupies the position where Herbig Ae/Be stars are placed, while in $U-B$ versus $B-V$ TCD it lies close to the MS locus; it could thus be either a reddened MS star or a Herbig Ae/Be member. 

Following Gutermuth et al. (2008) to classify young stellar sources, Riaz et al. (2012) used MIR IRAC data and found 2 Class\,I, 94 Class\,II, and 394 Class\,III or field stars in the region.  The figure 4(a) of Riaz et al. (2012) is plotted with the $(H-Ks)$ and $[4.5]-[8]$ TCD for the 490 sources. This shows both YSOs and the diskless sources to have similar NIR colors, but from their IRAC TCD the photospheric and the disk population at IRAC color $[4.5]-[8] \sim$~0.4~mag are readily distinguishable. They have noted that a few Class~III/field stars are mixed with Class\,II sources. Their IRAC TCD in figure.~4(b) shows different locales of Class~I and Class~II sources. The protostars (Class~I) are located in the top-right corner and exhibit the reddest in the $[3.6]-[5.8]$ color, whereas the Class~II sources are placed at $[4.5]-[8] \ge 0.5$~mag, $[3.6]-[5.8] \ge 0.4$~mag, and the Class\,III/field stars are found to be near $[4.5]-[8]$ and $[3.6]-[5.8] \sim 0.2$ to 0.3~mag.

\subsubsection{NIR and MIR TCDs}
To see the distribution of young variable sources we have plotted them in the NIR and MIR TCD (left panel) and MIR TCD (right panel) in Fig.~14.  To obtain these plots we have cross-matched the coordinates of variable stars with those from the Spitzer Galactic Legacy Infrared Mid-Plane Survey Extraordinaire (GLIMPSE), yielding MIR counterparts of all 88 variable stars. A few stars, namely Nos.~154, 238, 313, 1405, and 1406 do not have magnitudes at $[4.5]$ and other wavelengths. The $H-K$ versus $[4.5]-[8]$ TCD shows most young stellar sources to have $H-K \gtrsim 0.3$~mag whereas the $[4.5]-[8]$ and $[3.6]-[5.8]$ TCD shows a few young stellar objects to be positioned as field stars or other populations.

\subsubsection{IPHAS data}
To identify the young stellar sources with $H_{\alpha}$ emission, i.e., the indicator of  accretion disks, we have compared the present data with Table~3 for IPHAS photometry of Riaz et al. (2012). We got 29 common stars after the match that have IPHAS photometry, with stars No.~502, 576, 655, 679, and 979 having $H_{\alpha}$ emission with equivalent width greater than 10 ${\AA}$, judged by the $(r'- i')$ versus $(r'- H_{alpha})$ TCD for NGC\,6823 (Riaz et al. 2012).
The location of these five stars is shown with the red square in the present $(J-H)$ versus $(H-K)$ TCD. 
The $H_{\alpha}$ emission is found to be variable in nature, therefore, it is necessary to check the location of the 
objects in spectral type/color versus magnitude diagram to know their membership and nature (Mart\'in et al. 2000).
 Two stars Nos.~655 and 979 could be
considered as PMS objects which may possess circumstellar accreting disk. 
Barrado y Navascu\'es et al. (2001) showed that $H_{\alpha }$ emission
depends on the spectral type or color in the sense that $H_{\alpha }$
emission is found to be larger for cooler objects in a plot between
the $H_{\alpha}$ emission and $(I-J)$ color. The $(I-J)$ and $(I-K)$ colors
for star no. 655 is about 2.26 mag and 4.682 mag, respectively. In the case of star no. 979, we have
taken $I$ magnitude from Pigulski et al. (2000) to
determine its  $(I-J)$ and $(I-K)$ colors due to lack of $(V-I)$ color in present observations, yielding  $(I-J)$ and $(I-K)$ colors as
2.020 mag and 3.789 mag, respectively.
Stars Nos.~655 and 979 in particular satisfy both colors and $H_{\alpha}$ emission being greater than 10 ${\AA}$,  hence are young stars with accretion disks.
\subsubsection{$V$ vs $V-I$ CMD} 
Sixty one variable stars were detected in both $V$ and $I$ bands. Their $V$ magnitudes and $V-I$ color are given in Table~2, and Fig.~15 shows their $V$ versus $(V-I)$ CMD.  The PMS isochrones and evolutionary tracks for different masses are taken from Siess et al. (2000). The solid curve represents ZAMS by Girardi et al. (2002).  We determined the distance modulus of the cluster to be $(V-M_{V})=14.31$~mag by comparing the ZAMS of Girardi et al. (2002) for solar metallicity to the $V$ versus $V-I$ CMD, which corresponds to a distance of 2.59~kpc. The present estimate of distance matches well with those derived in earlier works of NGC\,6823. The isochrone of age 4~Myr also fits the data well.  The CMD is known to be contaminated by the foreground field stars (e.g. Guetter 1992; Pigulski et al. 2000; Bica et al. 2008). After analysis of CMD, Pigulski et al. (2000) and Riaz et al. (2012) noted two different populations in the cluster, one consisting of older, massive stars which are located near or on the ZAMS, while the other one being younger objects with ages less than 10\,Myr and are of lower masses ($\sim0.1$--0.4~M$_{\odot}$), that is, of PMS stars.  Pigulski et al. (2000) also concluded that stars lying in B region of their figure~11(a) are cluster stars of PMS nature, evolving towards the MS.  The present CMD containing variable stars also shows MS of the cluster to go up to around $V=16$~mag; location of variables in the CMD suggests the majority of these stars to be probable members. Most the fainter and redder stars lying between $(V-I)=\sim2$~mag and $\sim3$~mag could be PMS objects.  In this CMD, to maintain clarity we have not plotted star No.~1548 despite it being detected in the $I$ band, because it has $V-I$ color more than 6~mag.  Star No.~449 may be a possible PMS star but its placement in the $U-B/B-V$ and $J-H/H-K$ TCDs suggests a field star, even though it has proper motions in the range of probable cluster members.

\subsubsection{NIR CMDs}
The $J$ versus $(J-K)$ and $J$ versus $J-H$ CMDs for the present sample of variable stars are shown in Fig.~16. it is seen that the MS is almost vertical and clearly separated from the PMS objects/field stars as in the case of the $V$ versus $(V-I)$ CMD. Bica et al. (2008) described the same and from statistical cleaned CMD, they found that two populations are distributed separately where majority of PMS objects are faint and redder. They found the age of the cluster in the range from 2 to 7~Myr, a color excess $E(B-V)$=$\sim$0.86~mag, and $A_{ V}$ =$2.7\pm0.2$. In their work after fitting theoretical models, the absolute distance modulus of the cluster was found to be $(m-M)_{O}=11.5$~mag.

Although stars numbered Nos.~142, 402  826, 950, 924, 1025, 1066, 1087, 1298, and 1548 are nonmembers from the analysis of the Gaia data, we have considered them as possible PMS objects based on their positions in different CMD and TCDs. We note that there are 10 stars, namely, Nos.~240, 614, 623, 752, 965, 1235, 1352, 1500, 1525, and 1526, that are designated as nonmembers from proper motion and parallax and, these could be MS stars based on location in TCDs and CMDs. Star 1506 is located well away from the MS in $U-B/B-V$ TCD while it is lying on te MS in $V/V-I$, $J/H-K$ and $J/J-H$. The analysis of Gaia data also found it to be nonmember. This is a doubtful case for being an MS member. From Gaia data analysis we found stars Nos.~ 201, 239, 449, 619, 765, 861, 1151, 1168, 1191 and 1508 as possible or probable members but their locations in various TCDs and CMDs do not seem to be consistent with membership. Thus, these are considered field stars. 

We looked for the matches of our 8 previously identified PMS stars based on their 2MASS colors and their spectra taken at the 2.16 m telescope of Beijing Observatory (Hojaev et al. 2003). The spectra showed the strong H-alpha in emission, the SED in continuum and other features typical either for TTS or Herbig Ae/Be stars.
The cross-match yields 3 common stars namely 154, 655 and 679 for two of them we have already determined their features (their location on the diagrams, their proper motions and parallaxes).  In the present work, we have classified 655 as classical TTS and 679 as MS. There was doubt to consider 679 as PMS because in $U-B$ versus $B-V$ TCD it is lying on MS but in $J-H$ versus $H-K$ diagram it is placed in the location of classical TTS. Now, we determined the membership of star No 154. It might be probable member of the cluster as Herbig Ae/Be type star while earlier (Hojaev et al. 2003) it has been classified as classical TTS, though it has very different position in $V$ versus $V-I$ CMD to be PMS star but GAIA suggests it to be highly probable member, and in $J-H$ versus $H-K$ it is located where Herbig Ae/Be stars are found. Therefore, it could be considered as Herbig Ae/Be star.

We have considered members those stars which fulfill criteria like location in TCDs, CMDs, and have consistent kinematics.  Therefore, using proper motion data together with location in various TCDs and CMDs obtained from present and available photometric $UBVI$, $NIR$, and $MIR$ data we classify 25, 48, and 15, respectively,  as MS, PMS members of the cluster, and field stars. The classification of variables is given in Table 3.

\begin{figure}
\includegraphics[width=8cm]{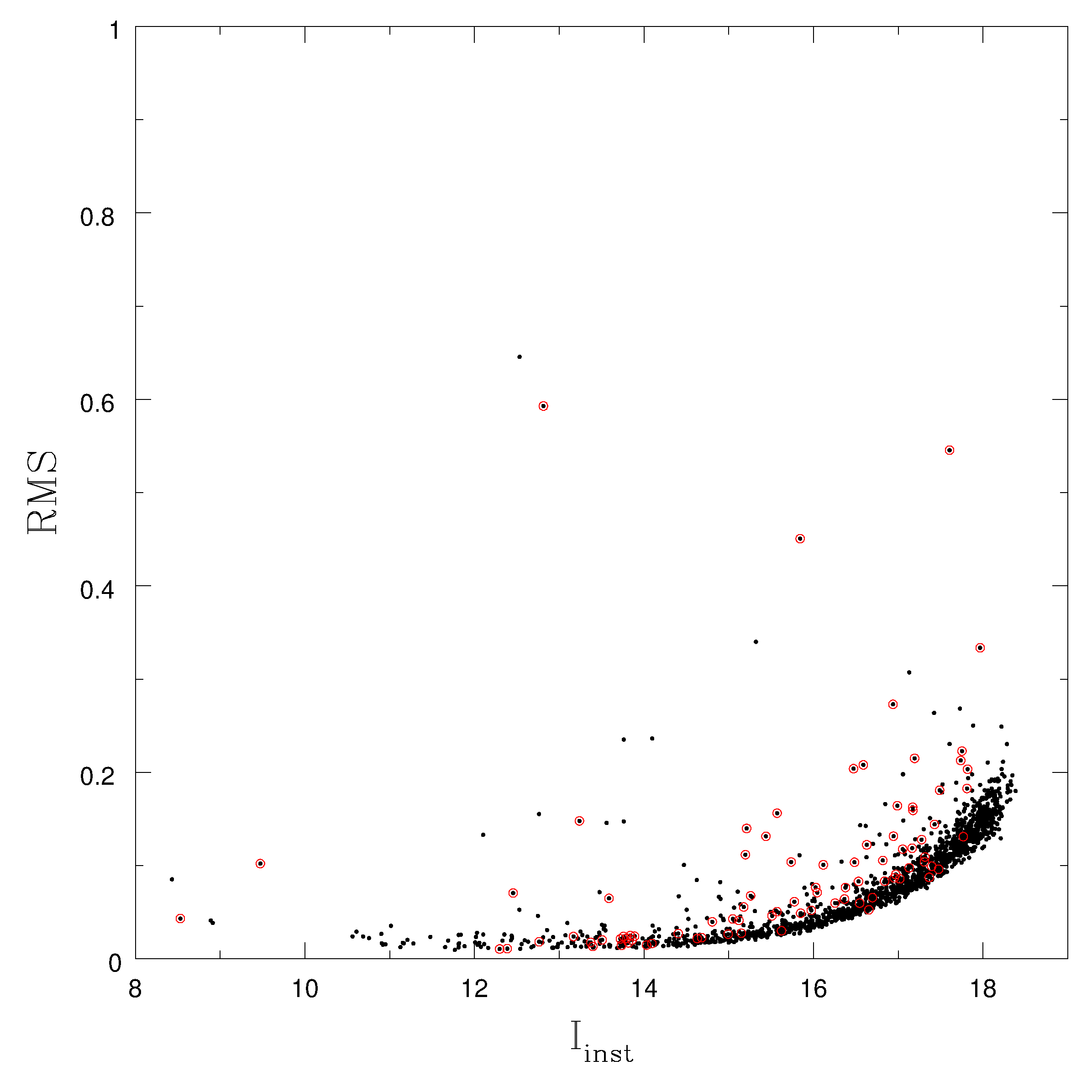}
\caption{Magnitude as a function of $rms$ value of each star detected  in $I$ band. Open circles represent variable stars identified in this work.
}
\end{figure}

\begin{figure}
\centering
  \includegraphics[width=9cm,angle=0]{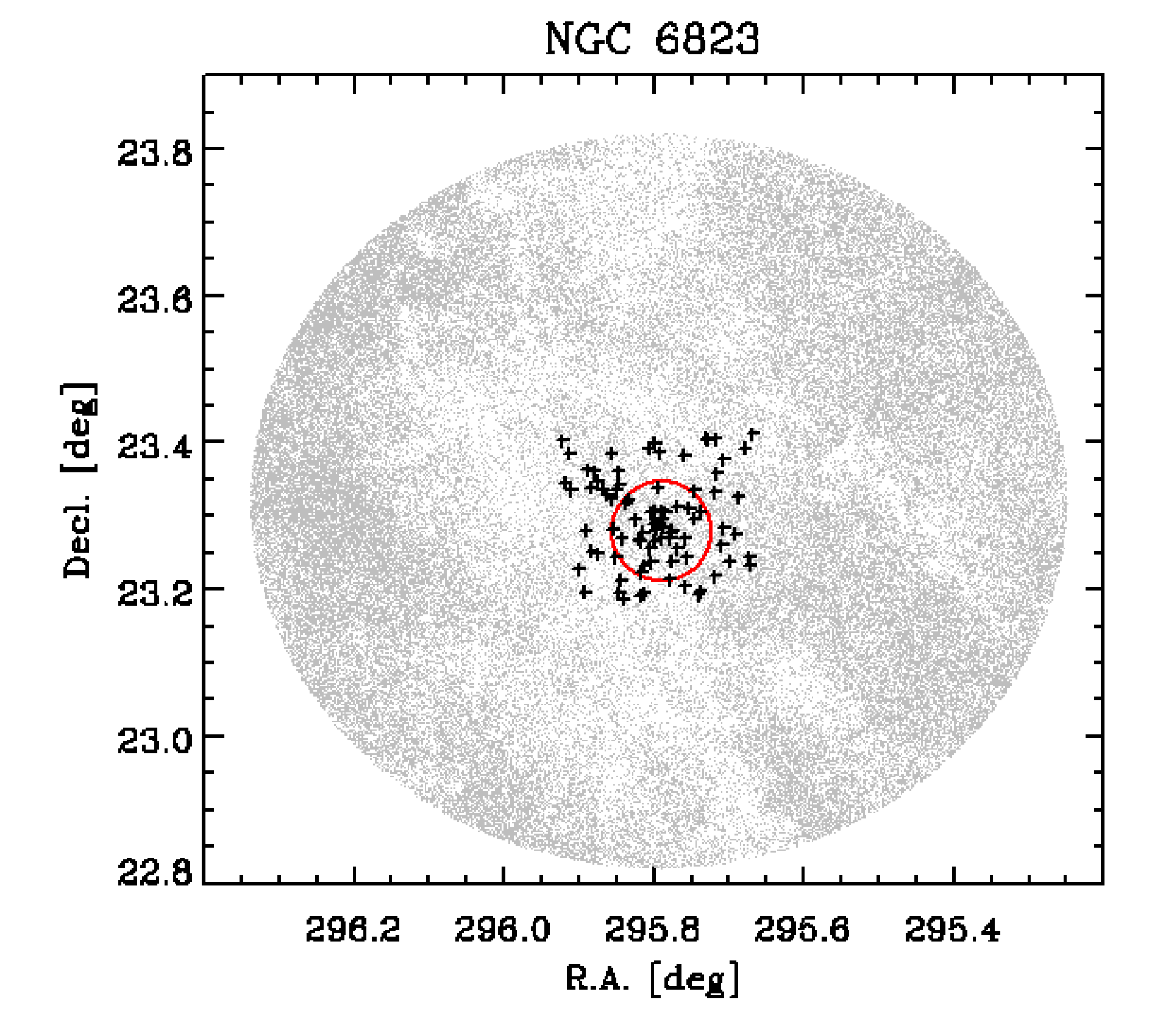}
  \caption{The sky positions of all the Gaia sources (in gray) within $30\arcmin$ toward NGC\,6823. 
  }
  \label{fig:rade}
\end{figure}

\begin{figure}
\centering
\vbox{
  \includegraphics[width=9cm,,angle=0]{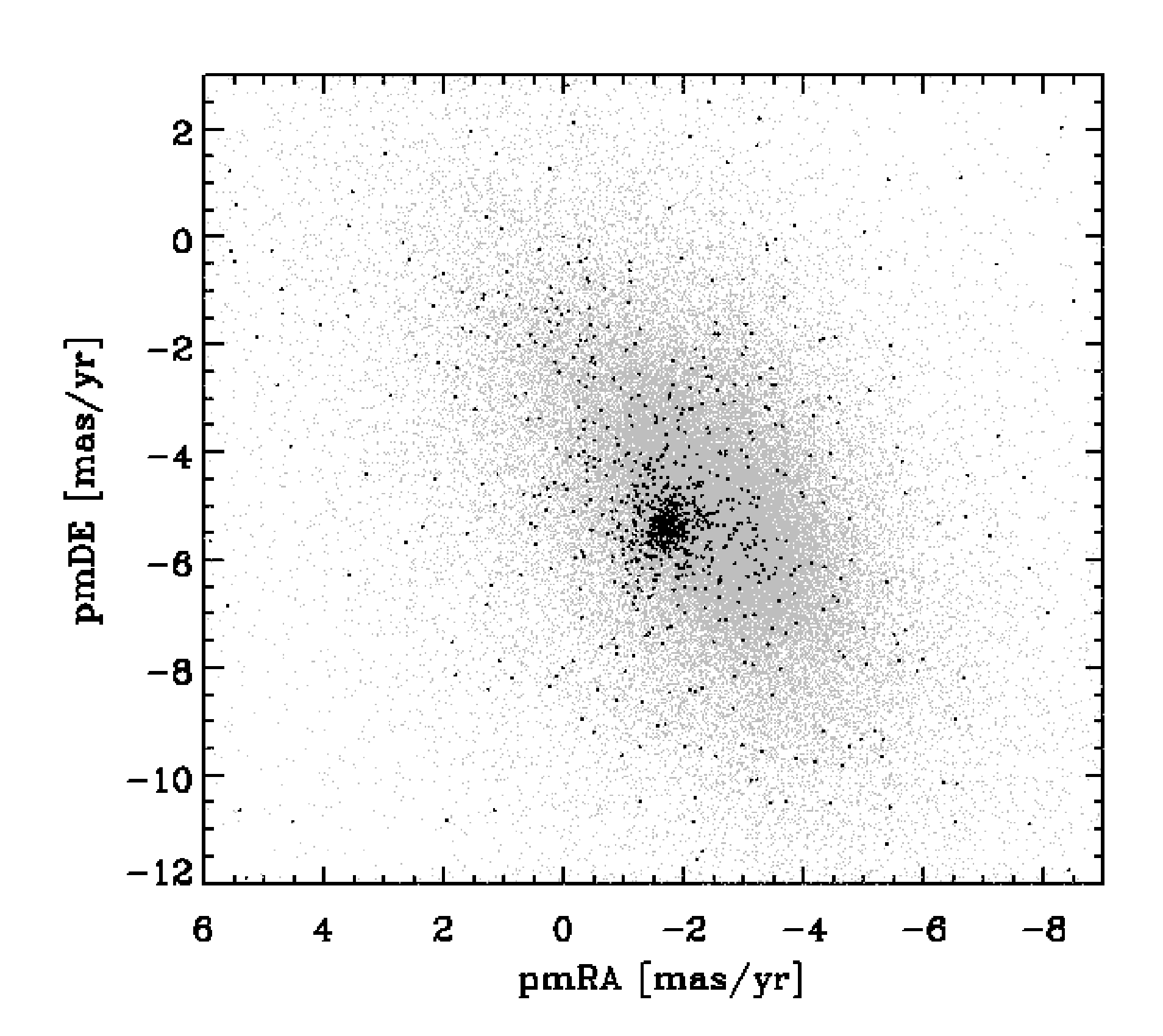}
  \includegraphics[width=9cm,angle=0]{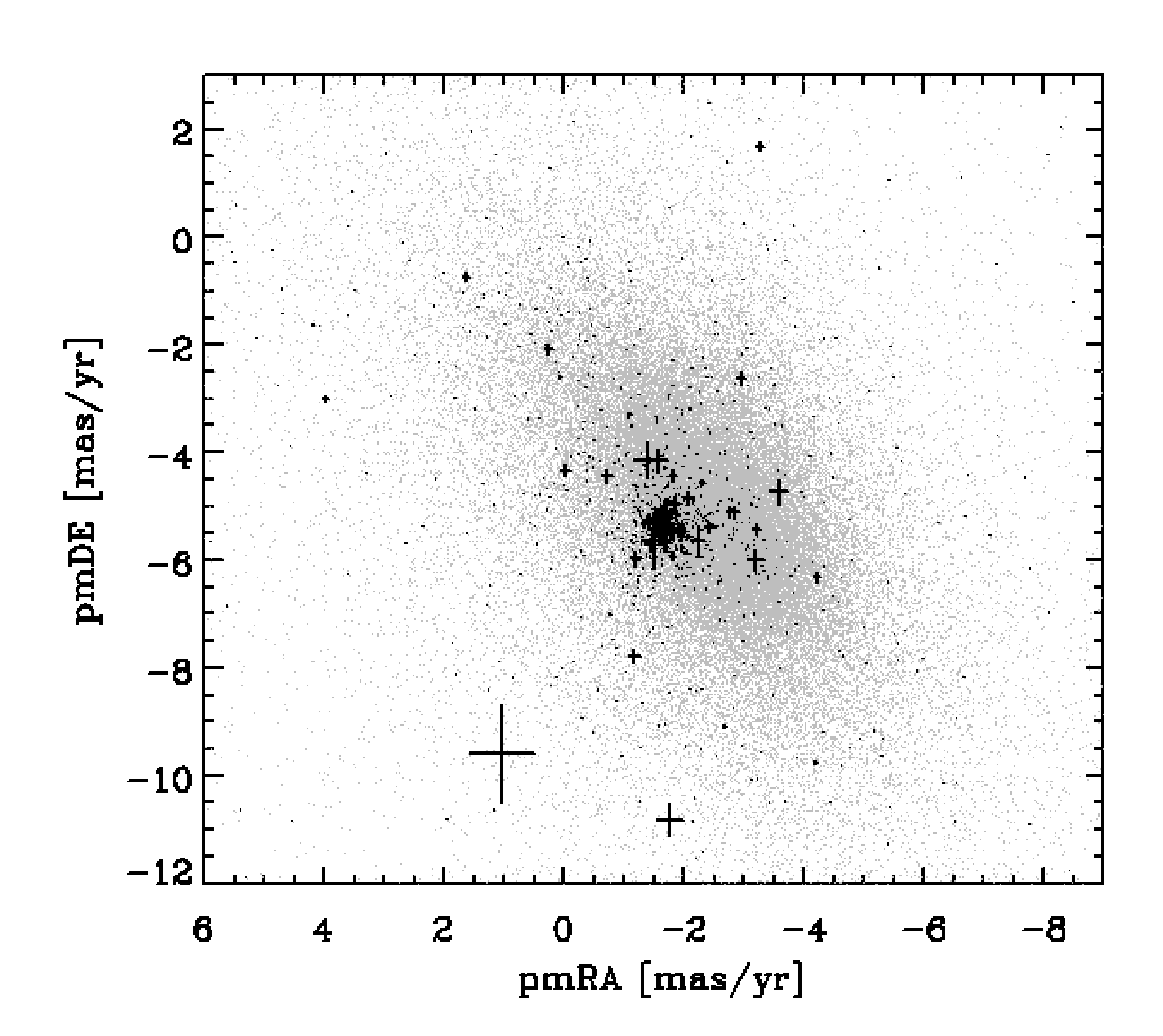}
}
  \caption{The upper panel represents proper motion of all the stars (gray) and those within $4\arcmin$ 
cluster region (black small circles, 1294 stars). The lower panel shows proper motions for
variable stars (in black with error bars). }
  \label{fig:pm}
\end{figure}

\begin{figure}
\centering
  \includegraphics[width=9cm,angle=0]{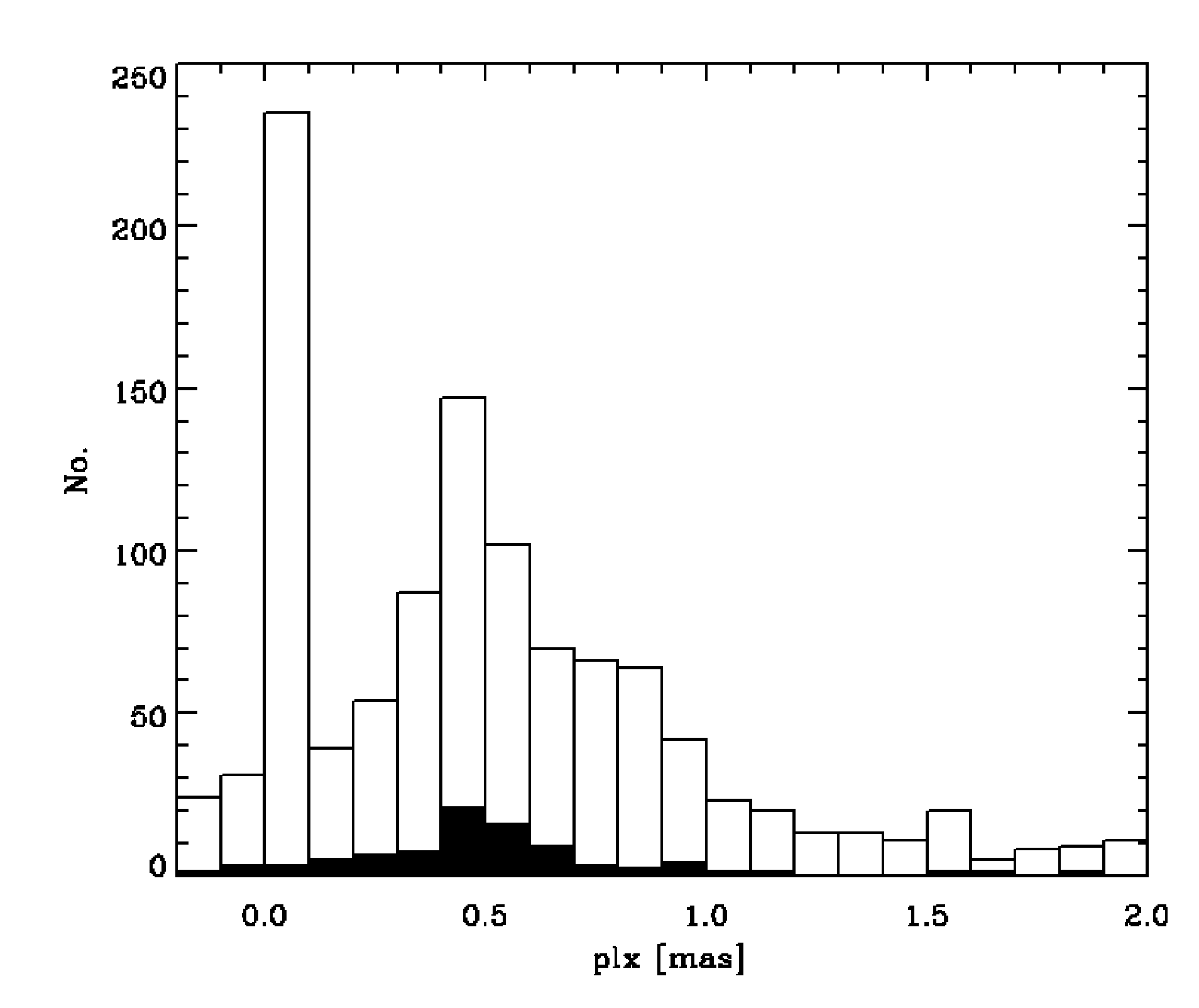}
  \caption{Histogram of parallaxes for stars within 4 arcmin, where histogram shaded with black is for variable samples identified in the present work. }
  \label{fig:plx}
\end{figure}

\begin{figure}
\centering
  \includegraphics[width=9cm,angle=0]{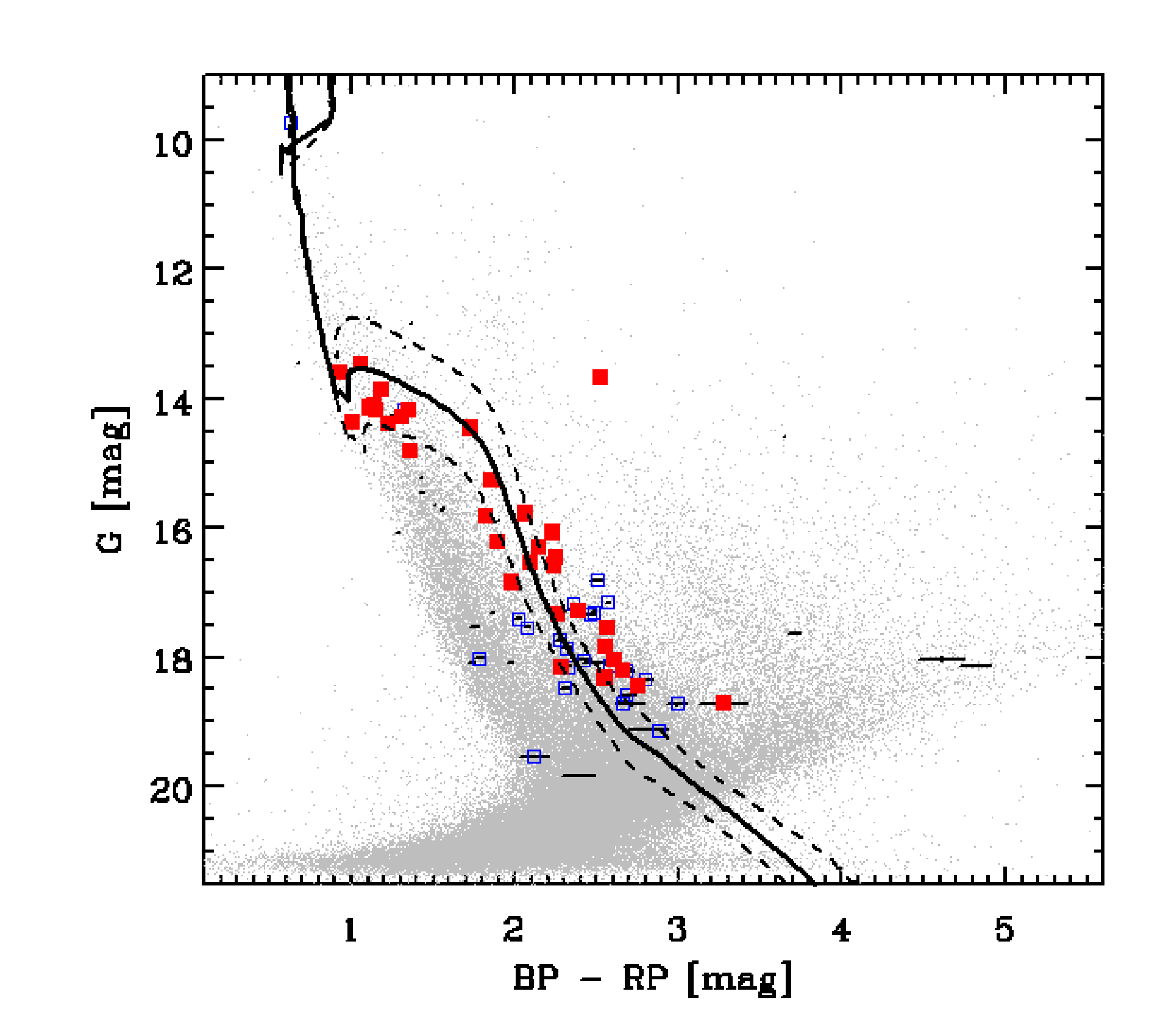}
  \caption{The $G$ vs $BP-RP$ CMD for the present sample of variable stars. The filled and open squares denote probable and possible cluster members, and dotted points are considered nonmembers. }
  \label{fig:cmd}
\end{figure}

\begin{figure}
\includegraphics[width=9cm]{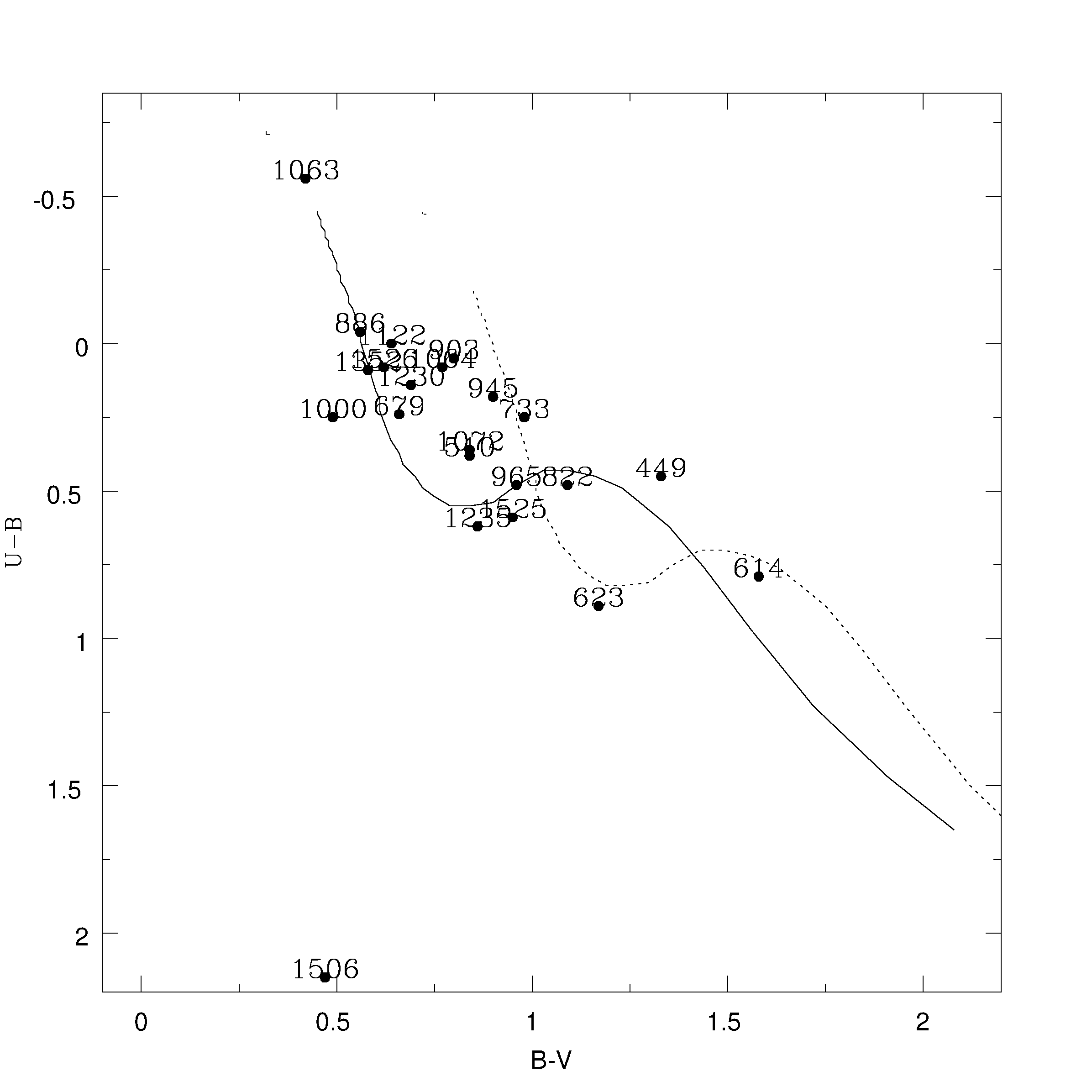}
\caption{$(U-B)/(B-V)$ TCD for variable stars identified in the present study. All the $UBV$ data are taken from
Massey et al. (1995). The continuous and dotted line represent the ZAMS (Girardi et al. 2002) which are shifted along the reddening vector for reddening $E(B-V)= 0.32$ mag and 0.45 mag.  
Triangles are those stars that are identified as MS variables. 
}
\end{figure}

\begin{figure}
\includegraphics[width=9cm]{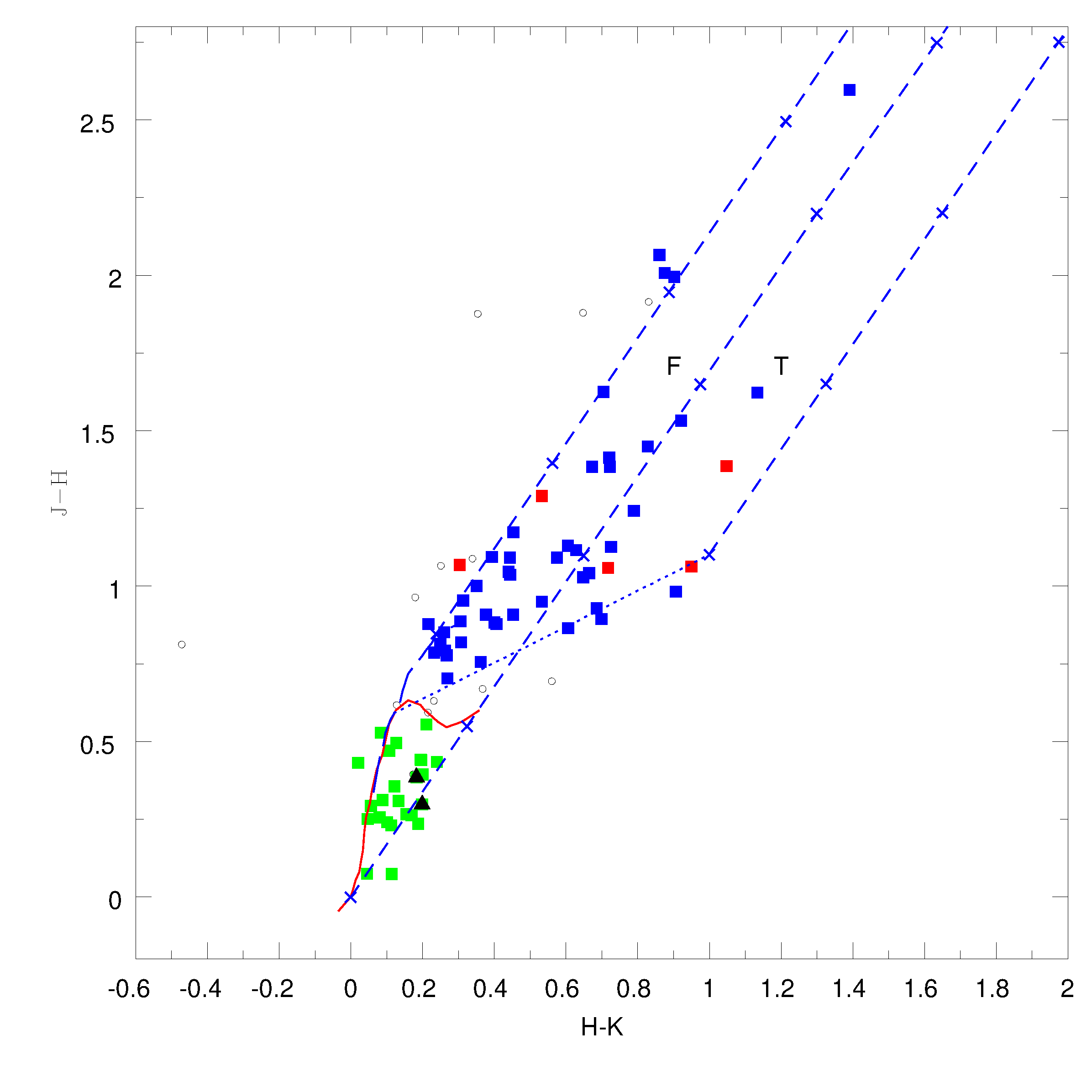}
\caption{$(J-H)/(H-K)$ TCD for variable stars detected in the field of
 NGC 6823. The $JHK$ data have been taken from the 2MASS catalog (Cutri et al. 2003). The continuous and long dashed lines show sequences for dwarfs and giants (Bessell \& Brett 1988), respectively. The TTS locus (Meyer et al. 1997) is shown by a dotted line.
The small dashed lines are reddening vectors (Cohen et al. 1981) and
an increment of visual extinction of $A_{V}$ = 5 mag is denoted by crosses on the reddening vectors.
Filled squares with blue colors represents PMS. The MS population are shown by green squares whereas open circles may be either MS members of the cluster or field stars. Triangles (black) represent two MS members BL 50 and HP 57.    
}
\end{figure}

\begin{figure}
\includegraphics[width=9cm]{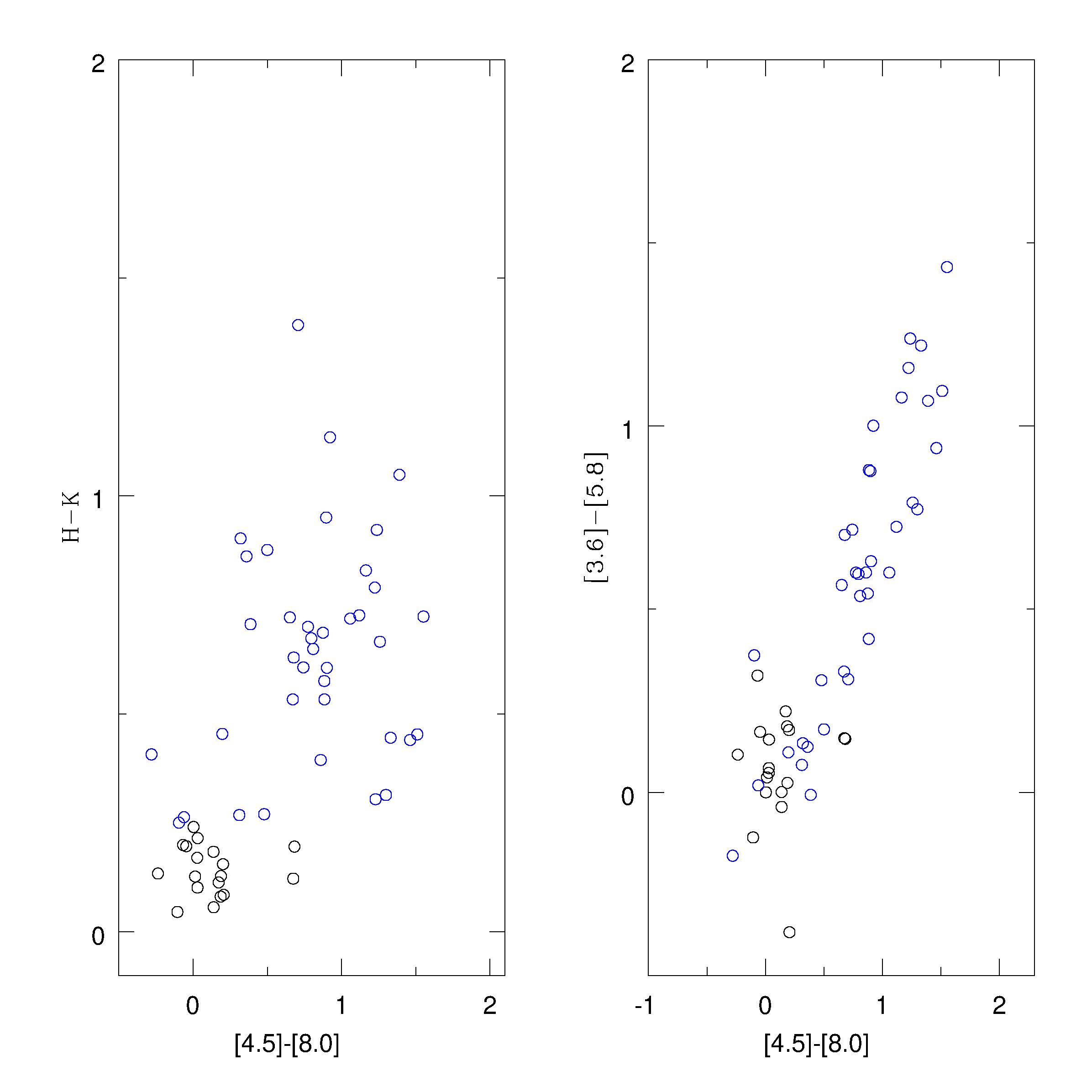}
\caption{$(H-K)$ vs $[4.5]-[8.0]$ and $[3.6]-[5.8]$ vs $[4.5]-[8.0]$ TCDs for variable stars detected in the field of
 NGC 6823. 
Blue circles are PMS young stellar sources while black circles are MS/field stars.
}
\end{figure}

\begin{figure}
\includegraphics[width=9cm]{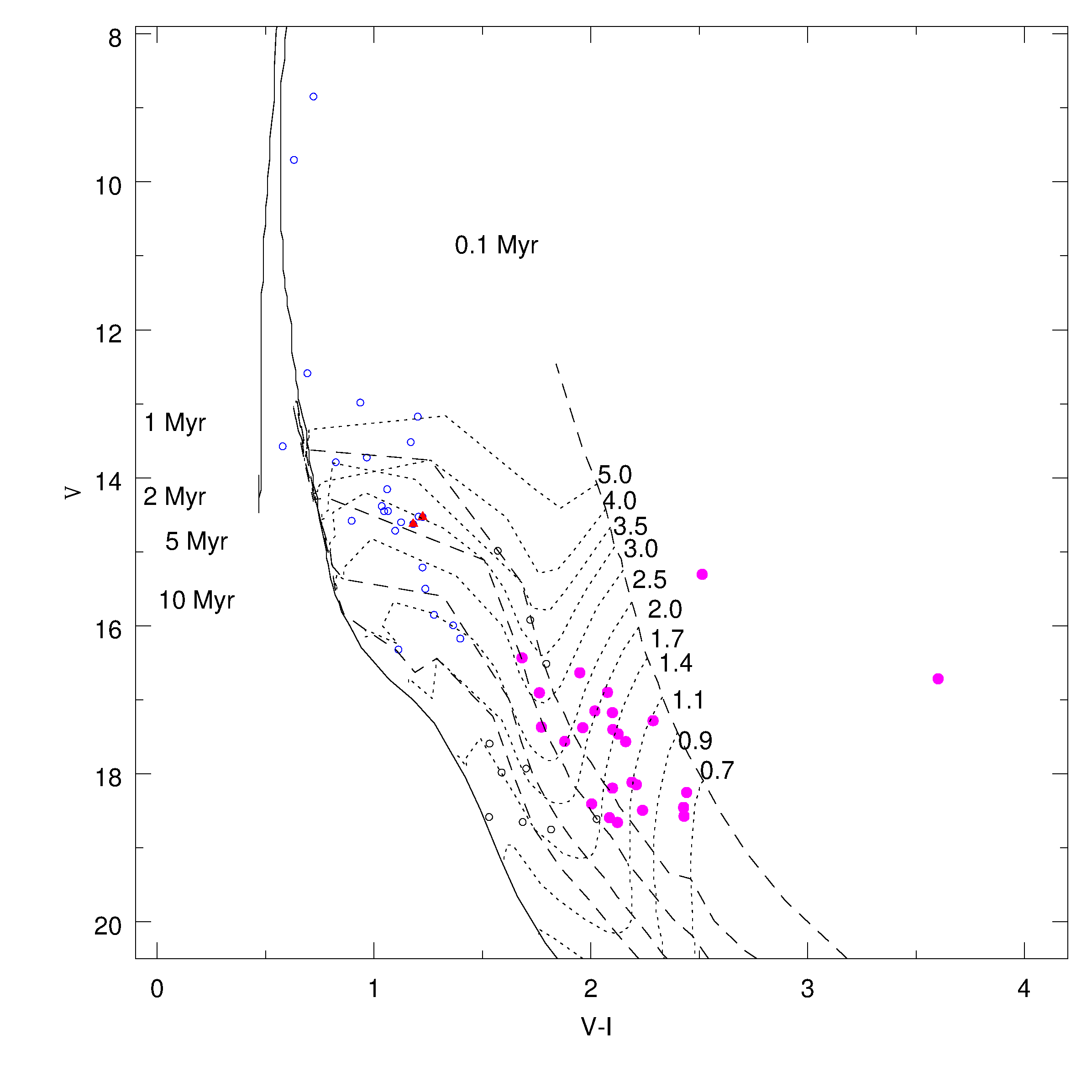}
\caption{$V/(V-I)$ CMD for variable stars in the region of the cluster NGC 6823.
The open circles (blue) are MS variables, and probable PMS variable stars are shown by filled circles (magenta). The open circles in black color are considered field stars.
The continuous curve is ZAMS by Girardi et al. (2002) while dashed lines
 are PMS isochrones taken for
0.1, 1, 2, 5, 10 Myrs (Siess et al. 2000).
The PMS evolutionary tracks for different masses ranging from 0.7 to 5.0~M$_{\odot}$ from Siess et al. (2000) are plotted with dotted curves. BL 50 and HP 57 are shown
by red triangles. 
}
\end{figure}

\begin{figure}
\includegraphics[width=9cm]{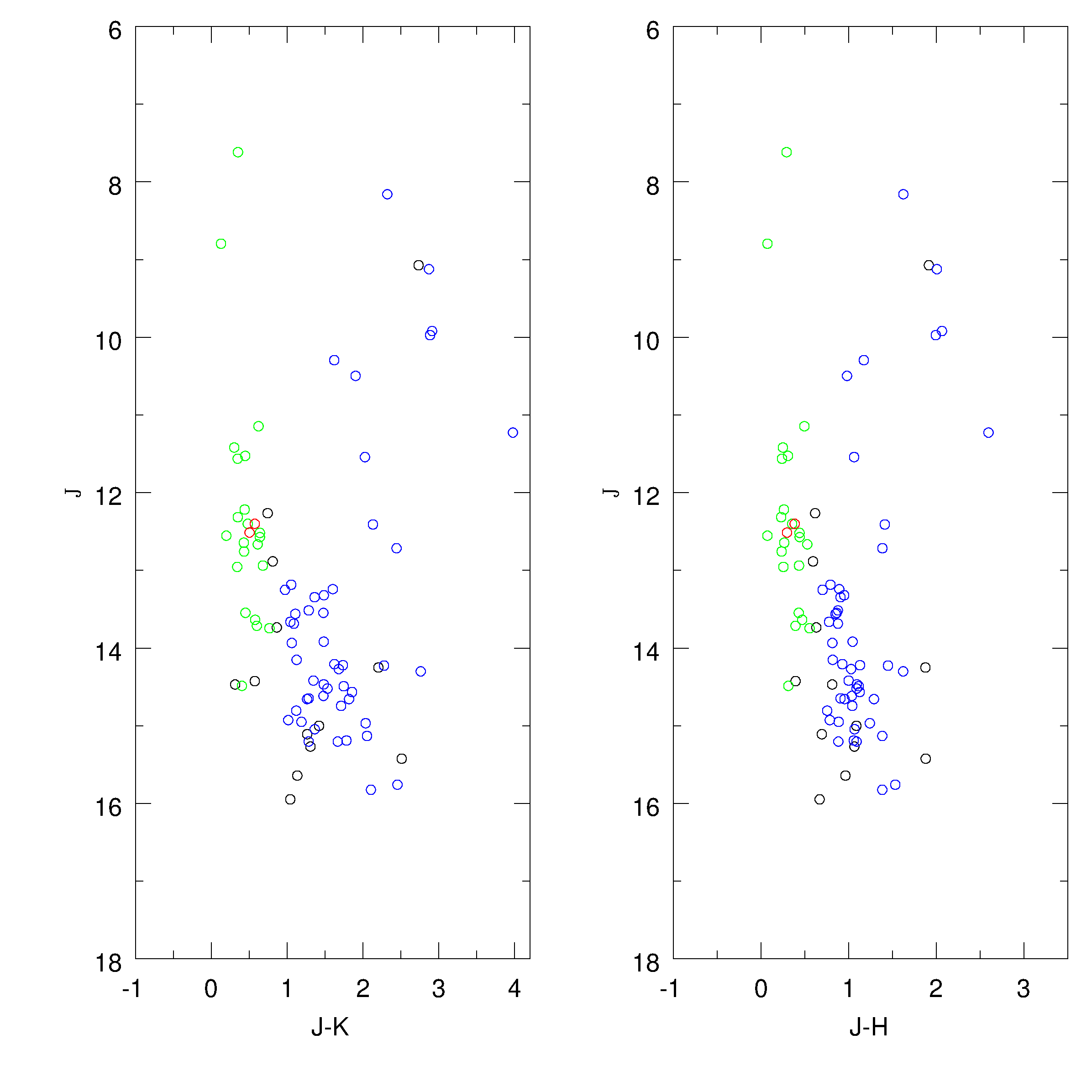}
\caption{$J/(H-K)$ and $J/(J-H)$ CMD for variable stars detected in the field of
 NGC 6823. The $JHK$ data have been taken from the 2MASS catalogue (Cutri et al. 2003). 
Circles (blue) and circles (green) represent MS and PMS, respectively. The Open circles in black color demonstrate the field stars.
The locations of stars No. 822 (BL 50) and 1007 (HP 57) are shown with open circle in red color.
}
\end{figure}

\section{Characteristics of variable stars}

The $\log (L/L_{\sun})$ vs $\log T_{\rm eff}$ diagram ($H-R$ diagram) for 21 members (MS variables) is shown as  Fig.~17. We are not able to locate four MS stars Nos.~240, 752, 1107 and 1500 in this plot due to lack of their $U$ and $B$ band data.  Here, the effective temperature and bolometric correction (BC)  have been determined from Toress relation (2010) using the intrinsic $(B-V)$ color.  The ${M}_{\mathrm{bol}}$ values of stars are obtained from the relation ${M}_{\mathrm{bol}}={M}_{V}+{BC}$, where $M_V$ is the absolute $V$-band magnitude. The luminosity was obtained from the relation $\mathrm{log}(L/{L}_{\odot })=-0.4({M}_{\mathrm{bol}}-{M}_{{\mathrm{bol}}_{\odot }})$, where ${M}_{{\mathrm{bol}}_{\odot }}$ is the bolometric magnitude for the Sun.  The MS variable stars have been classified according to their periods of variability, the shape of light curves, and their positions in the H-R diagram. We detected one star as $\beta$ Cep-type. Four stars Nos.~679, 886, 1122 and 1352 are located in the instability strip of slowly pulsating B type (SPB) stars. The positions in the cluster H-R diagram as well as the observed variability characteristics of nine stars allow us to conclude that these variables belong to the new class variables. One star based on its location in the $H-R$ diagram should be $\delta$ Scuti-type variable.

In the present study, we have detected 48 PMS stars as most likely cluster members in the PMS stage of evolution. Of these, 4, 8, and 36 stars are classified as Herbig Ae/Be stars, classical TTSs, and weak-lined TTSs, respectively. The amplitudes of weak-lined TTSs range from $\sim$0.05 to $\sim$0.2 mag, and most weak-lined TTSs vary with shorter periods of less than 1.0~days. The periods and amplitudes of classical TTSs are found to range from $\sim0.05$ to $\sim30$~days and $\sim0.2$ to $\sim0.7$~mag, respectively. The above results suggest that stars with disks, i.e., classical TTSs, exhibit relatively larger amplitudes than the weak-lined TTSs do, with the stellar variability in classical TTSs arising from the presence of the spots, hot and cold, on the stellar surfaces as found in the previous studies (e.g. Bouvier et al. 1993; Pandey et al. 2019). 

\subsection{Known variables}

In the CCD search for variable stars in NGC\,6823, Pigulski et al. (2000) demonstrated that all stars with spectral types later than A0 are PMS objects.  They detected two variable stars of $\delta$ Scuti type and these stars could be at the PMS stage of evolution and suggested that these objects can further be used to test the evolutionary changes in this class of variable stars. The CMD was used to compare and discuss the position of the two discovered $\delta$ Scuti stars with reference to the theoretical instability strip for PMS stars of this type.  They have also 13 other variables including one bright cluster eclipsing binary and an SPB candidate. 

Of 15 variables identified by Pigulski et al (2000), 14 were found to be variable in the present work. We could not detect variability in the star H8 (E88 or BL 4) (B0\,V:pe by Turner 1979), B1.5\,V by Massey et al. 1995), and B1\,V by Shi \& Hu (1999). Pigulski et al (2000) noted this stars as the brightest variable member in the observed cluster field and found to be a binary star where only one eclipse was detected in the $I$ band. 

Now we will describe the nature of all the known 14 variable stars individually.

Stars BL 50 (822) and HP 57 (1007) with periods 0.0718530 days, 
0.10114 days for BL 50 and 0.0785819 days, 
0.0644149 days for HP 57
were found to be most likely cluster PMS members by Pigulski et al. (2000). With their positions in the cluster CMD as well as the observed periods Pigulski et al. (2000) concluded that both objects could be $\delta$ Scuti variables. The present membership analysis, i.e.,  from kinematics and positions in various CMDs and TCDs, suggests both stars to be MS members. In the H-R diagram star No.~822 is positioned where new class variables are found (between the red edge of SPB and the blue edge of $\delta$ Scuti instability strip).  Star No.~1007 could not be placed in the H-R diagram due to unavailability of $UBV$ data. The present period of stars 822 is derived as 0.143 days and 0.084 days while the periodogram analysis gives period of 0.064 days for star 1007. The period derived for star 1007 is in good agreement with that derived by Pigulski et al. (2000). 
The location of these two stars were shown with red and black triangles in $V$ versus $(V-I)$ CMD and $J-H$
versus $H-K$ TCD, respectively. 

Star No.~903, a probable cluster MS member was discovered as the third pulsator (G\,51) by Pigulski et al. (2000). Its brightness varies with a period of 0.848~days with an amplitude about 0.03~mag.  The star was classified by Pigulski et al. (2000) as an SPB variable according to their variability characteristics.  

The brightness of star No.~886 (G52) found to be binary by Pigulski et al. (2000) varies with period of 0.61~days with an amplitude 0.03~mag. It is diagnosed as a member of the cluster from proper motion and its location in various photometric diagrams. The present estimates for period and amplitude are consistent with those reported in Pigulski et al. (2000).

Star No.~733 has proper motion values of $\mu_\alpha = -1.671$~mas/yr and $\mu_\delta= -5.453$~mas/yr, hence is a probable member of the cluster. Our analysis suggests possibly more than one period, with 0.143~d and 0.512~d. In Pigulski et al. (2000) it is H30, and they found its period of more than 3~days. 

The brightness of star No.~757 was found to be changing with one single period of 0.553~days. The present work classified this star to be a probable PMS cluster member. Pigulski et al. (2000) named it V2 and derived its period of about 1.24~days, commenting that the true period for this star corresponds to an alias frequency, and they found this star to be of PMS type source. The present observations confirm its variability and its PMS nature. The light curves and periodogram analysis manifest that it could be an eclipsing binary with primary and secondary depths being nearly equal.  Morales-Calderon et al. (2012) found six new candidate sources as PMS eclipsing binaries with multi-epoch data of about 2400 stars associated with the Orion Nebula Cluster, and it is stated that the PMS eclipsing binaries are valuable as they are in the stage of PMS evolution which is highly dynamic, therefore their detection is rare at this stage.

Star No.~924 may be a PMS variable with light curve varying with more than one period. The proper motion suggests a nonmember of the cluster but its position in TCDs and CMDs indicates a possible weak-lined TTSs. It is designated as V4 by Pigulski et al. (2000).

Star V5 of Pigulski et al. (2000) is numbered as No.~1061 in this work. We considered it a field star based on its location in CMDs and TCDs. Its brightness in $V$ and $I$ bands varies with a period of 0.438~days. The
variability of this star is confirmed in the present work. 

Star V8 (979) which could be a classical TTS based on its location in the $J-H$ versus $H-K$ diagram. It has a period of about 0.038~days. The kinematic data indicate it to be a possible member of the cluster of PMS nature. Pigulski et al. (2000) also
found it to be suspected PMS variable.

Star No.~753 was designated as V7 by Pigulski et al. (2000). The period of V7 could not be found by Pigulski et al. (2000) due to either irregular brightness or long-period variations.  In our analysis, this star is considered a probable member of the cluster according to the proper motion study. Its position in the CMD and TCDs suggests a PMS Class~II object. This star shows periodic brightness variation with its period and amplitude being 0.036~days and 0.225~mag, respectively.

Star No.~655 (V3 in Pigulski et al. 2000) is a periodic variable with two possible periods, of about 17~days and of 0.059~days. The location of this star in the $J-H$ versus $H-K$ TCD suggests a Class~II source while proper motion data also suggest cluster membership.  

Star No.~1087 is found to be nonmember based on its proper motion values. Its location in TCDs suggests a PMS source.  Its brightness changes periodically with a period of $\sim0.125$~days. In Pigulski et al. (2000) this star (V1) was the reddest object among their variable sample.  

Star No.~831 is referred to as V6 by Pigulski et al. (2000) and they found this star too red as a member of the
cluster.  We confirm this star, with a period of about 1~day, to be a field star.  

Star No.~623, E\,100, is a PMS object, though it was considered as a nonmember of the cluster in Pigulski et al. (2000) because its proper motion values were different from those of cluster members (Erickson 1971). They mentioned that this star might belong to the foreground population and it is of a late type object. The
present estimation of its membership using Gaia data also finds it to be nonmember. 

\begin{figure}
\includegraphics[width=9cm]{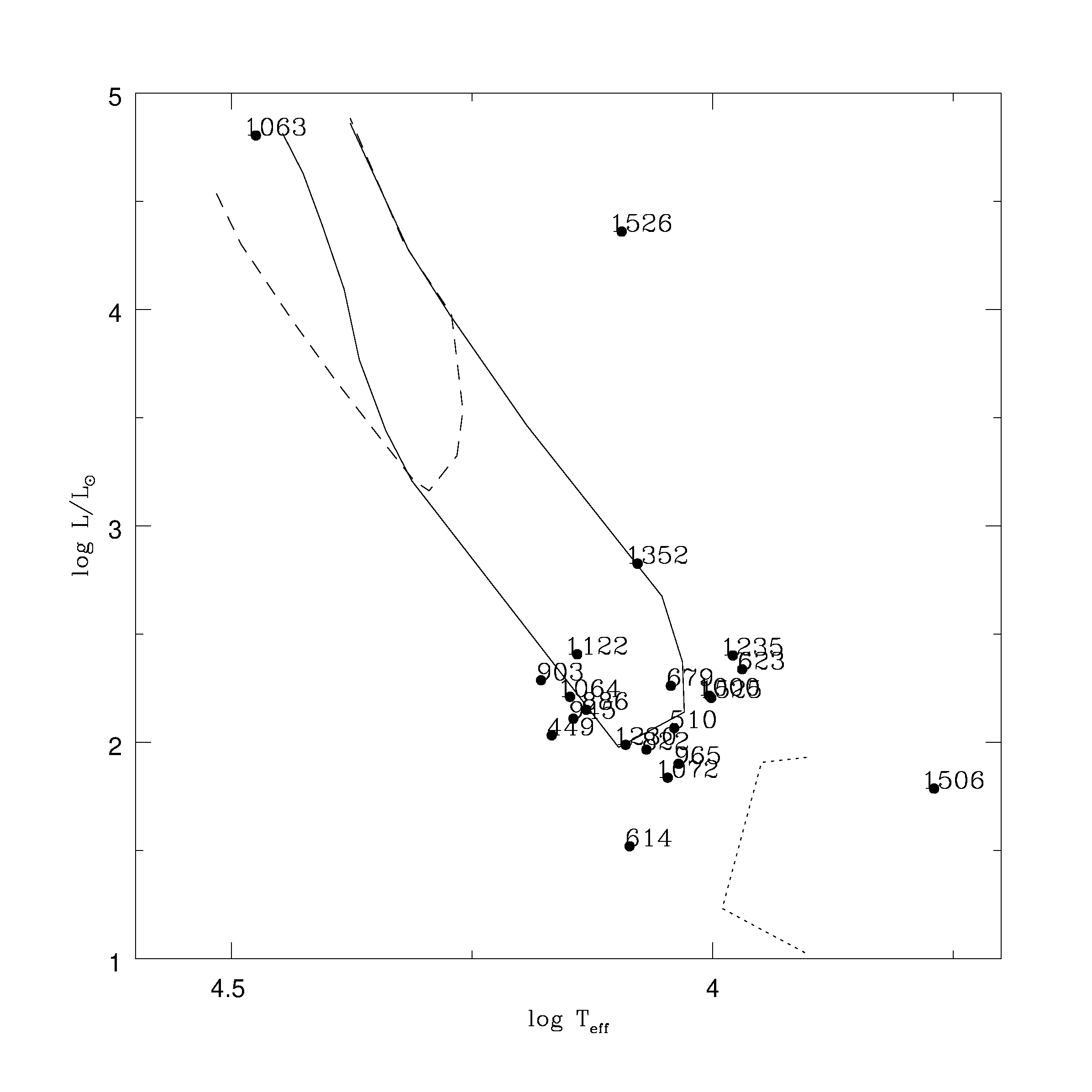}
\caption{ $\log(L/L_{\odot})/ \log T_{\rm eff}$ diagram for the probable MS variable stars identified in the present study.
The continuous curve represents the instability strip of SPB stars whereas dotted curve shows the instability region of $\delta$ Scuti stars. The dashed curve shows the location of $\beta$ Cep stars (cf. Balona et al. 2011).
}
\end{figure}

\begin{table}[h]
\caption{Period and amplitude of variable stars. Last column represents membership classification of stars along with their 
classification based on variability characteristics. The stars with asterisk are previously known variables. The cTTS, wTTS and HAe/Be are classical, weak-lined TTS and Herbig Ae/Be star, respectively.
}
\tiny
\begin{tabular}{lllll}
\hline
ID &          Period        & Period (TESS)  & Amp.    &    class.  \\
   &          (days)        & (days)         & (mag)   &            \\
\hline 
103&           1.919          &1.913    & 0.087   &   PMS, wTTS        \\
135&           0.041, 0.010   &-        & 0.336   &   PMS, cTTS \\
142&           0.504          &-        & 0.156   &   PMS, wTTS      \\ 
147&           0.940, 0.071   &-        & 0.080   &   PMS, wTTS      \\
154&           1.008          &-        & 0.580   &   PMS, HAe/Be    \\
177&           0.784          &-        & 0.129   &   PMS, wTTS    \\
201&           0.850, 0.845   &-        & 0.115   &   Field    \\
213&           0.253, 0.509   &-        & 0.167   &   Field    \\
238&           0.497          &-        & 0.166   &   Field    \\
239&           0.506, 0.969          &0.889, 2.429        & 1.028   &   PMS, wTTS    \\
240&          1.109, 0.067    &2.253    & 0.032   &   MS    \\
264&           0.546          &-        & 0.202   &   PMS, wTTS    \\
298&           0.357, 0.082   &-        & 0.107   &   PMS, HAe/Be    \\
369&          5.699           &2.400, 0.600, 7.041 & 0.288   &   PMS    \\
377&           0.0332         &-        & 0.103   &   Field    \\
385&          30.100, 0.958          &5.243,  3.960, 0.939      & 0.425   &   PMS, HAe/Be    \\
402&            0.804         &-        & 0.394   &   PMS, wTTS    \\
449&           3.852, 0.789   &5.297, 0.713, 4.084        & 0.037   &   Field    \\
452&            3.572         &-        & 0.203   &   PMS, wTTS    \\
478&           9.199, 0.898       &3.066,  2.622, 0.902        & 0.366   &   PMS, wTTS    \\
502&          1.138, 0.887    &-        & 0.238   &   PMS, wTTS    \\
510&           0.143, 0.030   &-        & 0.025   &   MS, New    \\
527&           0.671, 2.057   &2.045, 2.879        & 0.168   &   PMS, wTTS    \\
529&           0.047          &-        & 0.290   &   PMS, wTTS    \\
531&           0.755, 3.064          & 3.084, 6.232, 9.713         & 0.219   &   PMS, wTTS    \\
546&           0.806, 0.057   &-     & 0.044   &   Field    \\
561&          10.300, 0.909   &3.24        & 0.149   &   PMS, wTTS    \\
576&           0.882          &0.850, 3.715        & 0.237   &   PMS, wTTS    \\
614&           0.707          &0.705, 1.775        & 0.051   &   MS    \\
619&           0.852          &3.825, 0.850        & 0.030   &   Field    \\
623*&          0.044, 0.053   &0.153, 1.595       & 0.022   &   MS    \\
655*&         17.716, 0.030   &-       & 0.236   &   PMS, cTTS \\
679&           1.1242, 0.059, 0.564  & 0.565, 3.376        & 0.046   &  MS \\
706&           4.336, 0.561   &-        & 0.049   &   PMS, wTTS \\
731&           0.359, 0.099   &-        & 0.065   &   PMS, wTTS \\
733*&          0.512, 0.143   &-       & 0.029   &   MS \\
752&            0.153         &0.153, 10.770        & 0.257   &   MS \\
753*&          0.036          &-       & 0.225   &   PMS, cTTS \\
757*&          0.553          &-       & 0.125   &   PMS, wTTS \\
765&           0.112, 0.124   &-        & 0.133   &   Field \\
822*&          0.143, 0.084   &-       & 0.034   &   MS, New \\
826&           0.072          &-        & 0.139   &   PMS, wTTS \\
831*&          1.009          &1.147, 1.545, 1.095        & 1.463   &   Field \\
860&           8.517, 0.523   &-        & 0.197   &   PMS, cTTS \\
886*&          0.446, 0.618   &-       & 0.032   &   MS \\
903*&          0.848          &-       & 0.033   &   MS \\
924*&          3.206, 0.59, 0.759&-    & 0.123   &   PMS, wTTS \\
945&           0.662, 0.663, 1.965   &1.966, 6.000       & 0.026   &   MS \\
950&           0.504          &3.179, 2.442        & 0.265   &   PMS, wTTS \\
951&           1.392, 0.775   &-        & 0.064   &   PMS, wTTS \\
965&           2.661, 0.726   &2.65, 4.805        & 0.053   &   MS, New \\
979*&          0.036          &0.064, 5.002       & 0.185   &   PMS, cTTS \\
1000&         0.042, 0.486    &-      & 0.016   &   MS, New \\
1007*&        0.064           &0.527, 0.064, 4.818       & 0.037   &   MS \\
1025&         0.059           & -     & 0.600   &   PMS, wTTS \\
1061*&        0.438           &1.581, 6.269      & 0.099   &   Field \\
1063&         1.049, 0.028    &-      & 0.131   &   MS, $\beta$ Cep \\
1064&         0.653, 0.059,  0.395&0.804, 5.291  & 0.025   &   MS \\
1066&         0.0518, 0.082   &2.166, 1.203      & 0.013   &   PMS, wTTS \\
1072&         0.484, 0.032    &0.819, 11.649      & 0.025   &   MS New \\
1087*&       0.125            &-      & 0.103   &   PMS, wTTS\\
1094&         0.815            &-      & 0.081   &   PMS, wTTS \\
1122&         2.402, 0.705    &0.705, 11.649      & 0.039   &   MS,   SPB \\
1151&         0.769           &0.153, 4.834      & 0.359   &   Field \\
1155&        10.955, 0.100    &-      & 0.092   &   PMS, wTTS \\
1168&         0.526           &-      & 0.240   &   Field \\
1191&         0.924           &-      & 0.481   &   Field \\
1228&         0.902           &-      & 0.448   &   PMS, wTTS \\
1230&         0.386, 0.629    & 0.385, 1.755      & 0.019   &   MS, New \\
1235&         1.622, 3.267    &3.243       & 0.211   &   MS, New \\
1262&         1.215, 0.446    &-      & 0.058   &   PMS, wTTS \\
1266&         3.382           &-      & 0.148   &   PMS, wTTS \\
1268&        63.269           &5.240, 0.996      & 0.276   &   PMS, wTTS \\
1295&         0.028           &-      & 0.495   &   PMS, cTTS \\
1298&        0.983, 0.496     &-      & 0.438   &   PMS, cTTS \\
1317&         0.485           &-      & 0.671   &   PMS, cTTS \\
1352&         0.027, 0.336    &-      & 0.014   &   MS, SBP \\
1389&        0.111, 0.166     &-      & 0.086   &   PMS, HAe/Be \\
1405&         0.077, 0.064    &-      & 0.128   &   Field \\
1406&         0.140, 0.082    &-      & 0.123   &   Field \\
1459&         0.865           &-      & 0.172   &   PMS, wTTS \\
1500&         0.072           &-      & 0.046   &   MS \\
1506&        0.259, 0.491     &-      & 0.031   &   MS \\
1508&          0.027          &-      & 0.264   &   Field \\
1511&         0.059, 0.110    &-      & 0.032   &   Field \\
1525&        1.132, 0.063     &-      & 0.034   &   MS, New \\
1526&        0.058            &-      & 0.075   &   MS \\
1548&        0.986, 0.329     &-      & 0.174   &   PMS, wTTS \\
\hline
\end{tabular}
\end{table}
\subsection{Newly detected variables}

Now we present newly identified variables in this work.  Star No.~1235 is classified, on the basis of the shape of its light curve, to be an eclipsing binary, bearing similarity to that of an EA (Algol) type. In EA type eclipsing binaries, both stars are nearly spherical in shape, with an extremely wide range of periods from 0.2~days to 10000~days, and with a wide range of amplitude of variability.  Star No.~1235 has a period of 1.622~days and a variation amplitude of 0.211~mag. In the $H-R$ diagram, this star is found to be located in the region of new class variables.
More observations of this star are required to confirm its nature. This star is a member of the cluster based on locations in TCDs. However, Gaia data suggest a nonmember of the cluster.   

The light curves of star No.~449 in both $V$ and $I$ bands reveal it to be a short-period variable. Its periodogram in both $V$ and $I$ exhibits peaks around 3.2~days and 0.789~days. Gaia data suggest it to be a probable member.

The variability of the star No.~527 suggest that it is a periodic variable whose light varies with a period of 0.671~days.  It is found to be a probable member of the cluster from its proper motion measurements. Its location in CMDs and TCDs indicates a PMS object.

The brightness of star No.~531, a probable PMS member of the cluster, is found to vary with a period of 0.755~days or 3.065 days, with the variability characteristics consistent with TTSs. 

The proper motion values are not in favor of star No.~752 to be a cluster member, though in the $V$ versus $V-I$ CMD, it is located along the MS. The period is derived as 0.153~days and the amplitude is about 0.2~mag. The variability characteristics of this star is similar to a pulsating type star or eclipsing binary. After doubling its period its light curves show two minima which have almost equal depth.  It could be an EW type eclipsing (W Ursae Majoris eclipsing system). The EW type variables have periods of less than one day, with almost equal depths of primary and secondary minima.

Star No.~1508 has a period of $\sim0.027$~day with an amplitude of about 0.2~mag. This variable resembles that of the SX Phe type (Cohen \& Sarajedini 2012), which are similar to $\delta$ Scuti stars but pulsate with amplitudes up to 0.7~mag according to the variability types listed in the General Catalog of Variable Stars (GCVS).

\begin{table*}
\caption{The proper motion, parallax and photometry by Gaia.  
The last column refers to likely or possible membership for each variable star.
}
\tiny
\begin{tabular}{llllllllll}
\hline
ID  &    RA   &       DEC    &    $\mu_{ra}$  &  $\mu_{Dec}$ &  plx  &   gmag &  bpmag  &   rpmag  &  mem \\
     & degree &       degree &    mas/yr    &  mas/yr   &      mas &    (mag) & mag   &          &    \\
\hline
 103&  295.667936 & 23.412217 & -1.392$\pm$0.046& -5.264$\pm$0.077&   0.500$\pm$0.074& 17.288$\pm$0.003& 18.560$\pm$0.015& 16.172$\pm$0.006& 2 \\
 135&  295.729818 & 23.406842 & -1.335$\pm$0.050& -5.330$\pm$0.079&   0.605$\pm$0.085& 17.340$\pm$0.007& 18.463$\pm$0.025& 16.006$\pm$0.017& 1 \\
 142&  295.921757 & 23.403326 & -4.224$\pm$0.059& -6.330$\pm$0.096&  -0.125$\pm$0.102& 16.699$\pm$0.003& 21.455$\pm$0.111& 14.950$\pm$0.005& 0 \\
 147&  295.717480 & 23.404825 & -1.601$\pm$0.043& -5.533$\pm$0.068&   0.154$\pm$0.071& 17.151$\pm$0.003& 18.555$\pm$0.013& 15.988$\pm$0.005& 1 \\
 154&  295.729082 & 23.404078 & -1.583$\pm$0.012& -5.173$\pm$0.018&   0.411$\pm$0.019& 13.680$\pm$0.003& 15.006$\pm$0.006& 12.486$\pm$0.006& 2 \\
 177&  295.798833 & 23.398720 & -1.430$\pm$0.059& -5.249$\pm$0.097&   0.382$\pm$0.099& 17.841$\pm$0.003& 19.238$\pm$0.024& 16.679$\pm$0.007& 2 \\
 201&  295.806239 & 23.392781 & -1.829$\pm$0.065& -4.434$\pm$0.111&   0.764$\pm$0.121& 18.017$\pm$0.003& 18.879$\pm$0.021& 17.098$\pm$0.009& 1 \\
 213&  295.678068 & 23.391730 & -2.955$\pm$0.072& -2.616$\pm$0.122&   0.692$\pm$0.126& 18.086$\pm$0.003& 18.923$\pm$0.017& 17.191$\pm$0.008& 0 \\
 238&  295.856620 & 23.385850 & -1.810$\pm$0.048& -5.933$\pm$0.080&  -0.041$\pm$0.079& 15.379$\pm$0.005& 21.173$\pm$0.091& 13.658$\pm$0.008& 1 \\
 239&  295.792933 & 23.386486 & -1.545$\pm$0.133& -5.335$\pm$0.205&   0.314$\pm$0.186& 19.537$\pm$0.012& 20.487$\pm$0.072& 18.373$\pm$0.046& 1 \\
 240&  295.913449 & 23.384987 & -2.683$\pm$0.022& -9.082$\pm$0.035&   0.809$\pm$0.037& 16.070$\pm$0.003& 16.629$\pm$0.004& 15.346$\pm$0.004& 0 \\
 264&  295.758973 & 23.382855 & -1.839$\pm$0.081& -4.993$\pm$0.141&   0.316$\pm$0.146& 18.200$\pm$0.003& 19.690$\pm$0.028& 17.004$\pm$0.007& 1 \\
 298&  295.707379 & 23.376514 & -1.757$\pm$0.036& -4.931$\pm$0.055&   0.518$\pm$0.058& 16.844$\pm$0.005& 17.817$\pm$0.017& 15.833$\pm$0.012& 2 \\
 369&  295.889138 & 23.363276 & -2.759$\pm$0.058& -5.104$\pm$0.097&   0.180$\pm$0.097& 17.647$\pm$0.003& 19.962$\pm$0.039& 16.257$\pm$0.008& 0 \\
 377&  295.880082 & 23.361511 & -1.767$\pm$0.066& -5.422$\pm$0.102&   0.503$\pm$0.107& 18.157$\pm$0.007& 19.323$\pm$0.030& 17.036$\pm$0.021& 2 \\
 385&  295.847545 & 23.360887 & -1.713$\pm$0.025& -4.949$\pm$0.039&   0.454$\pm$0.041& 16.211$\pm$0.006& 17.138$\pm$0.021& 15.246$\pm$0.015& 2 \\
 402&  295.716202 & 23.358795 &  1.039$\pm$0.544& -9.595$\pm$0.908&  -1.529$\pm$1.016& 19.114$\pm$0.006& 20.359$\pm$0.063& 17.606$\pm$0.020& 0 \\
 449&  295.874874 & 23.347861 & -1.534$\pm$0.011& -5.289$\pm$0.018&   0.421$\pm$0.019& 14.456$\pm$0.003& 15.283$\pm$0.003& 13.559$\pm$0.004& 2 \\
 452&  295.919127 & 23.346023 & -1.632$\pm$0.091& -5.204$\pm$0.163&   0.187$\pm$0.154& 18.342$\pm$0.003& 19.687$\pm$0.034& 16.889$\pm$0.010& 1 \\
 478&  295.846924 & 23.343243 & -2.242$\pm$0.123& -5.663$\pm$0.287&   0.414$\pm$0.207& 18.715$\pm$0.003& 20.513$\pm$0.146& 17.239$\pm$0.011& 2 \\
 478&  295.847334 & 23.343217 & -1.835$\pm$0.094& -5.443$\pm$0.160&   0.513$\pm$0.166& 18.445$\pm$0.005& 19.556$\pm$0.035& 16.804$\pm$0.017& 2 \\
 502&  295.883889 & 23.339095 & -1.539$\pm$0.061& -5.143$\pm$0.098&   0.633$\pm$0.100& 17.859$\pm$0.005& 19.066$\pm$0.032& 16.755$\pm$0.015& 1 \\
 510&  295.795529 & 23.339088 & -1.736$\pm$0.010& -5.177$\pm$0.015&   0.483$\pm$0.016& 13.870$\pm$0.003& 14.370$\pm$0.003& 13.188$\pm$0.004& 2 \\
 527&  295.868122 & 23.335831 & -1.350$\pm$0.032& -5.299$\pm$0.052&   0.467$\pm$0.055& 16.580$\pm$0.003& 17.741$\pm$0.006& 15.496$\pm$0.005& 2 \\
 529&  295.910436 & 23.335234 & -1.692$\pm$0.169& -5.130$\pm$0.251&   0.688$\pm$0.245& 19.127$\pm$0.007& 20.413$\pm$0.056& 17.532$\pm$0.019& 1 \\
 531&  295.850017 & 23.335616 & -1.798$\pm$0.030& -5.427$\pm$0.046&   0.477$\pm$0.050& 16.447$\pm$0.004& 17.610$\pm$0.013& 15.360$\pm$0.009& 2 \\
 546&  295.746235 & 23.334619 &  1.642$\pm$0.054& -0.744$\pm$0.078&   0.666$\pm$0.079& 17.331$\pm$0.003& 18.235$\pm$0.008& 16.373$\pm$0.004& 0 \\
 561&  295.717802 & 23.332530 & -1.806$\pm$0.058& -5.112$\pm$0.077&   0.449$\pm$0.084& 17.330$\pm$0.003& 18.517$\pm$0.015& 16.256$\pm$0.006& 2 \\
 576&  295.862641 & 23.329178 & -1.923$\pm$0.092& -5.421$\pm$0.138&   0.003$\pm$0.150& 18.117$\pm$0.004& 19.508$\pm$0.028& 16.934$\pm$0.008& 1 \\
 614&  295.686941 & 23.325122 &  0.064$\pm$0.021& -2.606$\pm$0.028&   0.631$\pm$0.030& 15.454$\pm$0.003& 16.096$\pm$0.004& 14.665$\pm$0.004& 0 \\
 619&  295.856809 & 23.323021 & -1.705$\pm$0.018& -5.254$\pm$0.028&   0.468$\pm$0.030& 15.271$\pm$0.003& 16.180$\pm$0.004& 14.321$\pm$0.004& 2 \\
 623&  295.834392 & 23.322264 & -5.268$\pm$0.009&-19.204$\pm$0.013&   1.568$\pm$0.014& 12.832$\pm$0.003& 13.444$\pm$0.003& 12.086$\pm$0.004& 0 \\
 655&  295.837455 & 23.317270 & -1.973$\pm$0.042& -5.777$\pm$0.068&   1.110$\pm$0.079& 16.811$\pm$0.008& 18.136$\pm$0.032& 15.636$\pm$0.023& 1 \\
 679&  295.768320 & 23.313487 & -1.905$\pm$0.009& -5.483$\pm$0.013&   0.465$\pm$0.015& 13.468$\pm$0.003& 13.893$\pm$0.003& 12.833$\pm$0.004& 2 \\
 706&  295.752612 & 23.310267 & -1.994$\pm$0.035& -5.380$\pm$0.051&   0.518$\pm$0.052& 16.539$\pm$0.003& 17.599$\pm$0.007& 15.508$\pm$0.005& 2 \\
 731&  295.789104 & 23.307583 & -1.690$\pm$0.068& -5.348$\pm$0.086&   0.444$\pm$0.094& 17.553$\pm$0.003& 18.975$\pm$0.020& 16.406$\pm$0.007& 2 \\
 733&  295.798253 & 23.307303 & -1.671$\pm$0.016& -5.453$\pm$0.021&   0.447$\pm$0.022& 14.822$\pm$0.003& 15.418$\pm$0.003& 14.057$\pm$0.004& 2 \\
 752&  295.737297 & 23.306309 & -4.190$\pm$0.024& -9.756$\pm$0.033&   0.807$\pm$0.036& 15.715$\pm$0.004& 16.429$\pm$0.010& 14.875$\pm$0.009& 0 \\
 753&  295.798228 & 23.305611 & -1.439$\pm$0.091& -5.723$\pm$0.130&   0.527$\pm$0.132& 18.213$\pm$0.010& 19.645$\pm$0.052& 16.983$\pm$0.037& 2 \\
 757&  295.803663 & 23.305216 & -1.823$\pm$0.027& -5.343$\pm$0.036&   0.420$\pm$0.038& 16.075$\pm$0.003& 17.227$\pm$0.006& 15.000$\pm$0.006& 2 \\
 765&  295.785197 & 23.304768 & -1.971$\pm$0.055& -5.559$\pm$0.075&   0.582$\pm$0.084& 17.403$\pm$0.003& 18.407$\pm$0.020& 16.391$\pm$0.007& 1 \\
 822&  295.787697 & 23.296925 & -1.931$\pm$0.012& -5.372$\pm$0.017&   0.440$\pm$0.020& 14.172$\pm$0.003& 14.769$\pm$0.004& 13.413$\pm$0.004& 2 \\
 826&  295.825083 & 23.296002 &  0.000$\pm$0.000&  0.000$\pm$0.000&   0.000$\pm$0.000& 18.717$\pm$0.006& 19.528$\pm$0.088& 16.822$\pm$0.010& 0 \\
 831&  295.746221 & 23.296549 & -3.198$\pm$0.140& -6.012$\pm$0.190&   0.051$\pm$0.164& 18.030$\pm$0.034& 20.951$\pm$0.092& 16.343$\pm$0.113& 0 \\
 860&  295.798251 & 23.292496 & -1.763$\pm$0.053& -5.198$\pm$0.075&   0.296$\pm$0.086& 17.181$\pm$0.003& 18.444$\pm$0.012& 16.091$\pm$0.007& 1 \\
 886&  295.793856 & 23.290165 & -1.700$\pm$0.016& -5.457$\pm$0.017&   0.501$\pm$0.019& 14.140$\pm$0.003& 14.596$\pm$0.003& 13.490$\pm$0.004& 2 \\
 903&  295.800965 & 23.287961 & -1.443$\pm$0.018& -5.239$\pm$0.025&   0.444$\pm$0.025& 14.092$\pm$0.003& 14.566$\pm$0.003& 13.425$\pm$0.004& 2 \\
 924&  295.787347 & 23.285362 & -1.173$\pm$0.096& -7.786$\pm$0.132&  -0.663$\pm$0.160& 17.259$\pm$0.004& 18.378$\pm$0.013& 16.050$\pm$0.006& 0 \\
 945&  295.855106 & 23.282160 & -1.481$\pm$0.027& -5.241$\pm$0.040&   0.579$\pm$0.039& 14.176$\pm$0.003& 14.745$\pm$0.003& 13.423$\pm$0.004& 1 \\
 950&  295.706797 & 23.283426 & -3.581$\pm$0.166& -4.734$\pm$0.244&   1.034$\pm$0.261& 18.148$\pm$0.004& 21.250$\pm$0.095& 16.437$\pm$0.008& 0 \\
 951&  295.796964 & 23.282188 & -1.581$\pm$0.030& -5.247$\pm$0.043&   0.424$\pm$0.049& 16.307$\pm$0.003& 17.417$\pm$0.006& 15.267$\pm$0.005& 2 \\
 965&  295.891582 & 23.279985 &  0.251$\pm$0.066& -2.097$\pm$0.096&   1.647$\pm$0.102& 14.281$\pm$0.003& 14.802$\pm$0.004& 13.562$\pm$0.005& 0 \\
 979&  295.775108 & 23.280240 & -1.692$\pm$0.116& -5.685$\pm$0.169&   0.939$\pm$0.173& 18.478$\pm$0.004& 19.642$\pm$0.032& 17.336$\pm$0.011& 1 \\
1000&  295.814485 & 23.277802 & -1.142$\pm$0.009& -3.501$\pm$0.013&   0.635$\pm$0.014& 13.457$\pm$0.003& 13.714$\pm$0.003& 13.041$\pm$0.004& 0 \\
1007&  295.778155 & 23.276998 & -1.989$\pm$0.014& -5.533$\pm$0.017&   0.462$\pm$0.020& 14.277$\pm$0.003& 14.835$\pm$0.004& 13.530$\pm$0.004& 2 \\
1025&  295.690388 & 23.275981 & -1.408$\pm$0.245& -4.138$\pm$0.340&   0.725$\pm$0.434& 19.836$\pm$0.011& 20.939$\pm$0.089& 18.552$\pm$0.043& 0 \\
1061&  295.788730 & 23.269975 & -1.671$\pm$0.053& -5.426$\pm$0.070&   0.567$\pm$0.076& 17.298$\pm$0.003& 18.639$\pm$0.013& 16.154$\pm$0.005& 1 \\
1063&  295.778243 & 23.269561 & -2.431$\pm$0.095& -5.395$\pm$0.081&   0.282$\pm$0.090& 12.869$\pm$0.003& 00.000$\pm$0.000& 00.000$\pm$0.000& 1 \\
1063&  295.778279 & 23.270083 & -1.210$\pm$0.090& -5.981$\pm$0.149&  -0.091$\pm$0.141& 09.720$\pm$0.003& 09.906$\pm$0.003& 09.274$\pm$0.004& 1 \\
1064&  295.758524 & 23.270269 & -2.317$\pm$0.038& -4.554$\pm$0.053&   0.521$\pm$0.058& 14.183$\pm$0.003& 14.645$\pm$0.003& 13.493$\pm$0.004& 2 \\
1066&  295.843490 & 23.269270 & -3.124$\pm$0.021& -4.962$\pm$0.028&   0.291$\pm$0.030& 14.584$\pm$0.003& 16.863$\pm$0.004& 13.214$\pm$0.004& 0 \\
1072&  295.820520 & 23.269079 & -1.645$\pm$0.013& -5.282$\pm$0.018&   0.468$\pm$0.019& 14.395$\pm$0.003& 14.918$\pm$0.003& 13.690$\pm$0.004& 2 \\
1087&  295.798607 & 23.266730 & -3.210$\pm$0.068& -5.424$\pm$0.095&  -0.081$\pm$0.102& 16.758$\pm$0.005& 22.007$\pm$0.122& 14.938$\pm$0.008& 0 \\
1094&  295.817575 & 23.265043 & -1.438$\pm$0.023& -5.392$\pm$0.032&   0.428$\pm$0.033& 15.783$\pm$0.003& 16.830$\pm$0.005& 14.765$\pm$0.005& 2 \\
1122&  295.709131 & 23.260843 & -1.586$\pm$0.010& -5.419$\pm$0.015&   0.523$\pm$0.017& 13.588$\pm$0.003& 13.956$\pm$0.003& 13.029$\pm$0.004& 2 \\
1151&  295.805720 & 23.255733 & -1.697$\pm$0.114& -5.153$\pm$0.163&   0.173$\pm$0.158& 18.349$\pm$0.007& 19.909$\pm$0.042& 17.113$\pm$0.021& 1 \\
1155&  295.768235 & 23.255355 & -1.672$\pm$0.060& -5.173$\pm$0.080&   0.266$\pm$0.087& 17.543$\pm$0.003& 18.600$\pm$0.015& 16.523$\pm$0.007& 1 \\
1168&  295.883351 & 23.252588 & -1.697$\pm$0.098& -5.681$\pm$0.144&   0.519$\pm$0.147& 18.327$\pm$0.004& 19.726$\pm$0.041& 17.177$\pm$0.010& 2 \\
1191&  295.875000 & 23.248688 & -1.715$\pm$0.119& -5.193$\pm$0.172&   0.321$\pm$0.183& 18.703$\pm$0.006& 20.107$\pm$0.041& 17.449$\pm$0.014& 1 \\
1228&  295.850980 & 23.243847 & -1.660$\pm$0.116& -5.465$\pm$0.169&   0.200$\pm$0.170& 18.582$\pm$0.005& 20.077$\pm$0.048& 17.393$\pm$0.013& 1 \\
1230&  295.755855 & 23.244602 & -1.353$\pm$0.018& -5.260$\pm$0.023&   0.524$\pm$0.025& 14.356$\pm$0.003& 14.761$\pm$0.003& 13.759$\pm$0.004& 2 \\
1235&  295.672376 & 23.244490 &  4.176$\pm$0.009& -1.613$\pm$0.012&   0.757$\pm$0.013& 12.759$\pm$0.003& 13.223$\pm$0.003& 12.126$\pm$0.004& 0 \\
1262&  295.774562 & 23.237761 & -1.522$\pm$0.228& -5.636$\pm$0.521&   0.437$\pm$0.441& 19.291$\pm$0.006& 00.000$\pm$0.000& 00.000$\pm$0.000& 2 \\
1262&  295.774980 & 23.237889 & -0.719$\pm$0.097& -4.431$\pm$0.131&  -1.092$\pm$0.153& 16.379$\pm$0.003& 17.455$\pm$0.006& 15.228$\pm$0.007& 0 \\
1266&  295.802542 & 23.236774 & -1.624$\pm$0.068& -5.081$\pm$0.086&   0.554$\pm$0.093& 17.741$\pm$0.003& 18.960$\pm$0.022& 16.686$\pm$0.006& 1 \\
1268&  295.698738 & 23.237776 & -2.864$\pm$0.083& -5.120$\pm$0.113&   0.230$\pm$0.119& 15.750$\pm$0.004& 21.208$\pm$0.109& 13.997$\pm$0.007& 0 \\
1295&  295.812280 & 23.232347 & -1.564$\pm$0.182& -4.156$\pm$0.212&   0.354$\pm$0.225& 18.726$\pm$0.015& 20.052$\pm$0.046& 17.061$\pm$0.037& 1 \\
1298&  295.671363 & 23.233613 & -1.763$\pm$0.227&-10.821$\pm$0.301&  -1.850$\pm$0.289& 18.089$\pm$0.021& 18.880$\pm$0.071& 16.384$\pm$0.040& 0 \\
1317&  295.899959 & 23.227779 & -1.562$\pm$0.101& -5.431$\pm$0.123&   0.589$\pm$0.137& 18.039$\pm$0.010& 19.288$\pm$0.036& 16.868$\pm$0.028& 1 \\
1352&  295.816693 & 23.222874 & -0.419$\pm$0.008& -4.078$\pm$0.011&   1.831$\pm$0.012& 12.449$\pm$0.003& 12.760$\pm$0.003& 11.975$\pm$0.004& 0 \\
1389&  295.717974 & 23.217819 & -2.076$\pm$0.024& -5.544$\pm$0.032&   0.434$\pm$0.035& 15.834$\pm$0.004& 16.701$\pm$0.010& 14.881$\pm$0.008& 2 \\
1405&  295.778615 & 23.214104 & -0.016$\pm$0.083& -4.330$\pm$0.113&   0.999$\pm$0.120& 18.085$\pm$0.003& 19.069$\pm$0.014& 17.103$\pm$0.007& 0 \\
1406&  295.844691 & 23.213100 & -3.267$\pm$0.064&  1.679$\pm$0.081&   0.991$\pm$0.089& 17.530$\pm$0.003& 18.310$\pm$0.023& 16.559$\pm$0.011& 0 \\
1459&  295.758607 & 23.204733 & -1.710$\pm$0.090& -5.469$\pm$0.124&   0.343$\pm$0.133& 18.147$\pm$0.003& 19.354$\pm$0.022& 17.032$\pm$0.008& 1 \\
1500&  295.892106 & 23.195945 &  3.974$\pm$0.049& -3.001$\pm$0.071&   0.924$\pm$0.071& 15.646$\pm$0.003& 16.328$\pm$0.003& 14.820$\pm$0.004& 0 \\
1506&  295.737755 & 23.197158 & -1.684$\pm$0.022& -4.291$\pm$0.026&   0.674$\pm$0.028& 15.219$\pm$0.003& 15.857$\pm$0.004& 14.435$\pm$0.004& 0 \\
1508&  295.846040 & 23.195095 & -2.071$\pm$0.089& -4.858$\pm$0.124&   0.392$\pm$0.142& 18.061$\pm$0.005& 19.437$\pm$0.035& 16.831$\pm$0.013& 2 \\
1511&  295.813526 & 23.194842 &  4.169$\pm$0.025& 20.500$\pm$0.032&   3.135$\pm$0.035& 15.890$\pm$0.003& 16.829$\pm$0.004& 14.933$\pm$0.004& 0 \\
1525&  295.817542 & 23.191915 & -0.781$\pm$0.009& -7.011$\pm$0.012&   0.655$\pm$0.013& 13.206$\pm$0.003& 13.785$\pm$0.003& 12.468$\pm$0.004& 0 \\
1526&  295.740382 & 23.192673 & 59.603$\pm$0.010&-58.093$\pm$0.014&   9.066$\pm$0.015& 08.663$\pm$0.003& 08.977$\pm$0.003& 08.173$\pm$0.004& 0 \\
1548&  295.840568 & 23.186899 & -1.083$\pm$0.043& -3.305$\pm$0.056&   0.248$\pm$0.060& 13.737$\pm$0.003& 18.375$\pm$0.013& 12.116$\pm$0.006& 0 \\
\hline
\end{tabular}
\end{table*}

\section {TESS Light curves}

A few variables like No.~1235 and No.~752 have times series data from the Transiting Exoplanet Survey Satellite (TESS; Ricker et al. 2015). The high-quality light curves from the TESS can be used to understand stellar and planetary evolution and this data provide us opportunity to study the rotation of stars (Canto Martins et al. 2020). Here, we present folded light curves, exhibited as Fig.~18 for 32 stars which do not have flux contribution from nearby brighter sources to account for the low spatial resolution of TESS. TESS observes the sky in sectors with each sector observed for about 27~days.  The eleanor pipeline to extract times series data of objects from TESS images has been used, which is an open-source tool (Feinstein et al. 2019, https://archive.stsci.edu/hlsp/eleanor).
 We can use the eleanor package to create light curves for fainter objects for a more detailed or optimized analysis of individual objects (Feinstein et al. 2019). The eleanor uses TESS Full Frame Images (FFIs) to extract systematics-corrected flux for any given star observed by TESS. It takes TIC ID, coordinates (RA and DEC) of a star. First, the raw flux is calculated by aperture photometry as RAW$\_$FLUX that is background subtracted. This raw flux is then corrected for possible systematic effects, which creates a flux called CORR$\_$FLUX. For isolated stars, to obtain the corrected flux we have taken default apertures. The eleanor software also provides the option to define one's own aperture.  We have extracted light curves of all the detected in the present photometry. 

Out of 32 variables, there are 7 stars which are diagnosed as PMS and 14 as MS. 
The periods of all the 32 stars have been determined using the method described in the Section 2.1. The periods for 14 stars (103, 527, 531, 576, 614, 619, 679, 752, 945, 965, 1007, 1122, 1230
and 1235) are found to be in good agreement with that obtained from the present ground based optical data. 
The nature of star No.~752 and No.~1235 as mentioned earlier is confirmed from their TESS light curves; that is, star No.~752 shows unequal maxima, likely due to the O'Connell effect (O'Connell 1951), for which the maxima between eclipses in some eclipsing binaries are not found equal in brightness (Knote et al. 2022). 
The phased light curves of stars Nos.~369, 561, and 619 show brightness variations similar to Algol type eclipsing binaries. The light curves of stars 527, 531, 576, 965, 
1064, 1072 and 1122 were folded by doubling the value of their derived period. Three stars 527, 
531 and 576 of them are probable PMS stars while the remaining  4 stars are cluster members of MS type. The folded light curves and periods of 1064, 1072 and 1122 are similar to the variability characteristic of EW type variables. 
The light curve of star 576 seems to have properties of EA type variable. The stars 527 and 965 could be weak-lined TTSs based on their variability characteristics as these sources are of PMS nature and show periodic variability. The period of star 531 was derived as 3.084 days using TESS data while its period comes out to be 0.755 days from $V$ and $I$ band light curves. 
The variability characteristics for those stars whose periods determined from present $V$ and $I$ data do not match with that derived from TESS observations could be revisited
in the future observations of the cluster NGC 6823. The conflicts between periods for some cases may arise due to the contribution of flux from nearby stars in TESS data despite being selected isolated stars.
The present work identified 5 MS, 4 PMS stars and 1 field variable of eclipsing nature, two of which are confirmed eclipsing binaries and remaining are suspected ones. Their derived parameters are listed in Table 5. The masses and ages of two suspected PMS binaries
could not be obtained due to unavailability of their $V-I$ color. Since $UBV$ data of MS star no. 752 are not available, the temperature for this star has been obtained using theoretical models of Girardi et al. (2002) and present $V$ magnitude. 

\begin{table}[h]
\caption{The derived parameters of the confirmed/suspected eclipsing binaries. The last column 
refers to binary classification. 
}
\tiny
\begin{tabular}{llllllll}
\hline
ID &                Age    &   Mass        & ${M}_{\mathrm{bol}}$  &  $\log (L/L_{\sun})$ & $\log T_{\rm eff}$ & class. & Binary  \\
   &                Myrs   &   M$_{\odot}$ & mag &                       &                    &        &  class.             \\
\hline
369&               -       & -             & -  & -                    &-                    &  PMS   &  EA?          \\
561&               1.3     & 1.28          & -  & -                    &-                    &  PMS   &  EA?           \\ 
576&               -       & -             & -  & -                    &-                    &  PMS   &  EA?           \\      
619&               -       & -             & -  & -                    &-                    &  Field &  EA?           \\
752&               4.0     & 1.95          & -  &                 &3.915               &  MS    &  EW            \\
757&               0.4     & 1.50          & -  & -                    &-                    &  PMS   &  EW?           \\
1064&              4.0     & 3.40          & -0.794  & 2.211                &4.148               &  MS    &  EW?          \\
1072&              4.0     & 3.20          & 0.1394  & 1.837                &4.047               &  MS    & EW?           \\
1122&              4.0     & 4.59          & -1.286  & 2.407                &4.141               &  MS    & EW?           \\
1235&              4.0     & 6.75          & -1.272  & 2.402                &3.979               &  MS    & EA            \\
\hline
\end{tabular}
\end{table}

\begin{figure*}
\hbox{
\includegraphics[width=9cm, height=9cm]{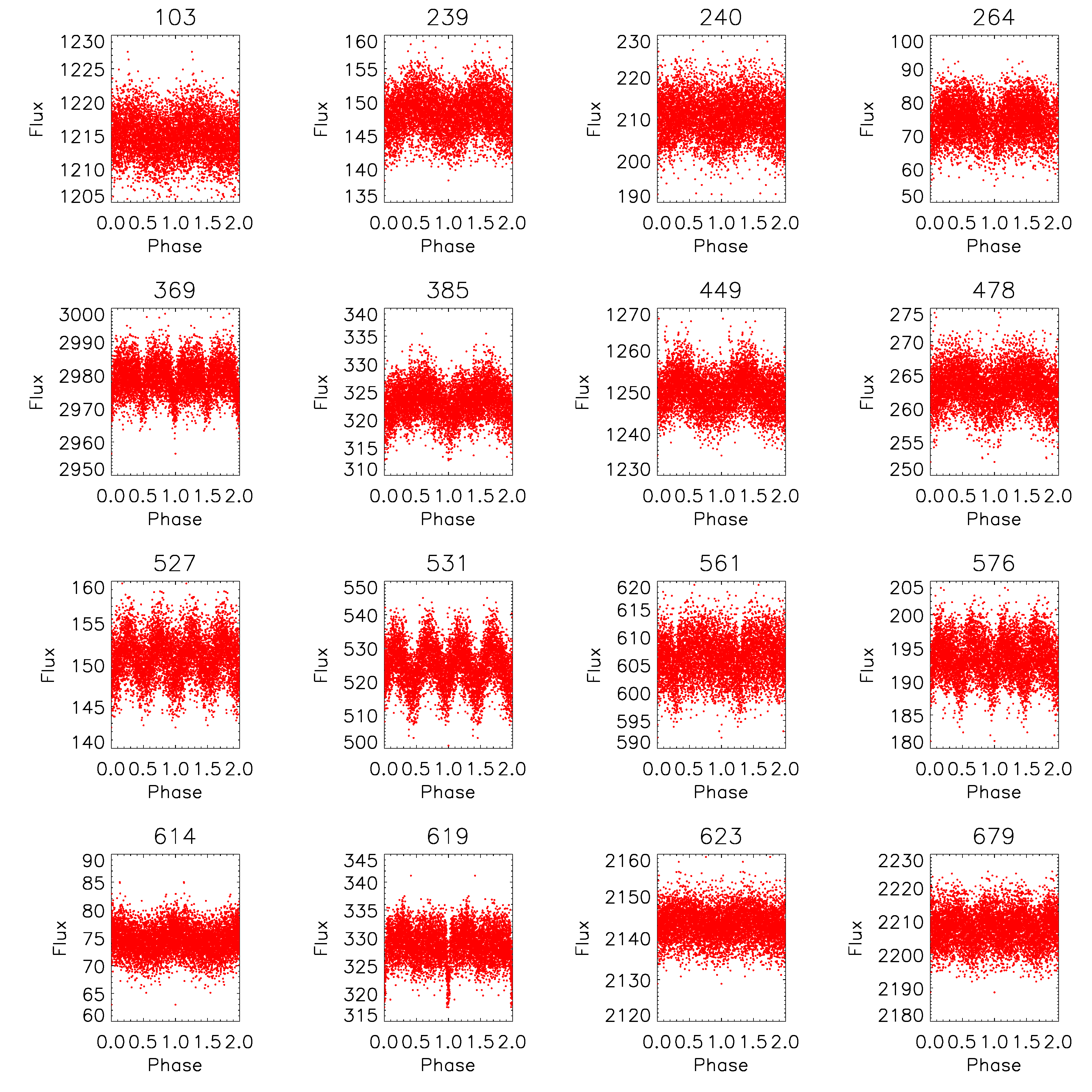}
\includegraphics[width=9cm, height=9cm]{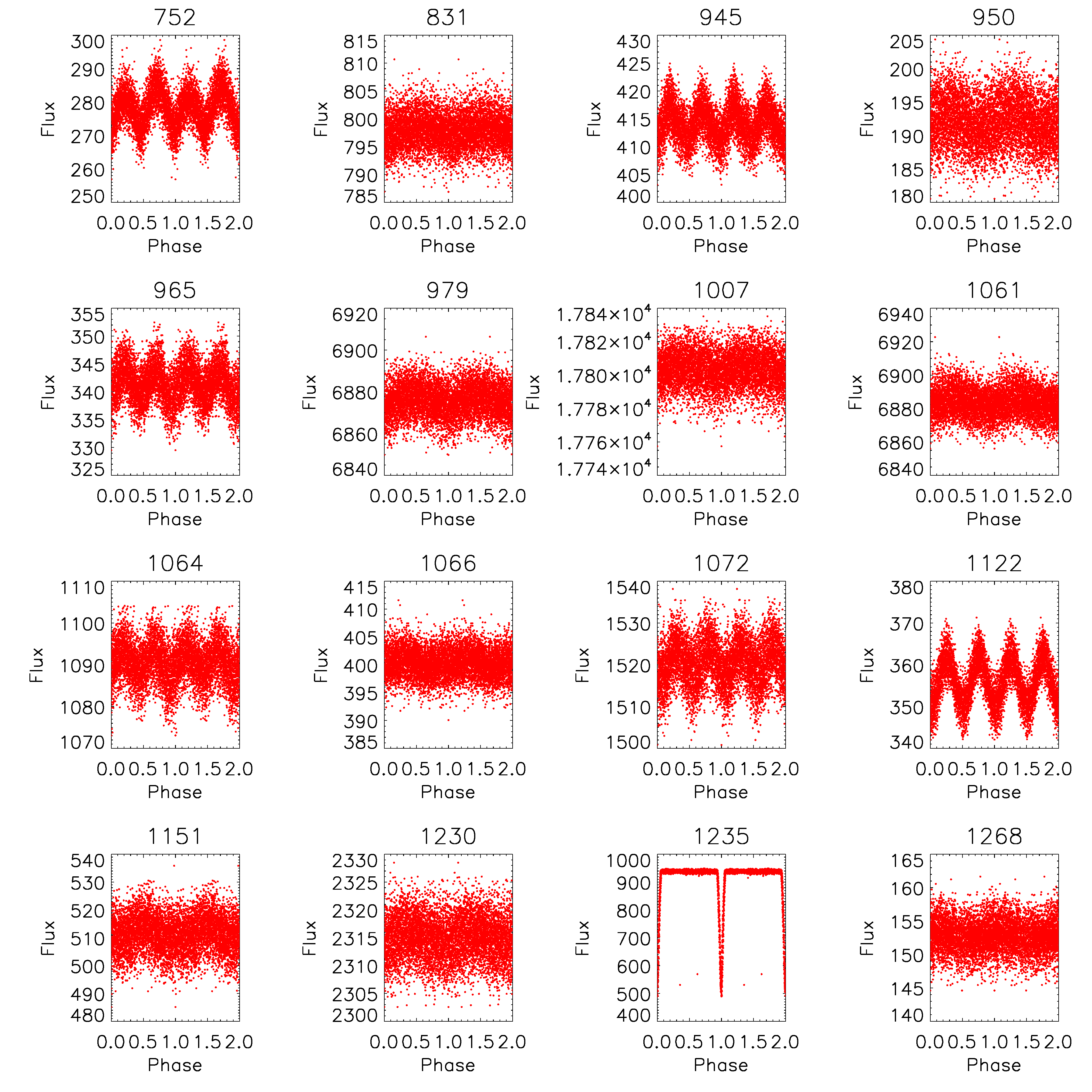}
}
\caption{The phased light curves of variable stars using TESS data.}
\end{figure*}

\begin{figure*}
\hbox{
\includegraphics[width=9cm, height=9cm]{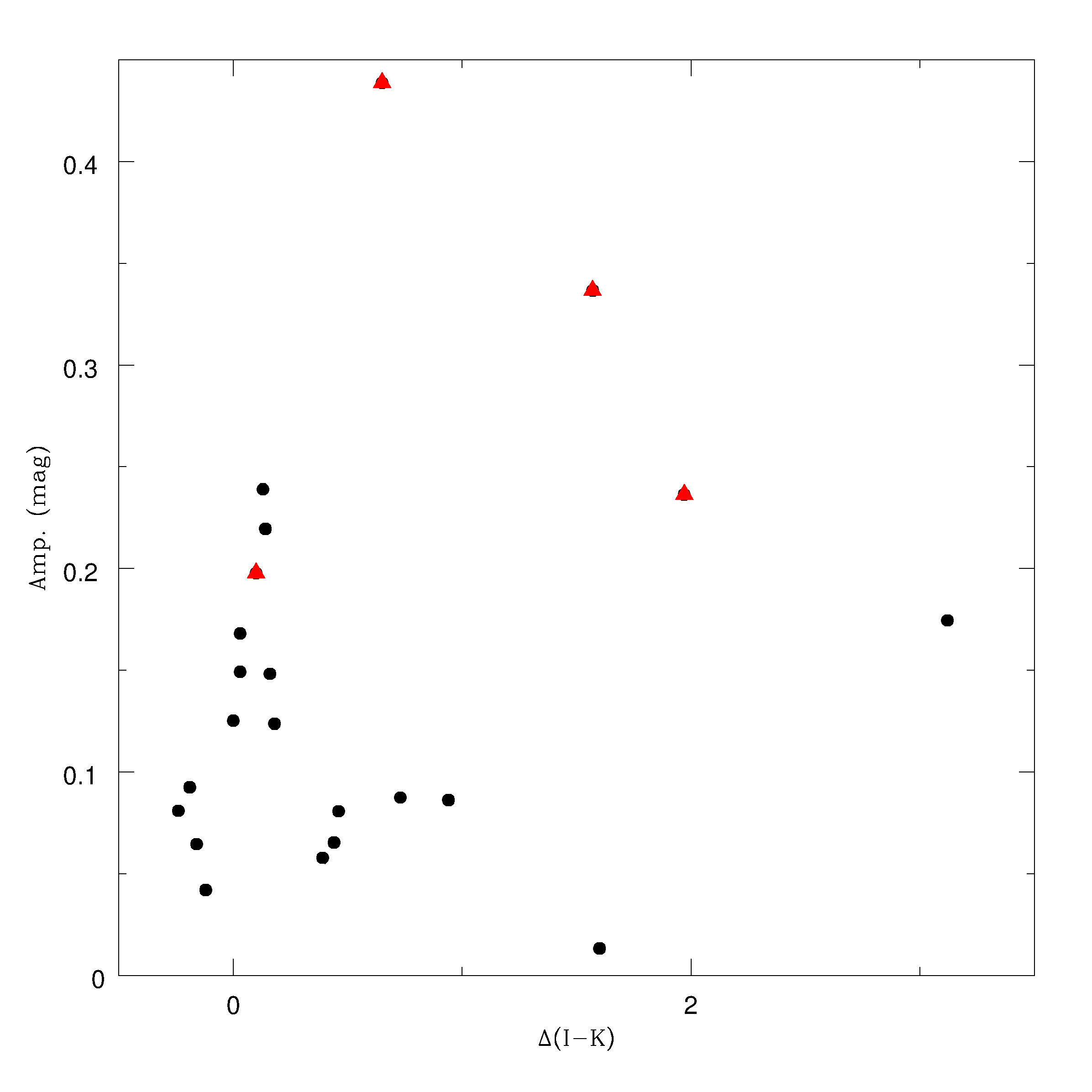}
\includegraphics[width=9cm, height=9cm]{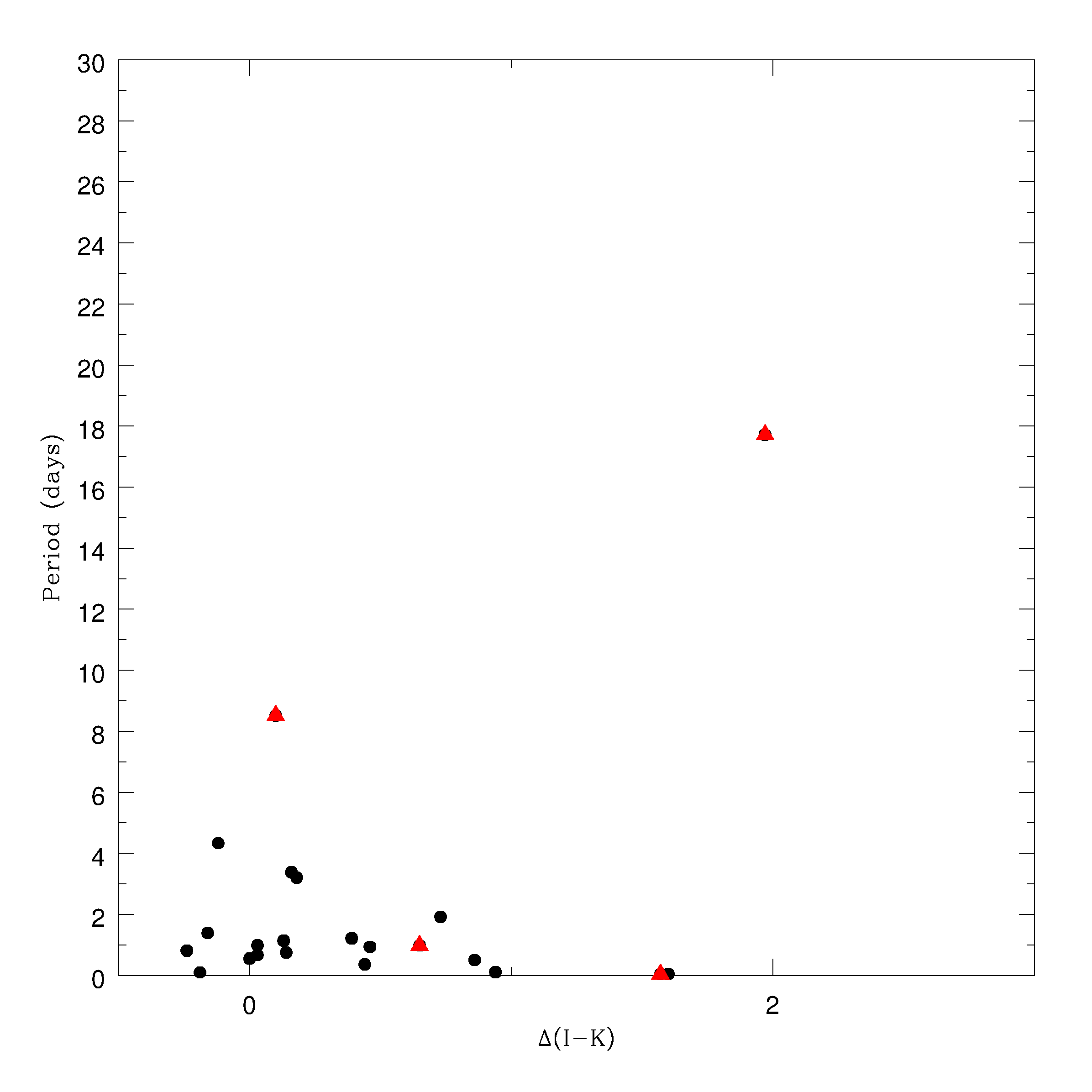}
}
\caption{Amplitude of variability and rotation period of TTSs with  $\Delta (I-K)$ is shown. 
}

\end{figure*}

\section{Correlation between circumstellar disks and variability }

Accretion onto the stellar surface creates hotspots that brighten the light curve up to 3~mag, whereas 
the magnetic field is responsible for cool and therefore dark spots.  Herbst et al. (1994) studied photometric 
variability of PMS stars in the Orion Nebula Cluster, and showed that slower rotators have larger IR excess 
than fast rotators, indicating disk locking, for which the angular momentum is transported through magnetic field 
lines from the central star to the circumstellar disk.  This supported the results by Edwards et al. (1993) for which 
low-mass young stars with accretion disks have periods more than 4~days, whereas stars without have periods ranging 
from 1.5 to 16~days. Rotation seems to be regulated after the disk is dissipated, as the star spins up while 
contracting towards the MS (Bouvier et al. 2007).  These results support the magnetic-disk model which controls PMS 
winds and angular momentum of young stellar objects during the PMS evolution. 

Models for a disk-star interaction (Ostriker \& Shu 1995, Shu et al. 1994, Ghosh \& Lamb 1979) are supported by 
the rotation periods of PMS objects in young open star clusters (Attridge \& Herbst 1992, Herbst et al. 2001, 2002, 2004). 
Kearns \& Herbst (1998) and Nordhagen et al. (2006) determined the rotation periods in two clusters. 
James et al. (2010) derived light-curve periods of sun-like sources in the young cluster NGC\,1039.  
Lamm et al. (2004) presented the rotation period of PMS objects, which supports the disk-locking mechanism in young stars. 
Broeg et al. (2006) measured rotational periods of young objects to understand the star formation scenario that 
the off-cloud young sources should rotate faster if these objects were ejected from the cloud.
They did not find significant period distribution off-cloud weak-lined TTS south of Taurus-Auriga with respect to weak-lined TTS inside the Taurus-Auriga molecular cloud. Godoy-Rivera (2021) studied stellar rotation and found that the distribution of period with mass in the case of open clusters gives important constraints to study angular momentum evolution and it is evident that spin down process depends on the mass.  The rotation periods of the members of cluster have been presented, which are found to be in range from 0.5~days up to 11.5~days (Meibom et  al. 2009, 2011).  Gondoin (2018) concluded that the stellar rotation evolution in open star clusters could be from loss of angular momentum, which occurs due to strong winds during the early evolution of young solar type stars. 

A disk-bearing YSO spins down due to magnetic braking (Koenigl 1991; Ostriker \& Shu 1995). The increasing disk fraction 
with rotation period in open clusters was reported by Cieza \& Baliber (2007).  As discussed above, to explore a correlation 
between the variability of classical TTSs with color excess, mass, and age, we have plotted amplitude of variability with $\Delta(I-K)$ excess in Fig.~19 (left panel), while the right panel of Fig.~19 shows the rotation period with $\Delta(I-K)$ excess.  Here the $\Delta(I-K)$ excess of PMS sources is determined using the following relation,

$\Delta(I-K) = (I-K)_{obs}-(A_{I}-A _{K})- (I-K)_{0}$, 

\noindent where $(I-K)_{obs}$ and $(I-K)_{0}$ are the observed and intrinsic colors of stars, whereas $A_{I}$  and $A_{K}$ denote the interstellar extinction in the $I$ and $K$ bands, respectively. To estimate the value of $(I-K)_{0}$ of YSOs, it was necessary to estimate their masses and ages. These values are available for only 22 PMS objects from the $V$ versus $V-I$ CMD after comparing with the theoretical models of Siess et al. (2000).  Of the 22 sources, there are 4 classical TTSs for which we could estimate the age and mass.  
The $A_{I}$ and $A_{K}$ are estimated using the relations given by Cohen et al. (1981) by adopting $A_{V} = 2.24$~mag. 
The ($I-K)_{0}$ value is obtained from the PMS evolutionary models of Siess et al. (2000) of a given mass and age.  
Fig.~19 (left panel) shows that a larger $\Delta(I-K)$ value for classical TTSs corresponds to a relatively larger amplitude 
of variability, consistent with those found in the literature, though we find no clear correlation 
between $\Delta(I-K)$ and rotation period.  
However one classical TTS No. 655 with larger $\Delta(I-K)$ excess is found to be rotating with a longer period. 

\section {Summary}
This work presents 88 variable stars in the young star cluster NGC\,6823. The association of detected variables to the cluster has been discussed with the Gaia kinematic data, and the optical and NIR TCDs and CMDs. 
The membership of previously known variables has also been discussed. We have detected 48 stars as PMS stars, of which 
eight are classified as classical TTSs while 36 and 4 as weak-lined TTSs and Herbig Ae/Be stars, respectively.
Three known variables H30, V2 and V8  are found to PMS variables as suggested by Pigulski et al. (2000) while two stars BL 50 and HP 57 previously detected as PMS $\delta$ Scuti pulsators are turned out to be MS members of the cluster from their proper motion, parallax values and positions on the TCDs and CMDs.   
  TTSs have periods ranging from 0.01~days to 30~days, and amplitudes of brightness variation from 0.05~mag to 0.7~mag, with the classical TTSs varying generally with larger amplitudes than weak-lined TTSs do.
It is noted that 3 of the 4 classical TTSs with larger values of the disk indicator ($\Delta(I-K)$) are found to have relatively larger amplitude variation.
The present results do not support the disk-locking mechanism,
 however one classical TTS having large $\Delta(I-K)$ is found to be rotating slowly.
 In addition, we have identified 25 stars to be MS variables (SPB stars, $\delta$ Scuti, $\beta$ Cephei and new class variable stars). Their variability has been characterized based on the period, amplitude, shape of the light curves, and location on the $H-R$ diagram.  Fifteen variable stars may belong to the field star population.   

\section{Acknowledgments}
We are thankful to Prof. E. L. Mart\'in for the valuable suggestions that improved scientific
content of the present work. 
Late Dr. A. K. Pandey facilitated this collaboration 
project as Director of ARIES during WPC's visit.  
He will be forever remembered. SL will always be grateful to him for all the support and encouragement.
We acknowledge the assistance of Michael Schwartz who 
managed the Tenagra Observatory in acquisition of the images 
of this study. ASH and JCP thanks Ministry of Innovation Development of Uzbekistan and Department of Science and Technology of India for financing the joint project (Project References: UZB-Ind-2021-99 \& INT/UZBEK/P-19).
This publication makes use of data products from the 2MASS, which is a joint project of the University of Massachusetts and the Infrared Processing and Analysis Center/California Institute of Technology, funded by the National Aeronautics and Space Administration and the National Science Foundation.
This paper includes data collected by the TESS mission. Funding for the TESS mission is provided by the NASA's Science Mission Directorate. 
We also acknowledge "Galactic Legacy Infrared Midplane Survey Extraordinaire" (GLIMPSE) Legacy Program for Spitzer IRAC data.
This work also used data from the European Space Agency (ESA) space mission Gaia. Gaia data are being processed by the Gaia Data Processing and Analysis Consortium (DPAC). Funding for the DPAC is provided by national institutions, in particular the institutions participating in the Gaia MultiLateral Agreement (MLA). The Gaia mission website is https://www.cosmos.esa.int/gaia. The Gaia archive website is https://archives.esac.esa.int/gaia.

\section*{Availability of data}

The data underlying this article will be shared upon request to the corresponding author.
The 2MASS data are available at https://vizier.u-strasbg.fr/viz-bin/VizieR?-source=II/246.
The Gaia and Spitzer IRAC data are obtained from https://gea.esac.esa.int/archive/ and  https://irsa.ipac.caltech.edu/cgi-bin/Gator/nph-scan?submit=Select\&projshort=SPITZER, respectively. We used the following links https://archive.stsci.edu/hlsp/eleanor and https://adina.feinste.in/eleanor/ to obtain TESS data.

\bibliographystyle{mn2e}

\begin{thebibliography}{36}
\bibitem[]{}Andre Philippe, Ward-Thompson Derek, Barsony Mary, 1993, ApJ, 406, 122 
\bibitem[]{}Appenzeller I., Mundt R, 1989A\&ARv, 1, 291A 
\bibitem[]{}Attridge Joanne M., Herbst William, 1992, ApJ, 398, 61 
\bibitem[]{}Bailer-Jones C. A. L., Rybizki J., Fouesneau M., Demleitner M., Andrae R., 2021, AJ, 161, 147 
\bibitem[]{}Barrado y Navascu\'es D., Zapatero Osorio M. R., B\'ejar V. J. S., Rebolo R., Mart\'in E. L., Mundt R., Bailer-Jones, C. A. L., 2001, A\&A, 377, 9 
\bibitem[]{}Bessell M. S., Brett J. M., 1988, PASP, 100, 1134
\bibitem[]{}Bica E., Bonatto C., Dutra C. M., 2008, A\&A, 489, 1129
\bibitem[]{}Bouvier J., Cabrit S., Fern\'andez M., Mart\'in E. L., Matthews J. M., 1993, A\&AS, 101, 485
\bibitem[]{}Bouvier J., Alencar S. H. P., Boutelier T., Dougados C., Balog Z., Grankin K., Hodgkin S. T., Ibrahimov M. A., Kun M., Magakian T. Yu., Pinte C., 2007, A\&A, 463, 1017 
\bibitem[]{}Broeg C., Joergens V.,  Fernandez M., et al., 2006, A\&A, 450, 1135
\bibitem[]{}Canto Martins B. L., Gomes R. L., Messias Y. S, 2020, ApJS, 250, 20
\bibitem[]{}Cohen J. G., Persson S. E., Elias J. H., Frogel J. A., 1981, ApJ, 249, 481
\bibitem[]{}Cohen R. E., Sarajedini A., 2012, MNRAS, 419, 342  
\bibitem[]{}Cutri R. M.  et al., 2003 , 2MASS All Sky Catalog of Point Sources, VizieR Online Data Catalog, University of Massachusetts and Infrared Processing and Analysis Center (IPAC/California Institute of Technology), 2246, 0 https://vizier.u-strasbg.fr/viz-bin/VizieR?-source=II/246
\bibitem[]{}Cantat-Gaudin T., Anders, F. 2020, A\&A, 633, A99. doi:10.1051/0004-6361/201936691
\bibitem[]{}Edwards Suzan, Strom Stephen E., Hartigan Patrick, Strom Karen M., Hillenbrand Lynne A., Herbst William, Attridge Joanne, Merrill K. M., Probst Ron, Gatley Ian, 1993, AJ, 106, 372
\bibitem[]{}Erickson R.R., 1971, Astron. Astrophys., 10, 270.
\bibitem[]{}Feinstein Adina D., Montet Benjamin T., Foreman-Mackey Daniel, 2019, PASP, 131, i4502
\bibitem[]{}Finkenzeller U, Mundt R., 1984, A\&A Supp, 55, 109
\bibitem[]{}Gaia Collaboration et al. 2018, A\&A, 616, 13 https://gea.esac.esa.int/archive/
\bibitem[]{}Girardi L., Bertelli G., Bressan A., Chiosi C., Groenewegen M. A. T., Marigo P., Salasnich B., Weiss A., 2002, A\&A, 391, 195
\bibitem[]{}Ghosh P., Lamb F. K., 1979, ApJ, 234, 296
\bibitem[]{}Godoy-Rivera Diego, Pinsonneault Marc H., Rebull Luisa M., 2021, arXiv210101183G
\bibitem[]{}Gondoin P., 2017, A\&A, 616, 154
\bibitem[]{}Gutermuth R. A.  et al., 2008 , ApJ , 674 , 336
\bibitem[]{}Guetter H.H., 1992, Astron. J., 103, 197
\bibitem[]{}Herbst W., Herbst D. K., Grossman E. J., Weinstein D. 1994, AJ,108, 1906
\bibitem[]{}Herbst W., Booth J. F.,  Koret D. L., et al., 1987, AJ, 94, 137
\bibitem[]{}Herbst William, Hamilton Catrina M., et al., 2002, PASP, 114, 1167
\bibitem[]{}Herbst W., Bailer-Jones C. A. L., Mundt R., Meisenheimer K., Wackermann R., 2002, A\&A, 396, 513
\bibitem[]{}Herbst W., Bailer-Jones C. A. L., Mundt R., 2001, ApJ, 554, 197
\bibitem[]{}Herbst W., Rhode K. L., Hillenbrand L. A., Curran G., 2000, AJ, 119, 261
\bibitem[]{}Hojaev A. S., Chen W. P., Lee H. T., 2003, A\&AT, 22, 799 
\bibitem[]{}Huang P. C., et al., 2019, ApJ, 871, 183
\bibitem[]{}James D. J., Barnes S. A., Meibom S., et al., 2010, A\&A, 515, 100
\bibitem[]{}Johnstone D., et al. 2018, ApJ, 854, 31
\bibitem[]{}Joy Alfred H., 1945, ApJ, 102, 168
\bibitem[]{}Kearns Kristin E., Herbst William, 1998, AJ, 116, 261
\bibitem[]{}Lada C.~J., 1987, Star Forming Regions, IAUS\,115, 1
\bibitem[]{}Lamm M. H., Bailer-Jones C. A. L., Mundt R., Herbst W., Scholz A. 2004, A\&A, 417, 557
\bibitem[]{}Knote M. F., Caballero-Nieves Saida M., Gokhale Vayujeet, et al., 2022, arXiv220604142K  
\bibitem[]{}Mart\'in E. L., Brandner W., Bouvier J, Luhman K. L., Stauffer J.,  Basri G., Zapatero Osorio M. R., Barrado y Navascu\'es D., 2000, ApJ, 543, 299
\bibitem[]{}Massey Philip, Johnson Kelsey E., Degioia-Eastwood Kathleen, ApJ, 1995, 454, 151
\bibitem[]{}Meibom S., Mathieu Robert D., Stassun Keivan G., 2009, ApJ, 695, 679
\bibitem[]{}Meibom S.; Mathieu Robert D., Stassun Keivan G., Liebesny Paul, Saar Steven H., 2011, ApJ, 733, 115
\bibitem[]{}Meyer M. R., Calvet N., Hillenbrand L. A., 1997, AJ, 114 , 288
\bibitem[]{}Morales E.~F.~E., Wyrowski F., Schuller F., et al., 2013, A\&A, 560, A76. doi:10.1051/0004-6361/201321626
\bibitem[]{}Morales-Calder{\'o}n M., et al., 2011, ApJ, 733, 50	
\bibitem[]{}Nordhagen Stella, Herbst William, Rhode Katherine L., Williams Eric C, 2006, AJ, 132, 1555
\bibitem[]{}Ostriker Eve C., Shu Frank H., 1995, ApJ, 447, 813
\bibitem[]{}O'Connell D. J. K., 1951, Publications of the Riverview College Observatory, 2, 85 
\bibitem[]{}Pandey J. C., Karmakar S., Joshi A., Sharma Saurabh, Bhushan Pandey S., Pandey A. K., 2019, RAA, 19, 7
\bibitem[]{}Pedrosa Antonio, 1997, IAUS, 182, 306
\bibitem[]{}Pigulski A., Kolaczkowski Z., Kopacki G., 2000, AcA, 50, 113
\bibitem[]{}Ricker George R., Winn Joshua N, Vanderspek R., et al.,  2015, Journal of Astronomical Telescopes, Instruments, and Systems, 1, 014003
\bibitem[]{}Rangwal Geeta, Yadav R. K. S., Durgapal Alok K., Bisht D., 2017, PASA, 34, 68
\bibitem[]{}Riaz B., Mart\'in E. L., Tata R., Monin J. -L., Phan-Bao N., Bouy H., 2012, MNRAS, 419, 1887
\bibitem[]{}Sagar R., Joshi U.C. 1981, Astrophys. Space Sci., 75, 465
\bibitem[]{}Siess L., Dufour E., Forestini M., 2000, A\&A, 358, 593
\bibitem[]{}Shu Frank, Najita Joan, Ostriker Eve, Wilkin Frank, Ruden Steven, Lizano Susana, 1994, ApJ, 429, 781
\bibitem[]{}Stone D. G., 1979, Astron. J., 96, 1389
\bibitem[]{}Shi H.M.,  Hu, J.Y. 1999, Astron. Astrophys. Suppl. Ser., 136, 313.
\bibitem[]{}Stetson P.~B, 1992, J. R. Astron. Soc. Can., 86, 71
\bibitem[]{}Stetson P.~B., 1987, PASP, 99, 191
\bibitem[]{}Torres G., 2010,  AJ, 140, 1158
\bibitem[]{}Turner D. G., 1979, JRASC, 73, 74
\bibitem[]{}Zahajkiewicz E., 2012, AN, 333, 1086
\end{thebibliography}

\end{document}